\documentclass[smallextended]{svjour3} 

\smartqed 
\usepackage{array}
\usepackage{colortbl}
\usepackage{blindtext}
\usepackage{NJDnatbib}
\usepackage{graphicx} 
\usepackage{url} 
\usepackage{amsmath}
\usepackage{amssymb}
\usepackage{xcolor}
\usepackage{multirow}
\usepackage[pass]{geometry}
\usepackage{datatool}
\usepackage{hyperref}
\DTLsetseparator{=}
\DTLloaddb[noheader, keys={thekey,thevalue}]{studyvariables}{variables.tex}

\newcommand{\nvar}[1]{\DTLfetch{studyvariables}{thekey}{#1}{thevalue}}
\newcommand{\nvarc}[2]{#2} % ToDo: check if \DTLfetch{studyvariables}{thekey}{#1} == #2

\newcommand{\new}[1]{{\color{black}#1}}

\renewcommand\bibsection{%
       \section*{\bibname\markright{\MakeUppercase{\bibname}}}}

\newcommand{\responsetoreviewer}[3][black]{%
\expandafter\gdef\csname #2\endcsname{#3}%
\label{rev:#2}%
{\color{black}{#3}}\color{#1}\xspace}
\makeatother

\begin{document}
\begin{filecontents*}{variables.tex}
projects=362
projects_with_dbms=202
slices_smallest_project=10
slices_largest_project=8,143
percentage_smallest_project=10
total_committs=3,128,740
total_slices=37,964
days_to_run=325
projects_with_orm=241
projects_with_dbms_history=234
rq3_no_removals=ClickHouse (11 projects), MicrosoftAzureCosmosDB (7 projects), Singlestore (4 projects), Kdb (3 projects), GoogleCloudFirestore (3 projects), Riak\_KV (2 projects), Ignite\_Sql (2 projects), MicrosoftAzureTableStorage (1 project), and Impala (1 project)
\end{filecontents*}

\title{Analyzing the Adoption of Database Management Systems Throughout the History of Open Source Projects}
\titlerunning{Adoption of DBMSs Throughout the History of OSS}

\author{Camila A. Paiva \and 
Raquel Maximino \and
Frederico Paiva \and
Rafael Accetta Vieira \and
Nicole Espanha \and 
João Felipe Pimentel \and
Igor Wiese \and
Marco Aurélio Gerosa \and
Igor Steinmacher \and
Leonardo Murta \and
Vanessa Braganholo}

\institute{Camila Paiva, Raquel Maximino, Frederico Tosta, Rafael Vieira, Nicole Espanha, João Felipe Pimentel, Leonardo Murta, Vanessa Braganholo \at
Instituto de Computação, Universidade Federal Fluminense, Brazil\\
\email{\{jpimentel,leomurta,vanessa\}@ic.uff.br}
\and
Igor Wiese \at Universidade Tecnológica Federal do Paraná, Brazil\\
\email{igor@utfpr.edu.br }
\and
Marcos Aurélio Gerosa, Igor Steinmacher \at
Northern Arizona University, USA\\
\email{\{marco.gerosa,igor.steinmacher\}@nau.edu }}

\date{Received: Feb. 2024 / Accepted: XXX}

\maketitle

\begin{abstract}
The appropriate selection of DBMSs (Database Management Systems) is relevant for the success of modern software applications. Relational DBMSs are popular for structured data management, while non-relational systems, such as NoSQL databases, have gained traction for handling unstructured data and scaling in dynamic environments. These varying DBMS characteristics have led to an increasing trend of combining multiple systems within a single application to meet diverse requirements. However, existing work does not analyze whether DBMS are replaced or used together in a broad scope.
This paper presents an empirical study on DBMS usage across \nvarc{rq2_projects}{362} popular open-source Java projects hosted on GitHub. Our analysis focuses on the most widely adopted DBMSs, both relational and non-relational, as ranked by the DB-Engines website. By examining DBMS integration patterns, stability, and migration trends, we aim to uncover insights into the factors driving DBMS choices in real-world applications.
We investigated DBMS popularity, usage stability, migration patterns, synergy among DBMS, and the role of Object-Relational Mappers (ORMs) in DBMS interactions. We applied heuristics to detect DBMS presence, tracked usage trends over time, and analyzed the coexistence and replacement of different systems. We also examined ORM frameworks to understand their impact on DBMS management and query-building practices.
Our findings reveal that MySQL and PostgreSQL are the most popular DBMSs, although some projects replace them with other DBMSs.
While certain popular DBMSs (e.g., Redis, MongoDB) usually stay in the project after they are introduced (and therefore their adoption is stable), others (e.g., HyperSQL) are frequently replaced as project requirements evolve. We also observed patterns of polyglot persistence, where multiple DBMSs coexist to handle varied data types. Notably, Informix is a relational DBMS designed to handle real-time data processing and is always used with other DBMSs. Additionally, we identified ORM usage trends that facilitate database interactions and mitigate migration complexities. These insights contribute to a broader understanding of DBMS adoption, providing valuable guidance for developers and architects in selecting and managing database infrastructure over time. %Educators can also leverage these insights to adjust their curricula according to industry trends. Finally, DBMS vendors may adapt their tools to improve the interoperability among DBMSs based on their synergies.

\end{abstract}

\section{Introduction}
\label{introduction}

The growth and complexity of data generated by modern applications have resulted in the development of numerous database management solutions. However, selecting the most appropriate Database Management System (DBMS) for a given project is increasingly challenging due to the diversity of available systems, each optimized for different use cases~\citep{gessert2017nosql}. With applications having diverse and evolving requirements, no single DBMS can adequately serve all needs~\citep{cattell2011scalable}, leading developers to often combine multiple DBMSs to meet various demands for reliability, scalability, and performance.

Understanding how DBMSs are adopted, replaced, or used in tandem over time is necessary for making informed decisions about database infrastructure. Relational DBMSs, known for their reliability, stability, and support, continue to be widely used, but non-relational DBMSs (e.g., NoSQL) have gained prominence for handling unstructured data, such as key-value, column-oriented, document-based, or graph-based information, in scalable environments~\citep{cattell2011scalable}. This variety of systems enables developers to tailor solutions to specific project needs, often leading to the coexistence of different DBMSs within a single application~\citep{cattell2011scalable, sahatqija2018comparison}.

Moreover, as applications evolve, their database requirements change in response to shifts in project scope, growth, and performance demands. It is, therefore, important to understand the dynamics of DBMS usage across the history of projects, as this can reveal important trends in how databases are adopted, maintained, and eventually replaced \citep{gessert2017nosql}. For example, a simple DBMS might suffice early in a project’s lifecycle, but as the project scales or new features are introduced, more robust or specialized systems may be required \citep{cattell2011scalable}. In some cases, multiple DBMSs may need to coexist within a single application to handle different types of data or workloads \citep{sahatqija2018comparison}. These transitions are not always straightforward, and an appropriate selection or migration to a specific DBMS can significantly impact a project’s success.

Object-Relational Mapping (ORM) frameworks facilitate the interaction between the application and the DBMS. ORMs abstract the differences between object-oriented programming languages and relational databases, helping developers manage complex data structures with fewer lines of code and minimizing the friction when databases are updated or replaced. \responsetoreviewer{RIICVI}{This not only simplifies the development process but also ensures that changes in the database layer are seamlessly propagated to the application layer, and vice versa, reducing the likelihood of errors and improving maintainability \citep{KeithMike2018PJ2i, kleppmann2017designing}}. As such, understanding how applications interact with DBMS through ORM frameworks helps anticipate the impact of eventual DBMS changes.

Previous research has explored various aspects of DBMS adoption to better understand its role in real-world applications. For instance, \cite{Lyu_2017} investigated DBMS usage in Android apps, identifying eight prominent systems, including SQLite, Oracle, and MongoDB, and noting that 40.8\% of the apps did not use any DBMS at all. Other studies have examined the relationship between application code and database schema over time \citep{qiu2013empirical, Goeminne_2014, Linares_2015}, while some \citep{scherzinger2020empirical, dimolikas2020study, vassiliadis2021profiles} have focused specifically on the evolution of database schemas throughout a project’s history. Additionally, research by \cite{Goeminne_2015}, \cite{yan2017}, and \cite{yang2018} addressed the performance, scalability, and replacement of ORM frameworks. While these studies provide valuable insights into specific contexts -- such as mobile apps, web applications, or schema evolution -- they tend to be limited in scope, concentrating on narrow subsets of DBMS usage or specific environments.

This research aims to comprehensively understand how DBMSs are utilized in practice across various software projects. Considering that the adoption of DBMS may change over time, we analyzed the usage of DBMSs in \nvarc{rq2_projects}{362} popular open-source projects developed in Java and hosted on GitHub. Our focus was on the 50 most widely adopted DBMSs, as ranked by the DB-Engines website \citep{db-engines_2022}, encompassing relational and non-relational data models. Through this examination, we sought to provide a detailed analysis of how DBMSs are integrated into applications, offering insights into their popularity, long-term stability, migration patterns, synergies among DBMS, and how applications interact with them. To guide this investigation, we formulated five research questions that progressively build on one another, each delving into a different facet of DBMS usage in real-world software development.

This study begins by addressing \textbf{which DBMSs are the most popular across open-source software projects} (RQ1). To answer this, we developed a set of heuristics (see Section~\ref{sec:databaseHeuristics}) to identify the presence of specific DBMSs within the source code of each project in our corpus. By counting their occurrences, we could rank the DBMSs based on their frequency of use, offering a comprehensive view of which systems dominate the landscape. Following this, we investigated \textbf{the stability of DBMS usage throughout the projects’ history} (RQ2). While RQ1 provided a snapshot, RQ2 delves into longitudinal data, tracking how DBMS usage persists or fluctuates over time. In cases where stability is disrupted, we extended our analysis to explore \textbf{the migration patterns of DBMSs} (RQ3). \responsetoreviewer{RIICVII}{Using sequential pattern mining, we uncovered recurring sequences in which certain DBMSs are replaced by others, either simultaneously (at the same commit) or over a defined small transition period (a window of 100 commits).}

In some cases, however, rather than being replaced, DBMSs coexist. To understand these instances, we examined \textbf{which DBMSs tend to have more synergy during the project history} (RQ4). \responsetoreviewer{RIICVIII}{By employing association rules, a data mining technique, we uncovered patterns of coexistence where multiple DBMSs are used in tandem within the same project, revealing polyglot persistence \citep{fowler2011polyglot}}. \responsetoreviewer{rIIcIX}{Finally, after gaining a deeper understanding of the adoption of DBMSs in the projects, we investigated in detail \textbf{how applications interact with these DBMSs}, both via Object-Relational Mapping (ORM) and through direct SQL queries (RQ5).} Given the variety of Object-Relational Mappers (ORMs) available, we analyzed each project’s source code to identify the ORMs in use, assess how many database-related files exist, and determine whether queries are built using query builders or raw SQL strings.

We found that MySQL and PostgreSQL are the most popular relational DBMSs (RQ1, RQ2) and the most popular overall, currently used in 55.9\% and 46\% of our corpus, respectively. Although around 10\% of the projects removed them at certain points in their history, both MySQL and PostgreSQL maintained or increased in popularity over time, as they replaced other DBMSs in 13.1\% and 23.5\% of the projects that used them. We also observed relational DBMSs that lost popularity, such as HyperSQL, which was removed from 23.5\% of the projects and only replaced other DBMSs in 13.6\% of the projects \new{(RQ3)}. This trend was further confirmed in the synergy analysis (RQ4), which revealed that HyperSQL was often combined with other DBMSs, like MySQL and MS SQL Server, early in projects. \responsetoreviewer{rIIcX}{However, towards the end of the project history, as analyzed by our methodology, MySQL, PostgreSQL, H2, and Oracle were more frequently found together, becoming the most common combinations of DBMSs we found.} The synergy analysis also revealed that Informix is used as a complementary DBMS, as it is always combined with other DBMSs, likely due to its capabilities in handling real-time data processing.
In the non-relational category, Redis and MongoDB emerged as the most popular and stable, with relatively few replacements, indicating greater stability in their adoption. Redis is currently used by 39.6\% of our corpus, replacing other DBMSs in 24.4\% of the projects that used it and being replaced by other DBMSs in 7.8\%, while MongoDB currently appears in 22.8\% of the corpus, replacing DBMSs in 27.5\% and being replaced in 7.8\% of the projects that used it.
\responsetoreviewer{rIIcXI}{Additionally, our analysis of ORM usage (RQ5) revealed that ORM frameworks are typically used in a small percentage of files, with EclipseLink requiring the fewest files to adopt it in a project.}

Our findings offer several implications for software engineers, educators, and DBMS vendors. For software engineers, the insights on DBMS popularity, stability, and synergy within specific application domains can aid in selecting DBMSs with more learning resources, robust communities, and minimized migration risks. Similarly, our results on ORM usage suggest that engineers may benefit from assessing the implementation effort required for each ORM solution, opting for simpler options based on project needs and constraints. Educators, meanwhile, could leverage these trends to align curricula with practices, preparing students to work with multiple DBMSs and ORM tools, as well as to navigate data migration tasks effectively. Lastly, DBMS vendors may use this information to enhance the interoperability of their products with other popular DBMSs and ORM frameworks, thereby broadening their applicability across diverse software ecosystems.

The remainder of this paper is organized as follows. Section \ref{sec:materials} details the corpus selection, while Section \ref{sec:method} presents the research methodology. Section \ref{sec:results} describes our findings, which are discussed in Section \ref{sec:discussion}. Section \ref{sec:validity} discusses the threats to the validity of our study. Section \ref{relatedWork} discusses related work. Finally, Section \ref{final_considerations} concludes our work and discusses some future work.

\section{Materials} 
\label{sec:materials}
 
To answer our research questions, we need three different corpora: first, we need a set of representative Java Open Source projects to serve as our main research subject; second, we need a set of DBMS that we would try to find in those projects (RQ1, RQ2, RQ3, RQ4); and third, we need a set of ORM that we would also try to find in those projects (RQ5). This section describes these three corpora. We start by describing the DBMS (Section \ref{sec:dbcorpus}) and the ORM corpus (Section \ref{sec:ORM}). Finally, Section~\ref{sec:corpus} describes how we selected the projects of our corpus.  

\subsection{Database Corpus \label{sec:dbcorpus} }

As our DBMS corpora, we selected 50 popular DBMSs listed in the February 2022 DB-Engines ranking \citep{db-engines_2022}, which ranks DBMSs according to their popularity. The list covers the top 13\% of the 383 DBMS listed in DB-Engines on that date. The reason we do not use the complete list is twofold. First, we had to carefully understand how to use each of the DBMSs in this list in Java projects. This is because we cannot search the projects for their usage if we do not know how they are used. Second, we believe the percentage of DBMS we chose is representative enough of the most popular DBMS.  
To produce the ranking, DB-Engines measures the popularity of a DBMS on the Web by using several criteria\footnote{\url{https://db-engines.com/en/ranking_definition}}: the number of search results on Google and Bing; frequency of searches in Google Trends; the frequency of technical discussions about the systems in Stack Overflow and DBA Stack Exchange; the number of job offers that mention that system in Indeed and Simply Hired; the number of profiles in which the system is mentioned in LinkedIn; and the number of posts on X (formerly Twitter) in which the system is mentioned. The popularity is then calculated by standardizing and averaging the individual parameters. According to DB-Engines, the obtained values only make sense when comparing two or more systems. If System $A$ has a higher score than System $B$, then $A$ is more popular than $B$.

For each system, DB-Engines provides a classification such as Relational, Key-Value, Search Engine, etc.). Some systems have more than one classification -- for example, Oracle is classified as relational and multi-model, while SQLite is classified only as Relational. Whenever a system has multiple classifications, we use the first one as our classification criteria. We used this classification to guarantee that we keep in our analysis only systems that are classified as DBMS (relational or non-relational). Thus, search engines such as ElasticSearch, Splunk, and Apache SolIR were discarded. We then looked at the remaining top 50 DBMS and their descriptions on their websites. Based on their descriptions, we discarded the ones that were data warehouses (such as Apache Hive and Google BigQuery), processing systems (such as Amazon Aurora and Presto), cache systems (such as EhCache), monitoring systems (such as Prometheus), and the ones that did not offer support for Java (such as dBASE). This resulted in the discarding of 10 DBMS in total. Note that we did not analyze the complete DB-Engines list -- we analyzed just the top DBMS to guarantee we would have 50 popular DBMS in total.  

From this list, we took the first 25 Relational DBMS and the first 25 non-Relational DBMS, to ensure a balanced analysis. Six of the DBMS were classified only as Multi-Model: Virtuoso, Dynamo-DB, Microsoft Azure Cosmos DB, MarkLogic, Ignite, and ArangoBD. For those cases, we looked at the specific models they support. Whenever they were listed as Relational and something else, we counted it on both categories. This occurred with Ignite and Virtuoso. The other four were classified as purely Non-Relational since the Relational Model did not appear among the models they support. 

Four DBMSs on our list of 50 (CockroachDB, MariaDB, Microsoft Azure SQL Database, and Sybase Adaptive Server Enterprise) are highly compatible with other systems: CockroachDB is compatible with PostgreSQL, MariaDB with MySQL, Microsoft Azure SQL Database with MS SQL Server, and Sybase Adaptive Server Enterprise is compatible with SAP Adaptive Server. Notably, Sybase and SAP Adaptive Server were merged following SAP's acquisition of Sybase in 2010, which integrated their DBMS offerings. This compatibility means that using any of these pairs of DBMSs in Java projects requires the same connection strings, imports, and drivers, which makes it hard to correctly identify which one is being used by a given project (Section \ref{sec:databaseHeuristics} explains exactly how we use this info to search for the DBMS in the source code of the projects in our corpus). Consequently, we decided to merge the following pairs in our analysis: CockroachDB with PostgreSQL, MariaDB with MySQL, Microsoft Azure SQL Database with MS SQL Server, and SAP Adaptive Server with Sybase. However, since Sybase was not on our original list of 50 popular DBMS, its synergy with SAP Adaptive Server does not impact the size of our database corpus. Since we counted the remaining ones as three DBMSs instead of six, we added three more relational DBMSs so that we ended up with a list of exactly 25 relational DBMSs in our analysis. The complete list is available at \url{https://tinyurl.com/y25cm6e5}.

%%RETIRAR-RETIRAR-RETIRAR-RETIRAR-RETIRAR-RETIRAR-RETIRAR-RETIRAR-RETIRAR-RETIRAR-RETIRAR-
%Although we found some DBMSs that were a fork of another, such as MariaDB, which is a fork of MySQL, we considered them separate DBMSs since they are considered as such by the ranking.

\subsection{ORM Corpus} 
\label{sec:ORM}

%Esse conteúdo estava an seção de heuristicas de ORM. Movi pra cá pra ficar consistente.

ORMs are widely used by developers due to the simplicity of the conceptual abstraction that they provide between the source code and the database system \citep{Johnson2005}. Java provides a standard API for ORMs, which is called the Java Persistence API (JPA). JPA is an official specification that describes a generic interface between any application and an ORM. There are many implementations of JPA, such as Hibernate\footnote{\label{Hibernate}\url{https://hibernate.org/}}, OpenJPA\footnote{\label{OpenJPA}\url{https://openjpa.apache.org/documentation.html}}, and EclipseLink\footnote{\label{EclipseLink}\url{https://www.eclipse.org/eclipselink/}}. These implementations share similar designs and functionalities, although they have implementation-specific differences.

A survey \citep{JavaTechnologyReport} published in 2020 shows that 86\% of Java developers use the Spring framework, and 51\% use persistence technologies like Hibernate, OpenJPA, or EclipseLink. In our work, we selected the most popular frameworks according to \cite{JavaTechnologyReport}: Hibernate\footref{Hibernate}, JPA, MyBatis\footnote{\label{MyBatis}\url{https://mybatis.org/mybatis-3/index.html}}, Spring\footnote{\label{Spring}\url{https://spring.io/}}, EclipseLink\footref{EclipseLink}, jOOQ\footnote{\label{jOOQ}\url{https://www.jooq.org/}}, and JdbcMapper\footnote{\label{JdbcMapper}\url{https://github.com/moparisthebest/JdbcMapper}}.

\subsection{Project Corpus} 
\label{sec:corpus}

Our goal was to select popular open-source applications written in Java. We focused on a single programming language because our method for determining a project's DBMS usage depended on searching for language-specific database-related constructs within the project's source code. The choice for Java comes from its popularity. At the time of writing of this paper, it was the third most popular programming language according to the TIOBE index\footnote{\url{https://www.tiobe.com/tiobe-index/}}. 

\paragraph{Projects Selection.} We used the GitHub GraphQL API (v4) to search for all public repositories that were not forks of other repositories, had at least 1,000 stars, were not archived, and received at least one push in the last three months. According to \cite{Kalliamvakou_2014}, avoiding forks is important to guarantee that the corpus contains only one repository per project. Since we are aware of some forks that are much more successful than the original forked projects, we checked our initial sample for forked projects that met our selection criteria and found none. Besides that, restricting the number of stars to at least 1,000 indicates that our corpus contains relevant and popular repositories \citep{Borges_2018}. Finally, avoiding archived repositories or repositories that did not receive pushes in the last three months ensures a certain degree of activity in all repositories of our corpus. We did not filter out mirror repositories as our analyses do not focus on GitHub-specific features. Thus, replicas of external repositories stored in GitHub are welcome. This search was performed on March 27, 2021, and returned 21,149 repositories. 

Afterward, we analyzed the metadata of these 21,149 repositories to perform additional filters on the number of contributors (10 or more) and the number of commits (1,000 or more). Filtering out repositories with less than ten contributors aims at avoiding personal or coursework projects in our corpus \citep{Kalliamvakou_2014}. Moreover, restricting the number of commits to 1,000 or more is an attempt to remove immature or short-term projects from our corpus. After applying these filters, our corpus was reduced to 6,708 repositories. 
GitHub classifies these 6,708 repositories as using ten different primary programming languages. We used this information to filter projects written in Java, which is the focus of this paper. This resulted in a corpus with 633 repositories.

Since the search was conducted in 2021, we updated the corpus before doing the analysis presented in this paper. The update was performed in September 2024. We removed projects that were inactive in the previous 90 days, that had been moved to other repositories outside GitHub, or that had been archived or deprecated since our first search in 2021. We also removed duplicated projects and projects that changed the main programming language to something other than Java. This resulted in the discarding of 185 projects. 

We then proceeded to remove the ones that were listed as database engines on the DB Engines website (11 projects). Examples of excluded projects due to that reason are Presto and HBase. The reason for that exclusion is the fact that we are looking to find out how open-source projects use databases, so inspecting the source code of a DBMS itself would introduce bias. For the same reason, we removed projects that correspond to the ORMs we investigated on RQ5 (9 projects). This resulted in a corpus of 428 projects.

We then manually inspected the projects. For each project, four authors examined the GitHub repository and the project web page (when available) to answer two questions: (i) Is this repository documented in English? (ii) Does this repository contain a software project? %(iii) Does this repository contain an end-user application? 
%(iii) What is the project domain? 
The answers to these questions were discussed between two authors and inspected and revised by two other authors. 
The first question aimed to guarantee that the authors could understand the documentation of the projects. The second question aimed to refine the primary programming language's automatic filter, removing projects that were merely documentation (for instance, projects that contain snippets of Java code to solve a set of problems). 
%The third question aimed at keeping only applications, removing projects that are not focused on end users, such as frameworks, libraries, programming languages, compilers, and interpreters. 
%The third question aimed at identifying the general application domain of the project. 
%
After answering these two initial questions and dismissing those projects that are not documented in English and not software projects, %and not end-user applications, 
our corpus was reduced to 390 repositories, which we cloned (in September 2024). 

\begin{figure}
\begin{center}
  \includegraphics[width=.7\textwidth]{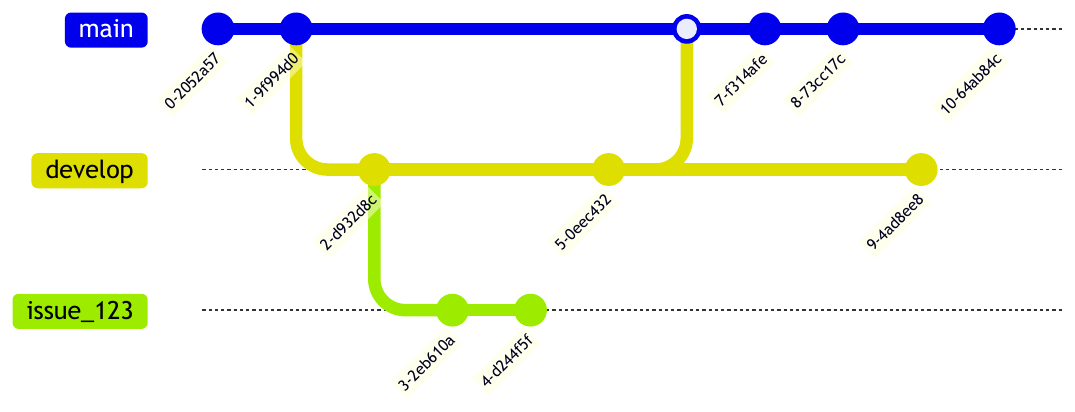}
\end{center}
  \caption{Example of the main line of development in blue, assuming that the \textit{main} branch is the default branch.}
\label{fig:main_line}
\end{figure}

After cloning the projects, we applied an additional filter to include only commits from the default branch. 
This branch is the one displayed when visiting the repository on GitHub and is checked out by default when the project is cloned. Since the default branch may include merge commits and our evolution-based research questions require a linear history, we traversed the branch using only the first parent of each merge commit. For instance, Figure \ref{fig:main_line} illustrates a project with three branches. Assuming the \textit{main} branch is the default, we would traverse through the first parent of each merge commit, selecting all commits highlighted in blue. These commits collectively represent the project’s main line of development and are used for answering RQ2, RQ3, and RQ4. We then proceeded to discard 28 projects that did not contain at least 1,000 commits in the main line of development. Consequently, our final corpus is composed of \nvarc{rq2_projects}{362} projects, which we used to answer our research questions (Section \ref{sec:method}). The complete list is available at our GitHub repository\footnote{\url{https://tinyurl.com/mr2t77st}}.

\begin{table}[t]
\centering
\caption{Characteristics of the selected projects for our corpus. \label{tab:corpus_describe}}
%\resizebox{3in}{!}{%
\begin{scriptsize}
\begin{tabular}{lrrrr} 
\hline
              & \multicolumn{1}{l}{\textbf{contributors}} & \multicolumn{1}{l}{\textbf{stargazers}} & \multicolumn{1}{l}{\textbf{~~ forks}} & \multicolumn{1}{l}{\textbf{~commits}}  \\ 
\hline
\textbf{min}  & 16& 1,041& 119& 1,014\\
\textbf{max}  & 3,532& 88,213& 40,308& 821,332\\
\textbf{mean} & 263& 7,280& 2,133& 8,642\\
\textbf{std}  & 365& 10,512& 3,754& 43,634\\
\textbf{25\%} & 83& 2,107& 578& 2,080\\
\textbf{50\%} & 148& 3,539& 1,006& 3,641\\
\textbf{75\%} & 276& 6,883& 2,040& 6,618\\
\hline
\end{tabular}
\end{scriptsize}
%}
\end{table}

Table~\ref{tab:corpus_describe} presents descriptive statistics for the characteristics of the projects selected for our corpus. For example, the ``commits" column provides an overview of the number of commits that belong to the main line of development. %of the analyzed projects. 
As of September 2024, the project with the fewest commits has 1,014, while the one with the most has 821,332 commits. On average, projects have 8,642 commits, demonstrating significant development activity in the projects. Table \ref{tab:corpus_describe} also shows statistics for the number of contributors, stargazers, and forks of the projects on our corpus. 

\paragraph{Projects Categorization.} From the original non-filtered set of \nvarc{rq2_projects}{633} projects, we took a large sample of 317 projects and manually categorized them by their application domains, since this was required to answer RQ1. This categorization was conducted by two of the authors working together, using open coding. We identified 19 main domains in this sample: Artificial Intelligence, Automation, Collaboration, Cryptocurrency, Data Management, Enterprise Resource Planning, File Management, Finances, Game, High-Performance Computing, Infrastructure Management, %Machine Learning, %(end-user applications in the machine learning domain), 
Media, Monitoring, Network, Personal Management, Program Analysis, Security, and Software Development, and Other. 

A smaller random sample of 30 of those projects was then given to ChatGPT 4-o (without the categories). The prompt we used was ``We need to categorize Java Open-Source projects according to the following categories: [list of our 19 categories]. Could you please provide the categorization for the following projects? [list of the URLs of the 30 sample projects]." We then compared the categories provided by ChatGPT with the ones made by the authors and measured the agreement using Cohen's Kappa \citep{kappa}. For six projects, ChatGPT indicated two categories instead of one. For those cases, we considered the suggestion to be consistent with the one produced by the authors whenever there was an intersection between the categories indicated by ChatGPT and the ones indicated by the authors. Whenever there was a disagreement, we carefully checked the classification, and in a few cases, we agreed with ChatGPT and changed the original classification accordingly. The final Cohen's Kappa is 92.6\%, which is considered an ``almost perfect" agreement \citep{kappa}. We thus proceeded to use ChatGPT to categorize the remaining of our corpus. Every time it suggested more than one category, we manually checked the project and chose the one we considered the most appropriate.

Table \ref{tab:domains} shows examples of projects of each of the domains and also the number of projects classified in each of the domains. For example, the bitcoin-wallet project is classified in the cryptocurrency domain\footnote{\url{http://www.github.com/bitcoin-wallet/bitcoin-wallet}}. The NewPipe project, on the other hand, belongs to the Media domain. Its description is: ``A libre lightweight streaming front-end for Android\footnote{\url{https://github.com/TeamNewPipe/NewPipe}}.''  The complete list is available in our repository\footnote{\url{https://tinyurl.com/mr2t77st}}.

\begin{table*}[t]
\centering
\caption{Samples of selected projects for our corpus and their domains, as well as the numbers of projects classified in each domain.}
%\resizebox{3in}{!}{%
\begin{scriptsize}
\begin{tabular}{lrl} 
\hline
\textbf{Domain} & \textbf{\# of Projects} & \textbf{Example}\\
\hline
%Application Container & docker-java/docker-java\\
Software Development  & 93& apache/netbeans\\ 
Data Management & 69& liquibase/liquibase\\
Infrastructure Management & 33 & apache/zookeeper\\ 
Program Analysis &25& pmd/pmd\\ 
Security & 21& i2p/i2p.i2p\\ 
Automation & 18 & SeleniumHQ/selenium\\
Network & 16 & google/nomulus\\ 
Game & 16 & jMonkeyEngine/jmonkeyengine\\ 
Monitoring & 12 & apache/skywalking\\
Media & 11 & TeamNewPipe/NewPipe\\ 
Enterprise Resource Planning & 9 & kiegroup/jbpm\\ 
Personal Management & 8 & Automattic/simplenote-android\\ 
Artificial Intelligence & 7 & kermitt2/grobid\\  
Collaboration & 7 & igniterealtime/Openfire\\
File Management & 5 & JabRef/jabref\\
Cryptocurrency & 5 & bitcoin-wallet/bitcoin-wallet\\
High Performance Computing & 3 & real-logic/agrona\\ 
Other & 3 & fossasia/pslab-android\\ 
Finances & 1 & killbill/killbill\\
\hline
\end{tabular}
\end{scriptsize}
%}
\label{tab:domains}
\end{table*}

\section{Methods}
\label{sec:method}

This section describes the research questions and the process we used to answer them. Section \ref{sec:questions} presents our research questions. Section \ref{sec:research-method} explains the method. Sections \ref{sec:databaseHeuristics}, \ref{sec:ormHeuristics}, \ref{sec:files}, and \ref{sec:queries} describe the heuristics we used to answer our research questions. Section \ref{sec:replacement} describes how we identify migration patterns to answer RQ3. Finally, Section \ref{sec:infrastructure} presents the infrastructure we built to run our analysis. %We analyzed \nvarc{rq2_projects}{362} Java project repositories from GitHub (as described in Section~\ref{sec:corpus}) and identified which DBMSs, among the 50 most popular DBMSs (Section \ref{sec:dbcorpus}), were used by the projects (RQ1, RQ2, RQ3, and RQ4) and how they are used (RQ5). To address the historical aspects of RQ2, RQ3, and RQ4, we sliced the projects and used data mining techniques.}

\subsection{Research Questions}
\label{sec:questions}

To guide our work, we investigate five research questions that look at different aspects of DBMS usage in Java Open Source projects. 

\noindent\textbf{RQ1: Which DBMS are the most popular across software projects?} In this question, we identify which DBMSs are most commonly adopted by the projects in our \textit{corpus}. We conceived a set of heuristics (see Section~\ref{sec:databaseHeuristics}) to identify each of the selected DBMSs in the source code of the projects in our corpus. Then, we counted their occurrences in the projects and ranked the most used DBMSs. %We also categorize the projects by domain, and the DBMS by data model to conduct further analysis. 

\noindent\textbf{RQ2: How stable are the DBMSs during the projects' history?} 
\responsetoreviewer{rIIcXII}{This question examines the stability of DBMS usage, which refers to the process of a DBMS remaining in continuous use throughout the project's history. While in RQ1, we look at a single snapshot of each project, in RQ2, we look for traces of DBMS occurrence in previous versions of each project.} 

\noindent\textbf{RQ3: Which DBMSs are frequently replaced by others?} 
In this question, we investigate DBMS migration patterns. We used sequential patterns mining, a data mining technique that finds patterns of sequential events over time. According to \cite{agarwal2013data}, this technique allows mining the set of frequent sub-sequences in a sequence or in a set of sequences. Using this technique, we aim to identify which DBMSs are commonly replaced by others simultaneously or in sequential time intervals. 

\noindent\textbf{RQ4: Which DBMSs are often used together?}
This question identifies the synergy between different DBMSs in the projects' history. We used association rules, a data mining technique that detects correlations among frequent item sets. According to \cite{agarwal2013data}, this technique generates frequent item sets, from which strong association rules in the form of $A \rightarrow B$ are defined. The analysis of associations enables the discovery of correlation rules, presenting statistical correlations between sets A and B. With this technique, we could detect DBMSs that are frequently adopted together.

\noindent\textbf{RQ5: How do the applications interact with these DBMS?} Just like with database systems, a variety of ORMs are available for use. To understand the interaction between the software and the database systems, we analyzed the source code of each project of our corpus to extract information about the ORM they use and find out which ones are most frequently used. We also investigate how many database-related files exist in each project and how queries are performed, i.e., using builders or pure SQL in a string.

\subsection{Research Method Overview} 
\label{sec:research-method}

To answer our research questions, we first conceived heuristics based on regular expressions for detecting which DBMSs are adopted by each project (Section \ref{sec:databaseHeuristics}),  which ORMs are used by each project (Section \ref{sec:ormHeuristics}), which files are affected (Section \ref{sec:files}), and how queries are performed (Section \ref{sec:queries}). We also built an infrastructure to automatically clone the projects, run the heuristics over each project, and populate a database with the obtained outputs (Section \ref{sec:infrastructure}). 

\begin{figure*}[t]
\centering
\includegraphics[width=1\textwidth]{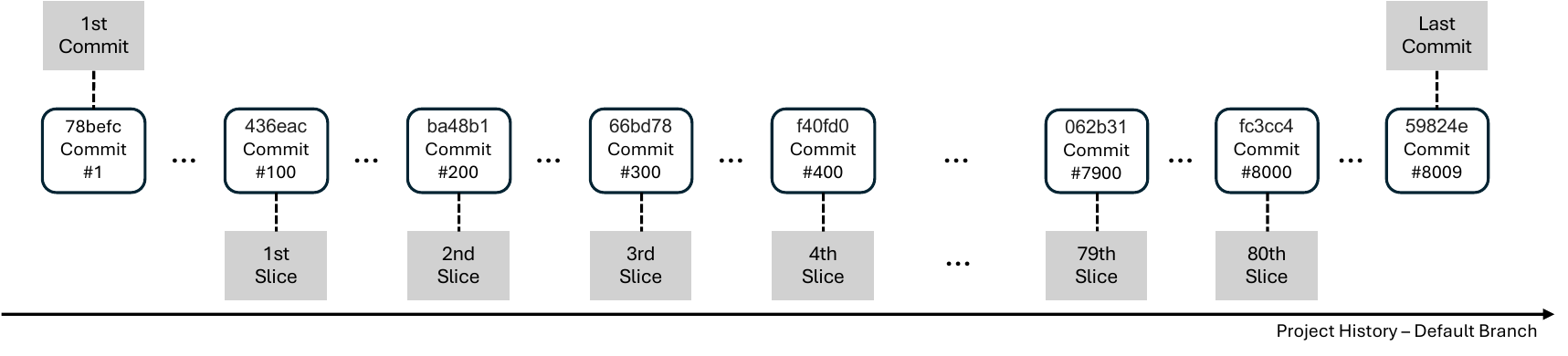}
\vspace{-5mm}
\caption{Slicing of the Skywalking repository commits' history on the main line of development.}

\label{fig:commits_skywalking}
\end{figure*}

For RQ1 and RQ5, we run the heuristics over the last version on the main line of development of each project. However, since some of our research questions (RQ2, RQ3, and RQ4) require an analysis of the history of the projects, we split each project into equal-sized slices in terms of commits on the main line of development. We then analyzed each of the slices of each project, looking for changes on added, kept, and removed DBMSs. In practice, we determined the total number of commits that belong to the main line of development and divided them into segments of 100 commits each. Therefore, the first slice does not correspond to the initial commit of the project but to the 100$^th$ commit of the project's main line of development history. Note that the last slice does not necessarily correspond to the project's last commit on the collection date. This may occur when the number of commits is not divisible by 100. In this case, we ignore some of the last commits to split the project history into slices of the same size. Also, the last slice may not reflect the end of the project since the history of the projects continued after we updated our data in September 2024.
As an example, consider the Skywalking repository. Its main line of development contains 8,009 commits, as shown in Figure \ref{fig:commits_skywalking}. When we slice it using slices of size 100, we get 80 slices (the last 9 commits are ignored). Each slice corresponds to 1.25\% of the project's history and is represented by its last commit, which is a snapshot of the repository at 1.25\%, 2.50\%, 3.75\%, ..., and 100\% of the project history.

Choosing the slice size is not an easy task. If it is too small, we may completely miss the presence of a DBMS. For example, if a DBMS was included in the project on commit $x$, and later removed on commit $y$, ($y > x$), but both $x$ and $y$ are within the limits of a single slice, we will completely miss it, leading to false negatives. Ideally, we could set the slice size to correspond to a single commit (that is, use 1 as the slice size). However, this would have two consequences. First, we would process lots of commits that bring no changes related to the use of DBMS. Second, the processing times of our analysis would significantly increase. Our heuristics take 9 seconds to execute on each slice, on average\footnote{We used a Ryzen 7735HS with 16GB RAM GDDR5, 512GB SSD nvme 4.0
to measure the performance of our analysis on a sample project.}. Taking the sum of the commits of all of the \nvarc{rq2_projects}{362} projects (\nvar{total_committs} commits on the main line of development) and the average time to analyze each slice (9s), we conclude that we would take \nvar{days_to_run} days to run our heuristics over the complete history of all projects of our corpus in case we used 1 commit per slice. After that, we would still need to run the mining and carefully analyze the results. As a compromise between lowering the false negatives and being able to run our analysis in a reasonable time frame, in this article, we chose to use slices of 100 commits. This means our smallest project has \nvar{slices_smallest_project} slices, and our largest project has \nvar{slices_largest_project} slices. This way, although we know that some false negatives may occur with short-living DBMS, we can still capture the governing patterns that include DBMS that existed in the project for at least \nvar{percentage_smallest_project}\% of the project history (considering the smallest project on our corpus). Furthermore, by discarding DBMSs that were introduced in one commit and removed just afterward, we are removing short-term tests or experimentations, which does not mean that the DB was "used" in the project.

After defining the method for slicing the history of the projects, we randomly selected five projects to manually validate the heuristics (see Section~\ref{sec:databaseHeuristics}) we used to automatically detect the DBMSs used by the projects. For this validation, each project was sliced into 10 slices. On each of these slices for each of the five projects, we used the heuristics to automatically detect the DBMSs and stored the results in the database we created for our analysis. Once we confirmed that the detection occurred properly (see Section \ref{sec:databaseHeuristics} on how the validation was made) in our five projects sample, we ran the heuristics in our corpus to answer the research questions RQ1 and RQ2. 

Using these results, we then generated a dataset suitable for the data mining tools used to answer the research questions RQ3 and RQ4 through association rules and sequential patterns. We first used the MLExtend library from \cite{raschkas_2018_mlxtend} to generate the association rules and the SPMF library from \cite{fournier2016spmf} to generate the sequential patterns. MLExtend has about 4.3k stars on GitHub, demonstrating popularity in the community. As for the SPMF repository, since it is not public, we cannot consider it in terms of stars, but we found around 1,000 research papers that cited or adopted the SPMF library\footnote{\url{https://www.philippe-fournier-viger.com/spmf/index.php?link=citations.php}}. Also, both libraries' authors are active researchers in the scientific community, having many papers published in the Data Mining and Machine Learning fields.

Aiming to validate the observed patterns and run complementary analyses, we developed a Pattern Counter tool\footnote{\url{https://patterncounter.readthedocs.io/en/latest/}}. Besides counting patterns from sequence lists, the tool generates support, confidence, and lift measures. We used this tool extensively to filter the results obtained with association rules and sequential patterns and to validate them (see Section~\ref{sec:infrastructure}).

Finally, aiming to understand the reasons for DBMS adoption and/or migration, we conduct a qualitative analysis using 10 randomly selected projects of our corpus. For those projects, we contrast our findings with commit messages, issues, pull requests, and the project's documentation. The results are reported in Section \ref{sec:qualitative}. 

\subsection{Database Heuristics} 
\label{sec:databaseHeuristics}

Our heuristics are a set of regular expressions that we use to search the source code of our corpus using \texttt{git grep}. To build them, we analyzed the official documentation of the top 50 DBMSs (selected as described in Section \ref{sec:dbcorpus}), looking for information about how they are accessed by Java programs. In conducting this careful analysis, our goal was to answer four questions: (i) What imports are required to use these DBMSs? (ii) Which drivers are needed? (iii) How is the connection established? (iv) Which libraries are needed? With the answer to each of these questions, we built a set of regular expressions to search for each DBMS in our corpus. Table \ref{tab:heuristics} shows examples of the heuristics we built by using these answers for two DBMSs: Oracle and Cassandra. Similar expressions were built for the other DBMSs in our database corpus. 

The first question aims to identify the required \textbf{imports} for using these DBMSs. Each DBMS has specific classes that must be imported to enable their usage in Java projects. This question allowed us to identify most non-relational DBMSs. For instance, Cassandra requires the statement \texttt{import com.datastax}, so we use this to build a regular expression in our heuristic for Cassandra. The expression we built looks for the word \texttt{import} followed by one or more white spaces, followed by \texttt{com.datastax}, as shown in row 5 of Table~\ref{tab:heuristics}. However, several relational DBMSs utilize Java Database Connectivity (JDBC), which means a generic import is used for most DBMSs of this type. For instance, Oracle, MySQL, and MS SQL Server are relational DBMSs that utilize JDBC. Therefore, we disregarded the import strings heuristic for relational DBMSs that used generic imports to avoid false positives, thus the empty row 1 of Table~\ref{tab:heuristics}.

The second question aims to identify how the project loads the \textbf{driver} to access each DBMS. The answer to this question gives us clues we use to search the source code. For example, for the Oracle DBMS, we built a regular expression that matches the following search strings: \texttt{oracle.jdbc.OracleDriver} and \texttt{oracle.jdbc.driver.OracleDriver}.

The third question aims to identify which \textbf{connection string} is used to connect to the DBMS. Each DBMS has its connection string, but in Java, relational DBMSs have certain features in common. Thus, when building our regular expression, we need to make sure we use patterns that are not substrings of others. For example, for the Oracle DBMS, instead of searching for \texttt{jdbc:}, which would be common to several relational DBMS, we search for \texttt{jdbc:oracle}.

Finally, the fourth question aims to identify a DBMS usage by looking for the \textbf{libraries} declared in the Maven descriptor. For example, we use the following search strings for the Oracle DBMS in our heuristic to find Oracle database usage: \texttt{ojdbc} and \texttt{com.oracle}.

\begin{table}[t]
\centering
\caption{Examples of heuristics that indicate the adoption of the Cassandra and Oracle DBMSs.}
\label{tab:heuristics}
\begin{scriptsize}
\begin{tabular}{c|c|c} 
\hline
\multirow{6}{*}{\textbf{Oracle}} & \textbf{Import} & - \\
\cline{2-3}
 & \textbf{Driver} & oracle\textbackslash{}.jdbc\textbackslash{}.driver\textbackslash{}.OracleDriver \\
            &    & oracle\textbackslash{}.jdbc\textbackslash{}.OracleDriver \\
\cline{2-3}
 & \textbf{Connection} &  jdbc:oracle  \\ 
\cline{2-3}
& \textbf{Libraries} & ojdbc \\
 &    &                com\textbackslash{}.oracle \\
\hline
\multirow{5}{*}{\textbf{Cassandra}} & \textbf{Import} & import\textbackslash{}s\{1,\}com\textbackslash{}.datastax\\
\cline{2-3}
 & \textbf{Driver} & 
                {[}\^\/a-zA-Z\textbar{}\^\/\textbackslash{}/``\textbackslash{}\_\#\textbar{}0-9]CassandraConnector[\^\/a-zA-Z\textbar{}\^\/\textbackslash{}/"\textbackslash{}\_\#\textbar{}0-9] \\
\cline{2-3}
 & \textbf{Connection} & - \\
\cline{2-3}
 & \textbf{Libraries} & \textless{}\textbackslash{}s*artifactId\textbackslash{}s*\textgreater{}\textbackslash{}s*cassandra\\
& & com\textbackslash{}.datastax\textbackslash{}.oss\\
\hline
\end{tabular}
\end{scriptsize}
\end{table}

We also discovered that some heuristics were common to multi-model DBMSs. For example, both the relational and the non-relational versions of the Virtuoso DBMS use the same \textbf{libraries} heuristics. Since we found 2 multi-model DBMSs that act as both relational and non-relational DBMSs (Ignite and Virtuoso), we separated the heuristics that are common to both models and combined them with the model-specific heuristics to define which of the models was used by each project that adopts such DBMSs. Thus, each of these two DBMSs has two sets of heuristics, one to identify its use as a relational DBMS and another for its use as a non-relational DBMS.

\responsetoreviewer{rIIcXIII}{We consider a given project to adopt a certain DBMS when at least one of the regular expressions of its respective heuristics produces one or more hits on the project's source code.} Going back to the Oracle example, if we find the string referring to its driver, \texttt{oracle.jdbc.driver.OracleDriver} or \texttt{oracle.jdbc.OracleDriver}, or the connection string, \texttt{jdbc:oracle}, or the strings referring to the libraries (\texttt{ojdbc} or \texttt{com.oracle}), we assume that Oracle is used in that project. The selected DBMS list with the respective heuristics is available in our replication package at our companion website\footnote{\url{https://tinyurl.com/y25cm6e5}}.

\noindent\textit{Validation.} We manually validated the heuristics for the top 10 DBMSs on a sample of five randomly selected projects of our corpus: Activiti, Che, Skywalking, Storm, and Pinpoint. The validation was conducted in 2022, and at that time, both the Che and the Pinpoint projects satisfied the criteria to be in our corpus. In the update we made in 2024, Che changed its main programming language, and Pinpoint did not receive pushes in the three months prior to our corpus update in September 2024, so they were discarded from our current corpus. Moreover, since the top 10 DBMSs included MySQL and PostgreSQL, we conducted their analysis in conjunction with their sibling DBMSs (MySQL/MariaDB and PostgreSQL/CockroachDB), as previously explained. 

For the validation, we analyzed the documentation of each project to determine which DBMSs they support. Then, we ran our heuristics and compared the DBMSs we identified with the ones used by the projects according to their documentation. Whenever we detected problems that were caused by our heuristics, we updated them to improve their precision and reflected those changes in the other heuristics as well (the ones that were not subjected to manual validation). For instance, we found situations where generic strings, such as \emph{DatabaseClient}, \emph{CosmoClient}, and \emph{MongoClient}, were present in the results. Although these examples could indicate the connection methods of different DBMSs, the application developer might have created a method or variable containing these strings as substrings. To mitigate this issue and focus specifically on identifying how the connection is made, we applied filters at the beginning and end of the search strings.

For the projects that mention JDBC connection support, we faced a challenge since this type of connection can be used generically without explicitly specifying the DBMSs. However, our heuristics are designed to be specific for each DBMS, allowing us to identify clues even when the documentation mentions a generic JDBC connection. Therefore, when we found a clue associated with a DBMS that was not specified in the documentation but is considered a possibility due to its usage of the JDBC mechanism, we considered it a successful identification. This means that despite the project mentioning a generic connection that could potentially obscure the specific DBMS being used, our DBMS-specific heuristics uncovered the clue associated with the DBMS.

\begin{table}[t]
\centering
\caption{Comparison between the results found by our Heuristics (H) and the project's documentation (D or d). Cells in red indicate false positives.}
\label{tab:validation}
%\resizebox{7in}{!}{%
\begin{scriptsize}
\begin{tabular}{cccccc} 
\hline
& \textbf{Activiti} & \textbf{Che} & \textbf{Skywalking} & \textbf{Storm} & \textbf{Pinpoint} \\
\hline
\textbf{Oracle}  & d & \textcolor{red}{H} & \textbf{Hd}  & d & \textbf{HD}  \\ 
\textbf{MySQL/MariaDB}   & d & \textcolor{red}{H} & \textbf{HD}  & \textbf{Hd}   & \textbf{HD}\\
\textbf{MS SQL Server} & d  &~ &\textbf{Hd}  & d & \textbf{HD} \\
\textbf{PostgreSQL/CockroachDB} & d  & \textbf{HD} & \textbf{Hd} & d  & \textbf{HD} \\
\textbf{MongoDB} &~ &\textcolor{red}{H}  &\textbf{Hd} & \textbf{HD} & \textbf{HD}  \\
\textbf{IBM DB2} & d &~ & d  & d &~\\
\textbf{Redis}   & ~ & \textcolor{red}{H} &\textbf{Hd}  & \textbf{HD} & \textbf{HD} \\  
%\textbf{\revise{Elasticsearch}} & ~ & \textcolor{red}{H}  & \textbf{HD}  & \textbf{HD}  & \textbf{HD} \\
\textbf{SQLite}  & d  &~ & d  & d  &~ \\
\textbf{MS Access}  & d  &~ & d  & d  &~ \\
%\textbf{Maria DB} & d &~ & \textbf{Hd} & d & \textbf{HD} \\
\textbf{Cassandra} &~ &~ &\textbf{Hd} & \textbf{HD} & \textbf{HD}  \\
\hline
\end{tabular}
\end{scriptsize}
\end{table}

Table \ref{tab:validation} compares the results obtained by our heuristics and those found manually in the projects' documentation. Each cell ($r$, $c$) in this table contains an \textbf{H} when the DBMS on row $r$ was found by our heuristics for the project at column $c$, and a \textbf{D} when the DBMS on row $r$ is specifically mentioned in the project's documentation. Also, a cell contains a \textbf{d} when the DBMS of row $r$ may or may not be used for the project on column $c$. This happens when the project documentation mentions the possibility of utilizing other DBMSs not explicitly mentioned. As an example, when the project mentions the possibility of using a JDBC connection mechanism commonly used by relational DBMSs, we assume the project is able to use any relational DBMS and thus mark the ones we found as \textbf{d}. 

Cells marked with \textbf{HD} or \textbf{Hd} means the DBMS was found by our heuristics and was mentioned, specifically or as a possibility, in the project documentation, characterizing the true positives. Cells marked with a single \textbf{d} are not considered false negatives since \textbf{d} are possibilities instead of obligations. Lastly, cells with a red \textcolor{red}{H} represent false positives since our heuristics detected the corresponding DBMS, but it was not mentioned in the project's documentation. 

For example, the Activiti Project documentation mentions Oracle, MySQL, MS SQL Server, PostgreSQL, IBM DB2, and H2 as examples of supported DBMSs, with H2 being the default option. It also states the support for Java Database Connectivity (JDBC), indicating that relational DBMSs are potential options for this project. Therefore, the fact that our heuristics did not find signs of usage of one of the eight relational DBMSs we used in our validation step was not considered a false negative. In the case of the Skywalking project, its documentation cites MySQL and H2 as existing implementations, and it mentions the user may ``implement [their] own." Thus, we considered that this project allows the implementation of other DBMSs, and thus the heuristic results are correct --- the heuristics found DBMSs that are not explicitly mentioned by the project's documentation, but we understand they are valid possibilities. Storm's documentation mentions the integration with JDBC, Cassandra, Redis, and MongoDB. Therefore, we considered relational DBMSs as possible options, and the results obtained by our heuristics were considered correct. All the DBMSs mentioned in the documentation for the Pinpoint project were found by our heuristics. Finally, for the Che project, we found evidence of Oracle, MySQL/MariaDB, PostgreSQL/CockrouchDB, MongoDB, and Redis. Given that PostgreSQL is the only DBMS mentioned in the project's documentation, the results indicate evidence that the other DBMSs diverged, so we considered them false positives.

As shown in Table \ref{tab:validation}, we found evidence of the adoption of 23 DBMSs in the 5 projects, with only 4 false positives. This corresponds to a precision of 82.6\% with 100\% recall. Therefore, we consider our heuristics adequate. % and sought mechanisms to mitigate possible failures.  

\subsection{ORM Heuristics}
\label{sec:ormHeuristics}

To implement the ORM detection heuristics, we analyzed the ORMs' official documentation looking for information regarding their usage in the Java programming language. We used this knowledge to build regular expressions, just as we did with the DBMS heuristics. 
Each ORM has its specific usage method. However, every ORM that implements JPA requires four main steps: (i) explicitly include a dependency through the dependency manager or by downloading and adding a library or package into the project, (ii) create a configuration file, (iii) define entities or create mappings, and (iv) implement queries. Given these four steps, we developed regular expressions to search for words, classes, tags, annotations, and imports that indicate the usage of the ORMs in our project corpus. 

The first step (i) includes all required \textbf{libraries} in the project. For this, developers can add a \emph{jar} file or use a dependency manager (Maven, for example) and add the dependency in the \emph{pom.xml} file. Each ORM has its own set of dependencies. For example, to use Hibernate, the dependency contains the string: \texttt{<groupId> org.hibernate}. When building our regular expressions, we have mapped different ways to include libraries for ORM in Java projects. 

The second step (ii) concerns the \textbf{configuration file} of the ORM. This file contains settings for an embedded database or object-relational mapping. As each ORM has its tags, the name of the tag, and the name of the file, we used this information to design our search heuristics. For example, to use Hibernate, the configuration file contains the string: \texttt{<\textbackslash{}hibernate-configuration\textbackslash{}>}

The third step (iii) identifies how to catalog project \textbf{entities}. An entity represents a set of information about a given system concept. Every entity has attributes, which are the information that references the entity \citep{Keith2006}. Each ORM has its own way of declaring entities. For example, some classes contain the annotation \texttt{@Entity} to use JPA. Hibernate uses the mapping metadata to determine how to load and store objects of the persistent class. For this, the tag: \texttt{<hibernate-mapping>} is used. We have mapped all these different ways of defining entities to generate the heuristics we used to search our project corpus.

Finally, the fourth step (iv) considers how each ORM handles \textbf{queries}. Each ORM has its own way of declaring queries. To use MyBatis, for example, the query statements can be defined in XML as \texttt{<select>}, \texttt{<insert>} or \texttt{<update>}, or as Annotations 
in Java files as \texttt{@Select}, \texttt{@Insert}, or \texttt{@Update}. So, we mapped these different ways to use it. 

Table \ref{tab:orm-heuristics} shows the heuristics we defined for the Spring ORM framework as an example. The complete set of heuristics is available at our companion website\footnote{\label{foot:heuristics}\url{https://tinyurl.com/3f8kp763}}. These heuristics help us search for footsteps left in the source code that indicate the presence of a given ORM.

\begin{table}[t]
\centering
\caption{Examples of heuristics that indicate the adoption of the Spring ORM.}
\label{tab:orm-heuristics}
\begin{scriptsize}
\begin{tabular}{c|c} 
\hline
\textbf{Build} & \textless{}\textbackslash{}s*groupId\textbackslash{}s*\textgreater{}\textbackslash{}s*org.springframework.boot                   \\ 
\hline
\textbf{Import} & import\textbackslash{}s\{1,\}org\textbackslash{}.springframework\textbackslash{}.data\textbackslash{}.repository  \\
\hline
\multirow{2}{*}{\textbf{Configuration}} & @SpringBootApplication\\ & www\textbackslash{}.springframework\textbackslash{}.org/schema/ \\ 
\hline
\multirow{3}{*}{\textbf{Queries}} & @Query \\ 
                 & @QueryHint\\ 
                 & @NamedEntityGraph \\
\hline
\end{tabular}
\end{scriptsize}
\end{table}

\begin{table}[t]
\centering
\caption{Comparison between the ORMs found by our Heuristics (H), the ``pom.xml" file (P), the project's  configuration file (C), and the project`s entities (E).}
\label{tab:validationORM}
%\resizebox{7in}{!}{%
\begin{scriptsize}
\begin{tabular}{cccccc} 
\hline
& \textbf{Activiti} & \textbf{Che} & \textbf{Skywalking} & \textbf{Storm} & \textbf{Pinpoint} \\
\hline
\textbf{Hibernate}  & HPC&~  &HC&~  &~   \\ 
\textbf{MyBatis}   & HPCE&~ &\textcolor{red}{PC}&\textcolor{red}{H}   & HPCE\\
\textbf{Spring} & HPC&~ & HPC& HC & HPC\\
\textbf{jOOQ} &~  &~ &~ &\textcolor{red}{H}  &~  \\
\textbf{JDBCMapper} &~ &~ &~ &~  &~\\
\textbf{EclipseLink}   & ~ &~  &~  &~  &~  \\ 
\hline
\end{tabular}
\end{scriptsize}
\end{table}

\vspace{6mm}\noindent\textit{Validation.} Similarly to what we did for the DBMSs heuristics, we conducted a manual validation for the ORM heuristics using the same 5 randomly selected projects. Table \ref{tab:validationORM} compares the results obtained by our heuristics and those found manually in the projects' source code. Each cell ($r$, $c$) in this table contains an \textbf{H} when the ORM on row $r$ was found by our heuristics for the project at column $c$, a \textbf{P} when the ORM on row $r$ is mentioned in the project's ``pom.xml'' files, a \textbf{C} when the ORM on row $r$ is mentioned in the project's configuration files, and \textbf{E} when it is mentioned in an entity declaration in the source code. Cells marked with \textbf{HPCE}, \textbf{HPC}, or \textbf{HC} means the ORM was found by our heuristics and was mentioned in the project source code, characterizing the true positives. The cell marked with a red \textcolor{red}{PC} represents a false negative since our heuristics did not detect an ORM that was mentioned in the project configuration files. Lastly, cells with a red \textcolor{red}{H} represent false positives since our heuristics detected the corresponding ORM, but it was not mentioned in the project's files. As shown in Table \ref{tab:validationORM}, we found evidence of the use of 11 ORMs in the five projects, with only two false positives. This corresponds to a precision of 80\% with 88.88\% recall.

\subsection{Files Needed to Use ORM}
\label{sec:files}

After applying our ORM heuristics, we identified the files that were related to the use of ORM. We called these files as \emph{DB-code}. Next, we grouped the files by their type. We call \emph{Java DB-Code} the files of type ``.java'', and \emph{XML DB-Code} the files of type ``.xml''. We perform this segmentation to evaluate the number of files needed to configure an ORM (XML DB-Code) and the number of files needed to use an ORM (Java DB-Code). 

Using fan-in \cite{SHenry_1981}, we discovered which files depend on or use the Java DB-Code files. We call this second group of files \emph{Dependencies}. Fan-in refers to the number of local flows into a file plus the number of data structures from which the file retrieves information. In the context of this paper, fan-in refers to the maximum number of files related to a Java DB-Code file. We calculated this number using the imported classes as a reference -- we searched for the class name of each Java DB-code file. For example, if the file \emph{JpaHelper.java} is returned when we use our search heuristics (thus, it is a Java DB-code file), we look for its class name using a regular expression \texttt{(\textbackslash{}s|[.,(]|\textasciicircum)JpaHelper(\textbackslash{}s|[.,(]|\$)} to find files that use this file. Finally, we define \emph{TotalDB} as the percentage of the total sum of the necessary files: (Java and XML) DB-Code and Dependencies Code.

\subsection{Heuristics on How Queries are Performed\label{sec:queries}}

Even when an ORM framework is in place, applications frequently need to run queries in the DBMS. The ORM framework automatically runs queries to build objects for the application or to save object changes into the database. However, global operations would suffer from performance bottlenecks if the ORM framework needs to materialize all the involved objects, as is the case in the well-known N+1 Select problem \citep{n+1}. To solve this problem, developers bypass the automatic query mechanism of the ORM and use pure SQL as strings, or query builders instead. For instance, running a SQL query to inform the mean salary of hundreds of thousands of government employees is way faster than materializing all objects that represent government employees to calculate the mean salary.   

Using pure SQL in a Java program simply requires JDBC. There is no need to add imports or libraries besides those of JDBC. Although using SQL to query relational databases has many advantages, there are still a few disadvantages. When dealing with pure SQL, there are always issues with escaping characters within literal strings, for example, spaces in the right place and parentheses matching up. Moreover, even after the code is debugged and thoroughly tested, it is often still very fragile. The slightest changes may throw things out of balance and require another round of testing and tweaking. A key aspect to consider here is that SQL queries as strings are treated as strings by the IDEs during compilation and demand external lint tools for syntax checking.

SqlBuilder \cite{sqlbuilder} is a library that aims to solve the problems of creating and generating SQL queries in Java programs. The library wraps SQL into Java objects that follow the builder pattern \citep{gamma1994design}. This strategy reduces the use of standard SQL inside strings.

In our work, we search our corpus to find out how queries to the database are performed. To build the heuristics for that (again, a set of regular expressions that we run using \texttt{git grep}), we analyzed the official documentation of the seven ORMs, looking for information on how to use pure SQL and Builder in the Java language. We looked for and analyzed the two possibilities of usage: (i) What is needed to use the Builder? (ii) What is needed to use pure SQL? 

Question (i) is aimed at guaranteeing that we understand what is required to use Builders in the seven ORMs. Each ORM has its specific way of use, so we searched for words, tags, and imports that were unique and referred to the use of those Builders within the application. 

To use the Builder, the developer must include the required \textbf{libraries} in the project. For this, the developers can add a jar file or use a dependency manager (Maven) and add the dependency in the \emph{pom.xml} file. Each ORM has its own dependencies. For example, to use JDBC the dependency contains the string: \texttt{<artifactId>JdbcQueryBuilder}, and thus our regular expression that matches this string is \texttt{<\textbackslash{}s*artifactId\textbackslash{}s*>\textbackslash{}s*JdbcQueryBuilder}. We mapped all these ways to include libraries for Builders in Java projects. The second requirement is to \textbf{import} the Builder in Java classes. Because of that, we search for classes that have some Builder import, as in the following example that uses the MyBatis framework: \texttt{import org.mybatis}. The regular expression we use to match this is \texttt{<\textbackslash{}s*artifactId\textbackslash{}s*>\textbackslash{}s*mybatis}.

Question (ii) is aimed at guaranteeing that we understand what is required to use pure SQL. Each type of SQL has its structure and its particularities, so we searched for words that were unique and referred to the use of those types of SQL queries within the application. For example, to find a SELECT query, we used the following regular expressions: \texttt{select\textbackslash{}s\{1,\}.*from\textbackslash{}s\{1,\}.*}. This expression matches any complete SQL search structure, regardless of the source tables or the \emph{where} clause. This heuristics is case insensitive. In the table available at our companion website\footnote{\label{foot:heuristicsSQL}\url{http://tinyurl.com/3acawe3k}}, we provide a list of SQL/Builders heuristics we adopted in our searches. 

We manually validated the heuristics for the Builders and SQL on a sample of five randomly selected projects of our corpus (the same we use for the other two types of heuristics). We tried to identify unmapped libraries. For that, we searched for classes that did some kind of data manipulation, using terms like insert, update, and delete. For the Activiti and Pinpoint projects, no evidence of unmapped Builder was found. The Che project uses the Javax library. Javax is a JPA library, so we made sure to update our JPA ORM heuristics (Section \ref{sec:ormHeuristics}) to include it. The Skywalking project uses IoTDB. In its documentation, we found the following: ``Apache IoTDB (Database for Internet of Things) is an IoT native database that offers different ways to interact with the server, to insert and query data.'' In the case of the IoTDB, we decided not to include it in our heuristics, as it is not a Builder library. The Storm project uses the Storm/Trident integration for the JDBC library. As the library is project-specific, we also decided not to include it in our heuristics.

\vspace{6mm}\noindent\textit{Validation.} We conducted a manual validation for the Builder and SQL heuristics using the same five randomly selected projects we used to validate the Database and ORM heuristics. Table \ref{tab:validationQueryBuilder} compares the results obtained by our heuristics with those identified manually in the projects' source code. Each cell ($r$, $c$) in this table contains an \textbf{H} if the Builder/SQL in row $r$ was detected by our heuristics for the project in column $c$ and confirmed as a true positive after manually checking the source code and configuration files. A red \textcolor{red}{\textbf{H}} represents false positives. We had no false negatives. We found evidence of SQL usage in all projects, with one false positive in a single project. For Builders, we found evidence of their use in two projects, with no false negatives or false positives. This corresponds to a precision of 87.7\% and a recall of 100\%.

%, and was found on the project . For the SQL row, each column ('SQL', $c$) has a \textbf{Q} if we found at least one SQL query as a string on the source code of the project of column $c$, and an \textbf{F} if the SQL in row $r$ is mentioned in the project's files but does not refer to queries, a \textbf{P} when the Builder on row $r$ is mentioned in the project's ``pom.xml'' files, and \textbf{I} when it is mentioned in an import declaration in the source code. Cells marked with \textbf{HQ} or \textbf{HBI} mean the Builder e SQL was found by our heuristics and was mentioned in the project source code, characterizing the true positives. The cell marked with a red \textcolor{red}{F} represents a false positive since our heuristics detected the corresponding heuristic, but it was not an SQL file. We found evidence of SQL usage in all projects, with one false positive in a single project. For Builders, we found evidence of their use in two projects, with no false negatives or false positives. This corresponds to a precision of 87.7\% and a recall of 100\%.}

\begin{table}[t]
    \centering
    \caption{Validation of the Query and Builder Heuristics. 
    }
    \label{tab:validationQueryBuilder}
    \begin{tabular}{lcccccc}
        \hline
        & \textbf{Activiti} & \textbf{Sky} & \textbf{Strom} & \textbf{Che} & \textbf{Pinpoint} \\
        \hline
        \textbf{SQL} & H & H & H & H & \textcolor{red}{H} \\
        \textbf{Builder} & H & H & & & \\
%        \textbf{SQL} & HQ & HQ & HQ & HQ & HQ\textcolor{red}{F} \\
%        \textbf{Builder} & HBI & HBI & & & \\
        \hline
    \end{tabular}
\end{table}

\subsection{Migration Patterns}
\label{sec:replacement}

When examining two DBMSs $X$ and $Y$, we can identify a potential migration pattern (RQ3) when we look at different slices in the history of the same project. This replacement occurs when DBMS $X$ exists in a specific slice, then in a subsequent slice, $X$ does not exist, but instead, DBMS $Y$ appears, and $Y$ is kept in a subsequent slice. We can formalize this replacement using the following rule:
\begin{center}
    \textbf{$X \rightarrow Y_{In} X_{Out} \rightarrow Y$} 
    %or \textbf{$X$ -> $X_{Out}$ $Y_{In}$ - >$Y$}
\end{center} 
where $X$ and $Y$ represent the DBMS, the suffix $_{In}$ represents the DBMS was added, the $_{Out}$ suffix represents the DBMS was removed, the absence of $_{In}$ and $_{Out}$ represents the permanence of the DBMS, and $\rightarrow$ separates the slices.

For instance, consider the sequential pattern below. This notation signifies that $PostgreSQL$ was present in a particular slice, and in a subsequent slice, $PostgreSQL$ was replaced by $Oracle$, which was kept in a subsequent slice.

\begin{center}
\textbf{$PostgreSQL \rightarrow Oracle_{In} PostgreSQL_{Out} \rightarrow Oracle$}
\end{center}

Another way to perceive a DBMS replacement is when DBMS $X$ exists in a specific slice, then in a subsequent slice, DBMS $Y$ enters, and in a later slice, DBMS $X$ is removed, while DBMS $Y$ is kept. The following equation formalizes this situation: 

\begin{center}
    \textbf{$X\rightarrow Y_{In} \rightarrow X_{Out} Y$}
    % or \textbf{$X$ -> $Y_{In}$ -> $Y$ $X_{Out}$}
\end{center} 

As an example, consider the sequence below. This indicates that Oracle was present in a particular slice, PostgreSQL entered in a subsequent slice, and in a following slice, Oracle was removed while PostgreSQL remained.  

\begin{center}
    \textbf{$Oracle \rightarrow PostgreSQL_{In} \rightarrow Oracle_{Out} PostgreSQL$}
\end{center} 

This formulation allows us to capture and analyze the replacement patterns among DBMSs over time. By identifying such replacements, we gain insights into the dynamics of DBMSs' usage and their transitions within the dataset.

\subsection{Infrastructure} 
\label{sec:infrastructure}
Most steps of our research were automatized to reduce error-prone and time-consuming manual executions and to increase the reproducibility and auditability of the results. We first implemented a series of scripts and Jupyter notebooks for collecting, filtering, and analyzing projects’ metadata from Git\-Hub, as described in Section \ref{sec:corpus}. Then, we implemented a script to automatically clone the repositories of our corpus.

The execution of each heuristic over the projects in our corpus was also automatized. The script we implemented first checks the existing heuristics and decides whether a new execution is necessary. This step populates a relational DBMS with information about the projects, the heuristics, and the execution of each heuristic for each project. Finally, we implemented a web application that allows us to manually validate the results. This application shows the pending matches and, for each match, the output generated by \texttt{git grep}. This application allows us to analyze the results and confirm whether they are valid. This app was especially useful in the heuristics building and validation phases. 

To identify migration patterns (RQ3), we used sequential pattern mining techniques. We adopted the Prefixspan algorithm, the most popular pattern-growth algorithm for sequential pattern mining \citep{han2001prefixspan}. Specifically, we used the implementation provided by the SPMF library. Given our objective of discovering subsequences in sequential datasets \citep{fournier2017survey}, we considered this algorithm as the most suitable choice. We aimed to identify the DBMS that was most frequently replaced over time, treating each project as a sequence composed of $n$ items, where $n$ is the number of slices of that project, defined as explained in Section \ref{sec:research-method}. Each item represents a slice of the project history, indicating the DBMSs that were added, kept, or removed from the project. Consequently, by combining the sequence records from all projects, we created a sequential dataset that served as the input for the PrefixSpan algorithm. We also established three flags, namely Init, In, and Out, to indicate the addition or removal of a DBMS. The rules we used to generate the input file are defined as follows: 

\begin{itemize}
\item The existence of an \textbf{Init} flag indicates that the DBMS occurs in that project since its first slice. For example, to denote that Oracle appears in the first slice of a given project, we add \emph{Oracle$_{Init}$}.
\item The existence of an \textbf{In} flag indicates the first occurrence of the DBMS in any slice other than the first. For example, if the first appearance of SQLite in a given project occurs in its third slide, we add \emph{SQLite$_{In}$}.
\item The existence of the DBMS in the previous slice and in the current slice indicates it was kept from one slice to the other. We denote this by using the name of the DBMS. For example, if MariaDB was present in the third slice of a given project and also appears in the fourth slide, we add \emph{MariaBD}.
\item The existence of an \textbf{Out} flag indicates the DBMS was removed from the project in that slice. For example, if Redis was present in the fifth slice of a given project but was not found in the sixth slice, we add \emph{Redis$_{Out}$}.
\end{itemize}

\begin{figure}[t]
\centering
\includegraphics[width=0.75\textwidth]{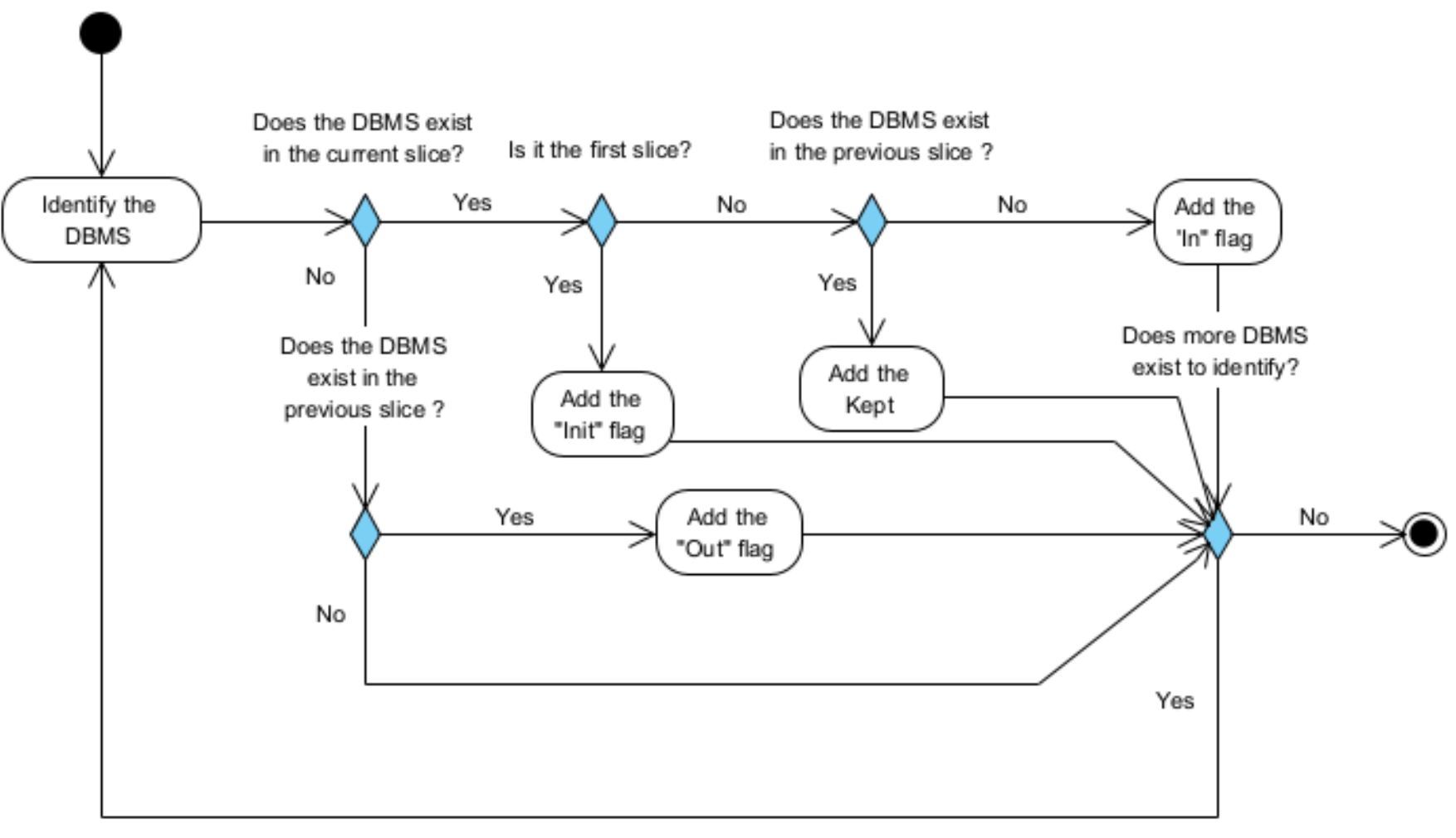}
\caption{Activity Diagram for identifying the DBMSs in a given project slice.}
\label{fig:Diagram_Input2}
\end{figure}

Figure \ref{fig:Diagram_Input2} shows an activity diagram illustrating how we identify if a given DBMS was added, kept, or removed in a given slice, as explained above. We emphasize that this activity diagram describes only part of the process implemented by our script that generates the data entry file required by SPMF. As an example, consider the Zendesk/Maxwell project and assume it was sliced into 10 equal-sized slices (in reality, this project was sliced into 37 slices of size 100, since it has 3790 commits -- we use 10 slices here for didactic purposes). For this project, the flags would be as shown below.

\begin{center}
  MySQL$_{Init} \rightarrow$ MySQL $\rightarrow$ MySQL $\rightarrow$ MySQL $\rightarrow$ MySQL $\rightarrow$ MySQL $\rightarrow$ Redis$_{In}$ MySQL $\rightarrow$ Redis MySQL $\rightarrow$ Redis MySQL $\rightarrow$ Redis$_{Out}$ MySQL
\end{center}

In this notation, slices are separated by an arrow ($\rightarrow$). In this example, MySQL was present in the first slice (MySQL$_{Init}$) and was kept as the only DBMS until the sixth slice (MySQL). In the seventh slice, Redis appears for the first time (Redis$_{In}$), and MySQL is kept (MySQL). In the eighth and ninth slices, Redis and MySQL are still present (Redis MySQL). Finally, in the last slice, Redis was removed (Redis$_{Out}$) while MySQL was kept (MySQL).

For our analysis, we also needed to count the occurrences of a given DBMS in a sequence, allowing for both the validation of the mined results and the execution of additional analyses. For that, we have developed a tool called Pattern Counter\footnote{https://patterncounter.readthedocs.io/en/latest/}. Pattern Counter allows counting patterns in a sequence of items using rules and variables. The tool uses the same input file we generated to work with the SPMF library. In addition, Pattern Counter provides filtering capabilities, allowing data extraction through various combinations of parameters.

Finally, to mine association rules, we adopted the Apriori algorithm, a popular algorithm for extracting frequent item sets, proposed by \cite{agrawal1994fast} in 1994 and implemented by the MLExtend library. We then use it to find the synergy between the DBMSs, i.e., which DBMSs are used together in the history of a given project (RQ4). This algorithm requires a dataset that indicates all items (i.e., DBMSs in our case) that are present in each transaction (i.e., a project in our case). Consequently, we created three datasets containing the results of the heuristics discovered at three stages of the projects' history (beginning, middle, and end) to capture the evolving DBMS adoption in the projects over time. To achieve this, we developed Python scripts for building and pre-processing datasets, using scripts from the Pandas library for the pre-processing. Afterward, we applied this algorithm to each dataset to generate the correlations between DBMSs.

The complete experimental package, containing the data and scripts used in this research and the reproducibility instructions, are available at \url{https://gems-uff.github.io/db-mining}.

\section{Results}
\label{sec:results}
%Link avaliação qualitativa: https://docs.google.com/spreadsheets/d/1tqaPlFT7CHKhFTm3_2EwlCXURLzAf2OCzoMP8QvXahI/edit?usp=sharing

In this section, we report the results according to our research questions. We also present the results of a qualitative analysis in Section \ref{sec:qualitative}.

\subsection{(RQ1) Which DBMS are the most popular across software projects?} 
\label{RQ1}

In order to answer RQ1, we divided our investigation into three aspects: DBMS, Database Models, and Project Domains. 

~

\begin{figure}
\centering
\responsetoreviewer{rIIcXIVrqIpI}{
\includegraphics[width=\linewidth]{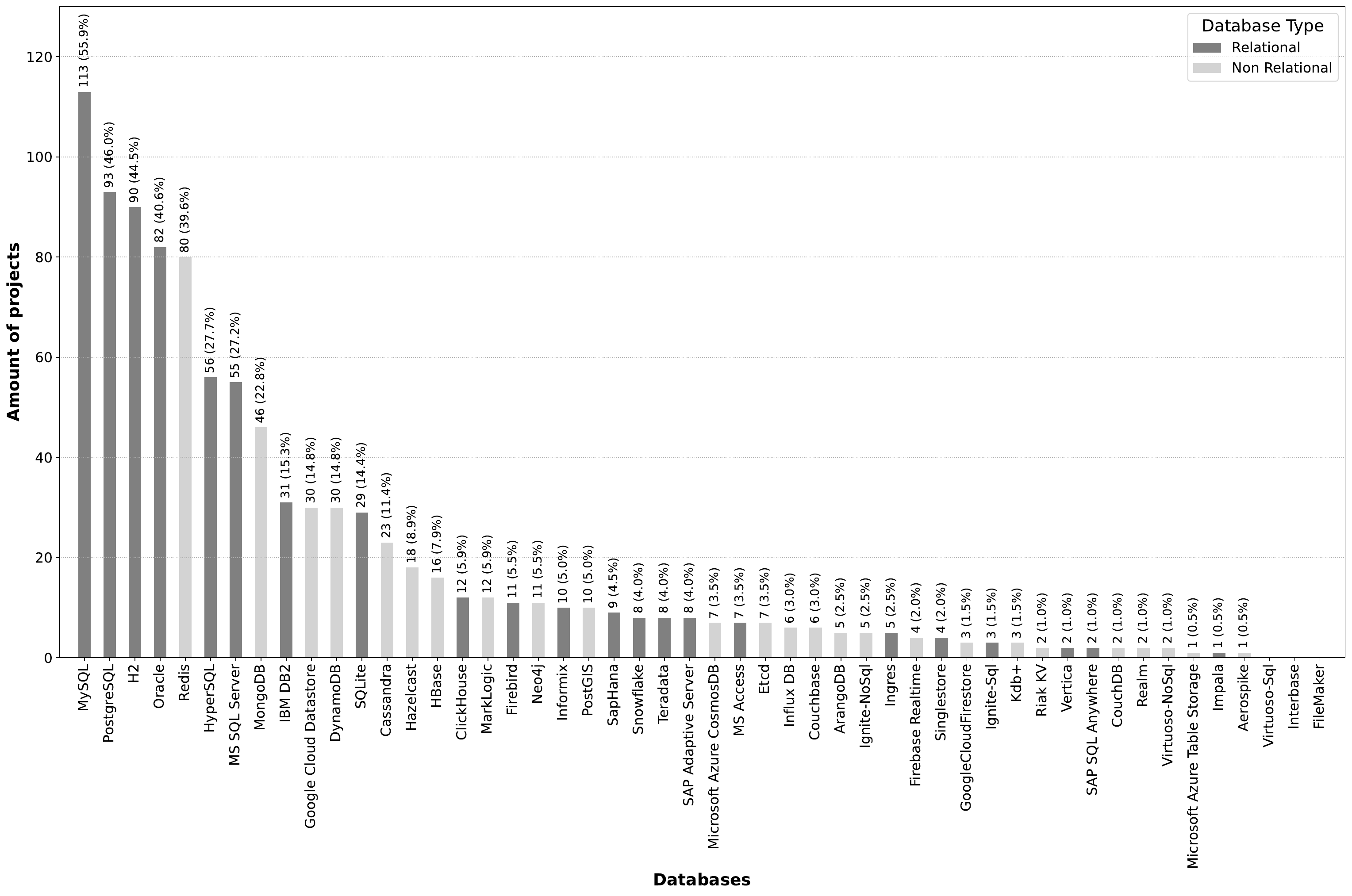}
}
\caption{\responsetoreviewer{rIIcXIVrqIpII}{Most popular DBMSs in our corpus, classified by data model.}}
\label{fig:rq1}
\end{figure}

\paragraph{DBMS.} Figure \ref{fig:rq1} shows our findings regarding the DBMS aspect. We found that Java projects use \nvarc{rq1_dbms_multimodel_combined}{46} different DBMSs. Among the DBMSs, \nvarc{rq1_multimodel}{two} of them are multi-model, and \nvarc{rq1_multimodel_both}{one} uses both models. Since we consider them to be different in our analysis, we found evidence of the use of \nvarc{rq1_dbms_total}{47} different DBMS. However, out of the \nvarc{rq2_projects}{362} projects that belong to our corpus, we found evidence of DBMS usage in just \nvarc{rq1_projects_with_dbms}{202}. We observed that, in our corpus, MySQL is the most popular DBMS. MySQL is present in \nvarc{rq1_mysql_total}{113} of the \nvarc{rq1_projects_with_dbms}{202} projects in which we found evidence of DBMS usage, representing \nvarc{rq1_mysql}{55.9\%} of the projects with DBMS. According to the DB-engine \citep{db-engines_2022} ranking (as of October 2024), MySQL is the second most popular DBMS. PostgreSQL comes in second place in our corpus. It is present in \nvarc{rq1_postgresql_total}{93} projects (\nvarc{rq1_postgresql}{46.0\%}). H2 comes in third, present in \nvarc{rq1_h2_total}{90} projects (\nvarc{rq1_h2}{44.6\%}). Oracle comes next, appearing in \nvarc{rq1_oracle}{40.6\%} (\nvarc{rq1_oracle_total}{82} projects) of the projects. Finally, in fifth place, Redis was discovered in \nvarc{rq1_redis}{39.6\%} (\nvarc{rq1_redis_total}{80}) of the projects. It is the only non-relational database found among the top 5 DBMSs of our corpus. In the DB-Engines ranking, Oracle, MySQL, PostgreSQL, and Redis (in this order) are among the top-10 DBMSs in terms of popularity. The discrepancy in our findings is H2, which appears in 49\textsuperscript{th} place in the DB-Engines ranking, but in second place in our results. However, this is natural since DB-Engines is not specific for DBMSs that are used in Java, and also because the ranking does not reflect use but popularity instead. 

Aerospike, Impala, and Microsoft Azure Table Storage appeared in just one project each. Out of the 50 DBMSs we analyzed, Interbase, FileMaker, and Virtuoso-SQL showed no evidence of being used. Another important fact is that \nvarc{rq1_no_dbms}{160} projects showed no evidence of the use of any of the 50 DBMS we searched for in our corpus. To verify that this result was correct, we reviewed the official documentation of ten randomly selected projects, Lottie-Android, FXGL, PDFBox, POI, Antlr4, Rocketmq, Curator, Struts, Kafka, and Jenkins, and concluded that they really do not use a DBMS.

%\begin{figure}
%\centering
%\includegraphics[width=\linewidth]{Imagens/rq1.png}
%\caption{Most Popular DBMS in Java projects.}
%\label{fig:rq1}
%\end{figure}

~

\paragraph{Models.} To investigate the Models aspect, we grouped our findings by database model (relational and non-relational). Figure \ref{fig:rq1} shows the occurrences of relational and non-relational DBMSs in our project corpus. Using our database heuristics to search the project corpus, we found evidence of the use of \nvarc{rq1_relational_dbms_found_total}{22} of the \nvarc{rq1_relational_dbms}{25} relational databases (\nvarc{rq1_relational_dbms_found}{88.0\%}) we searched for.

For non-relational databases, the DBMSs for which we found more usage evidence were Redis (in \nvarc{rq1_redis_total}{80} projects) and Mongo DB (in \nvarc{rq1_mongodb_total}{46} projects). Using our database heuristics to search the project corpus, we found evidence of the use of all of the \nvarc{rq1_non_relational_dbms}{25} non-relational databases we searched for (\nvarc{rq1_non_relational_dbms_found}{100\%}). Given that we found evidence of DBMS usage in \nvarc{rq1_projects_with_dbms}{202} projects, we observed that the adopted DBMSs are distributed as follows: \nvarc{rq1_relational_total}{166} projects used relational DBMSs (\nvarc{rq1_relational}{82.2\%}), \nvarc{rq1_non_relational_total}{138} projects used non-relational DBMSs (\nvarc{rq1_non_relational}{68.3\%}), and \nvarc{rq1_multimodel_ext}{eight} projects used multi-model DBMS (Ignite and Virtuoso – \nvarc{rq1_multimodel}{4.0\%}). Thus, from this perspective, relational DBMSs are more popular than non-relational DBMSs in our corpus. We also found a non-negligible intersection of \nvarc{rq1_model_intersection_total}{103} projects where both models are adopted (\nvarc{rq1_model_intersection}{51.0\%}). 

\begin{figure}
\centering
\includegraphics[width=\linewidth]{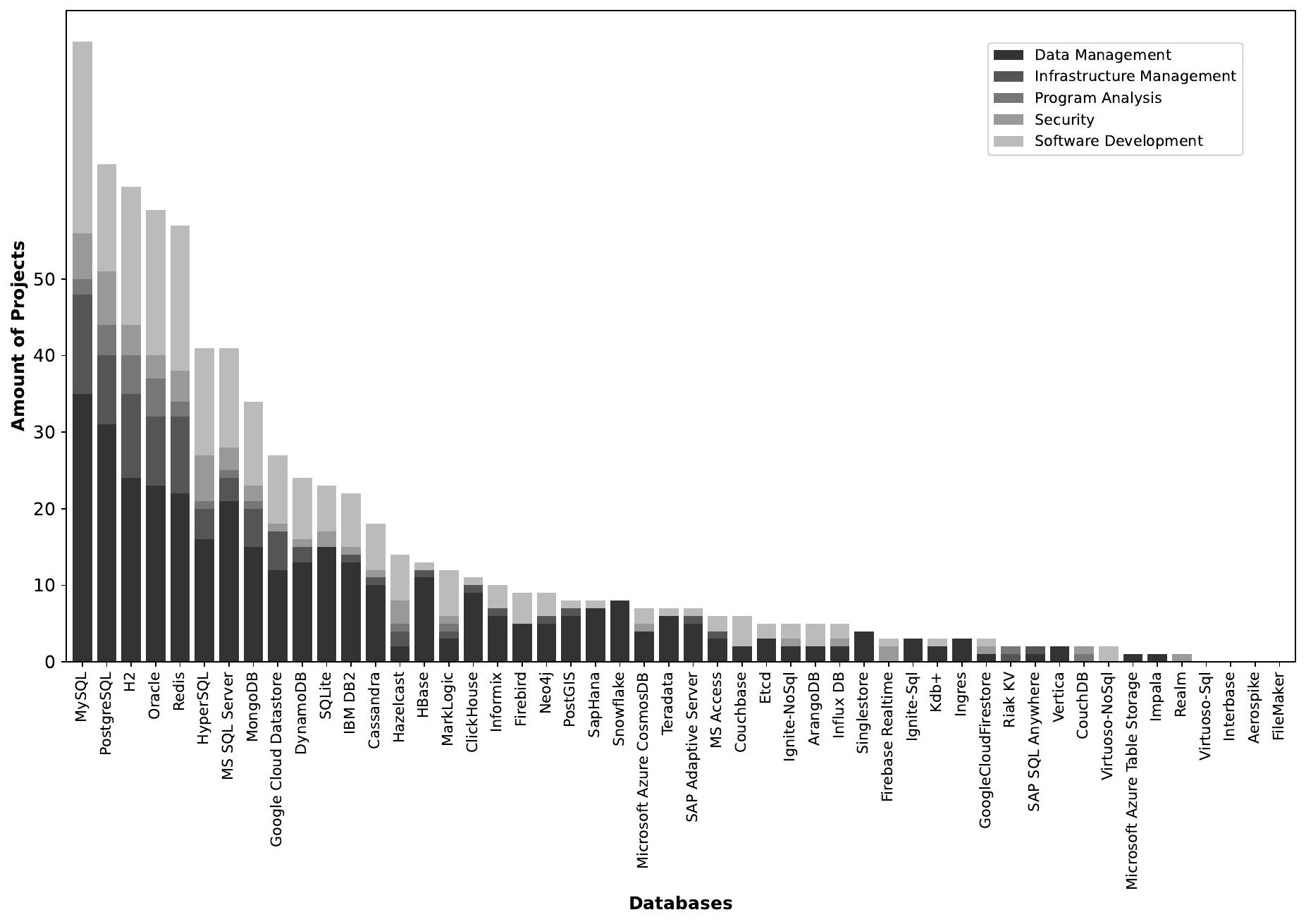}
\caption{DBMS by the top 5 application domains: Software Development, Data Management, Infrastructure Management, Program Analysis and Security.}
\label{fig:Top-5-domains}
\end{figure}

~

\paragraph{Domains.} To identify if there is a relation between database systems and project domains, we linked the DBMS discovered for each project and its domain. The results for the five most popular domains in our corpus are shown in Figure \ref{fig:Top-5-domains}. We observe that the most used DBMSs among the top-5 domains are MySQL, PostgreSQL, H2, Oracle, and Redis, following the pattern described above in DBMS analysis. We can see in Figure \ref{fig:Top-5-domains} that the number of projects in the top 5 domains per DBMS has a long tail, i.e., there is a large number of DBMSs that are rarely used and a small number of DBMSs with high usage rates by the projects in the top 5 domains.
Projects in the Software Development domain use \nvarc{rq1_software_development_dbms}{35} different DBMS. The most used are MySQL, Redis, Oracle, H2, and PostgreSQL. Most of these databases share some features, such as SQL support and in-memory capabilities. MySQL, Oracle, H2, and PostgreSQL are relational databases, while Redis is a key-value store. These DBMS vary in terms of scalability and persistence, reflecting the diverse needs of software development projects.

\vspace{+5mm}

\begin{footnotesize}
\begin{centering}
  \setlength{\fboxrule}{0.1em}
  \setlength{\fboxsep}{1em}
  \fbox{%
    \parbox{0.95\linewidth}{%
    \textbf{\\RQ1: Which DBMS are the most popular across software projects? }\\
    
    \textbf{Answer: } Relational DBMSs are more popular in our corpus than Non-Relational DBMSs. We found evidence of usage of all the \nvarc{rq1_non_relational_dbms}{25} non-relational DBMSs we searched for against only \nvarc{rq1_relational_dbms_found_total}{22} of the \nvarc{rq1_relational_dbms}{25} relational DBMSs we searched for. MySQL is the most popular relational DBMS in our corpus. It is the most used DBMS in the Software Development, Data Management, and Infrastructure Management domains. As for non-relational DBMS, the most used is Redis. It is also the most used non-relational DBMS in the Data Management, Software Development, and Infrastructure Management domains, indicating that both MySQL and Redis are frequently employed within the same domains.
    }%
}    
\end{centering}
\end{footnotesize}

\subsection{(RQ2) How stable are the DBMSs during the projects' history? } 
\label{RQ2}

In this question, we investigate the stability of a DBMS throughout the projects' history, considering the established heuristics. It is important to highlight that we consider that a project adopted a DBMS if it appeared in any segment of the project's history, even if it was no longer present in the final segment.
%It is important to highlight that once we find the adoption of a DBMS, even though it was discontinued from the project for any reason, we quantified it. %For this reason, the values presented in our results may not coincide with those collected in the characterization study that preceded the historical analysis.

\begin{figure}%[!htbp]
\centering
\responsetoreviewer{rIIcXIVprojectsperDBpI}{
\includegraphics[width=1.0\textwidth]{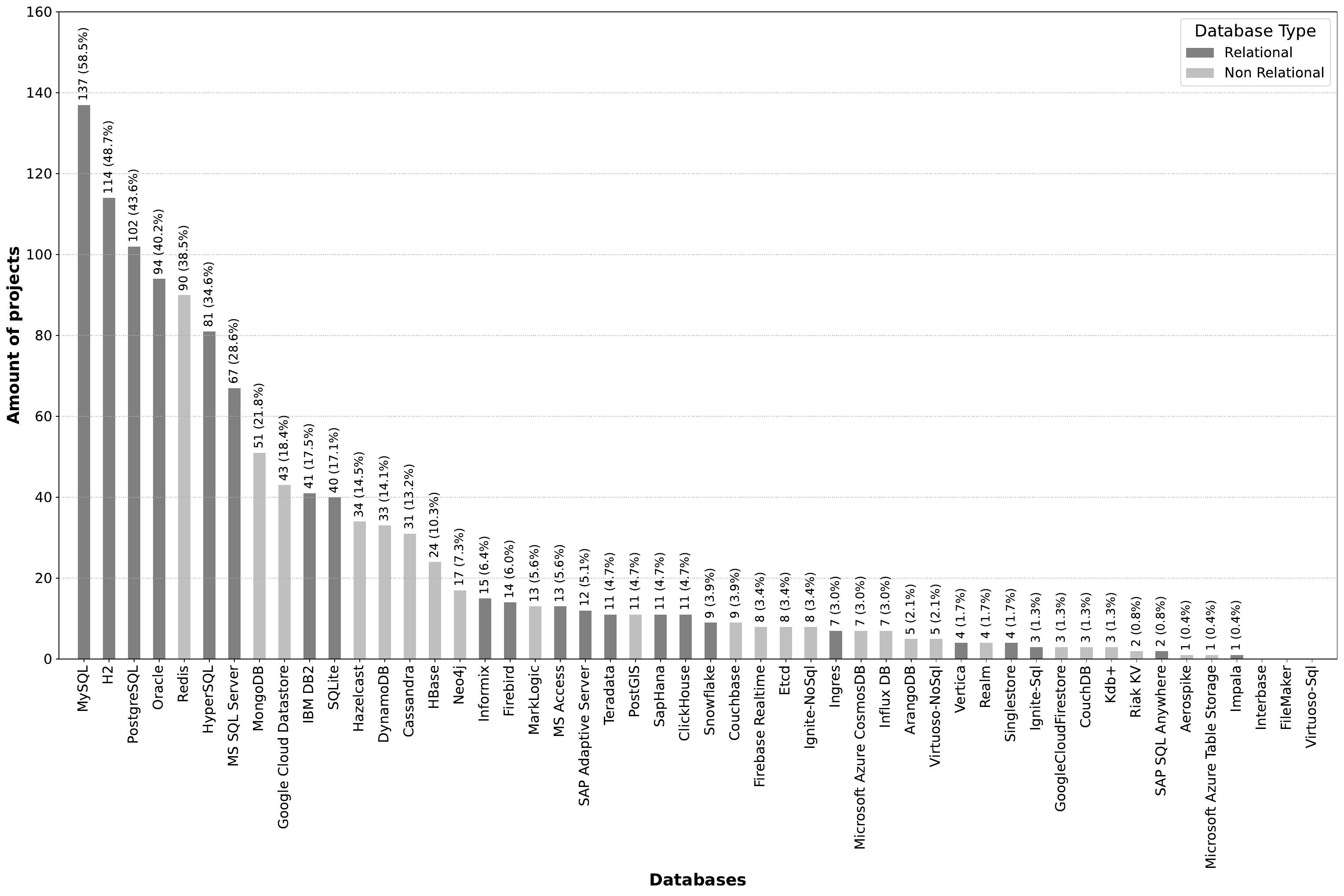}
}
\caption{\responsetoreviewer{rIIcXIVprojectsperDBpII}{Most Popular DBMSs in the history of Java projects.}}
\label{fig:projectsperDB}
\end{figure}

Figure~\ref{fig:projectsperDB} presents the most popular DBMSs in the projects of our corpus, considering the historical analysis performed across all equal-sized slices in terms of the commits of each project. As described in Section \ref{sec:databaseHeuristics}, we applied 52 sets of heuristics for the 50 surveyed DBMSs since some are multi-model. We found evidence of the adoption of \nvarc{rq2_dbms_multimodel_combined}{46} DBMSs in \nvarc{rq2_projects_with_dbms_history_total}{234} of the \nvarc{rq2_projects}{362} projects in our corpus. Compared to the results we obtained for RQ1 (in which we look only at the last version of each project), we note a larger number of projects with evidence of DBMS usage (\nvarc{rq2_projects_with_dbms_history_total}{234} for RQ2 versus \nvarc{rq1_projects_with_dbms}{202} for RQ1). This means that \nvarc{rq2_projects_with_dbms_diff_total}{32} projects in our corpus stopped using a DBMS at some point or switched to one that is not on our list of 50 DBMSs. We intend to further investigate this matter in future work. 

The top positions are occupied by MySQL, which appears in \nvarc{rq2_mysql_total}{137} projects, followed by H2, appearing in \nvarc{rq2_h2_total}{114} projects. PostgreSQL came in third, present in \nvarc{rq2_postgresql_total}{102} projects, then Oracle in \nvarc{rq2_oracle_total}{94} projects, and Redis in \nvarc{rq2_redis_total}{90} projects. Thus, MySQL was present in about \nvarc{rq2_mysql}{58.5\%} of projects that use a DBMS, H2 occurred in about \nvarc{rq2_h2}{48.7\%}, PostgreSQL in about \nvarc{rq2_postgresql}{43.6\%}, Oracle was present in about \nvarc{rq2_oracle}{40.2\%}, and Redis in about \nvarc{rq2_redis}{38.5\%} of these projects. The top-5 DBMS are the same when we compare the results with the ones in RQ1, however, there is a change of position between H2 and PostgreSQL.

As can be noted, following the same results as RQ1, the relational model is the most commonly used (\nvarc{rq2_relational}{85.0\%} of the projects that use a DBMS, use a relational DBMS). Still, the non-relational (NoSQL) model is also present.  We also found evidence of Ignite being adopted by 3.8\% of the projects in both categories: \nvarc{rq2_ignite_sql_ext}{one} project uses it just as a relational database, and \nvarc{rq2_ignite_nosql_ext}{six} projects use it as a non-relational database, and \nvarc{rq2_ignite_both_ext}{two} projects use it as a multi-model database.

We also observed that MySQL, PostgreSQL, and Oracle are among the most used DBMSs in the DB-Engines ranking \citep{db-engines_2022}, even considering the different contexts between this ranking and our research. On the other hand, H2 (Hypersonic 2) is our second most adopted DBMS and occupies the 49$_{th}$ position in this ranking (as of October 2024). The fact that H2 is among the top-2 DBMSs in our study might have happened due to the simplicity of integration of H2 in Java projects\footnote{http://www.h2database.com/html/history.html}.

Given the \nvarc{rq2_dbms_multimodel_combined}{46} DBMSs found, we observed that the adopted DBMSs are distributed as follows: \nvarc{rq2_relational_dbms_found_total}{22} relational DBMSs, \nvarc{rq2_non_relational_dbms_found_total}{25} non-relational DBMSs, and one multi-model DBMS (Ignite) that appears in both categories. This information reinforces the increasing trend of adoption of non-relational models among Java projects. According to \cite{davoudian2018survey}, non-relational DBMSs are not designed to replace relational DBMSs but as a solution to the gaps regarding the need for scalability and availability that certain distributed applications require. 

\begin{figure}%[!htbp]%[width=0.6\textwidth]
%\align{left}
\centering
\includegraphics[width=0.4\textwidth]{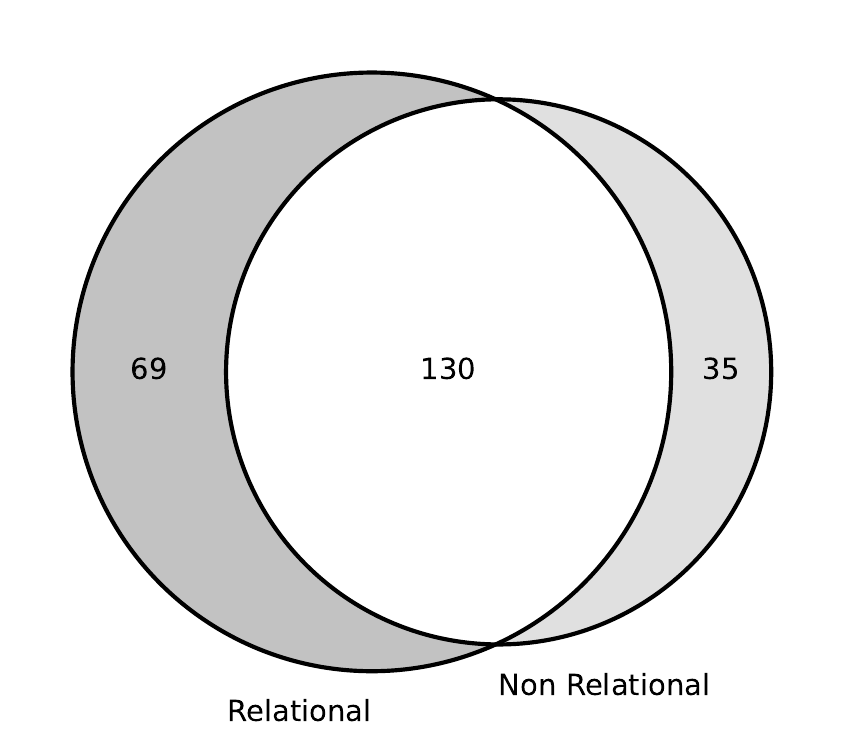}
\caption{Adoption of DBMS in the history of the projects, clustered by their data model.}
\label{fig:venn_datamodels}
\end{figure}

The Venn Diagram shown in Figure~\ref{fig:venn_datamodels} shows the number of projects classified by the type of data model they adopt.
Out of the \nvarc{rq2_projects_with_dbms_history_total}{234} projects in which we found evidence of the use of DBMSs, \nvarc{rq2_relational_projects_total}{69} adopt only the relational model, and \nvarc{rq2_non_relational_projects_total}{35} adopt only the non-relational model. However, we found an intersection where both models are adopted (\nvarc{rq2_intersection_projects_total}{130} projects). Thus, \nvarc{rq2_intersection_projects}{55.6\%} of these projects adopt both data models, while \nvarc{rq2_single_projects}{44.4\%} use only one of the models. The understanding that both models can complement each other is reflected in about half of the projects that use a DBMS in our corpus.

We did not find evidence of usage of \nvarc{rq2_no_usage_ext}{three} of the 50 DBMSs that we searched for: Virtuoso-SQL, Interbase, and FileMaker. This is the same result that we found for RQ1. Interbase and Virtuoso are not very popular DBMSs, which may explain their absence in our findings. In fact, in the DB-Engines ranking \citep{db-engines_2022}, they appear at the 76$^{th}$ and 72$^{nd}$ positions, respectively. Note that they were included in our analysis among the 50 popular DBMSs since lots of DBMSs that are listed before them in the ranking were discarded due to several reasons, as explained in Section~\ref{sec:dbcorpus}. Regarding the absence of FileMaker, we have not identified any explanation from the available data, as it is a well-established Database Management System, holding the 20$^{th}$ position in the aforementioned ranking. We believe the absence of a free version is the key to understanding its absence in our corpus since license costs can be a deterrent for OSS projects. 
 
\begin{figure}%[!htbp]%[width=0.5\textwidth]
\centering
\responsetoreviewer{rIIcXVdomaindbmodelspI}{
\includegraphics[width=1.0\textwidth]{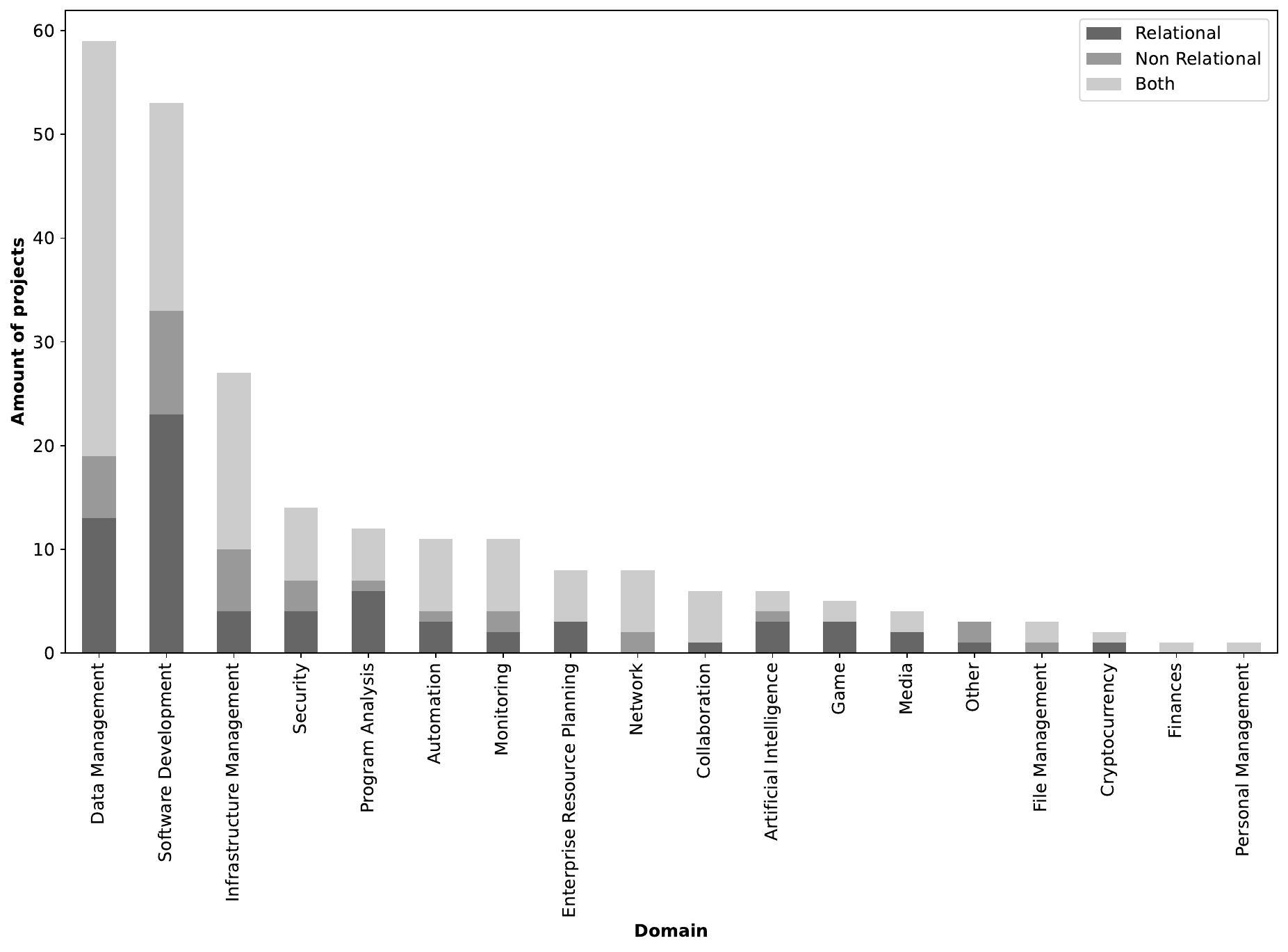}
}
\caption{\responsetoreviewer{rIIcXVdomaindbmodelspII}{Distribution of DBMS models by project domains, when considering the history of the projects.}}
\label{fig:domain_dbmodels}
\end{figure}

We also analyzed the adoption of DBMSs in the history of the projects, clustering the projects by their domains and the DBMSs by their data model. Figure~\ref{fig:domain_dbmodels} presents the results. The intention of this analysis was to discover which data models are used the most in the various project domains of our corpus. We found \nvarc{rq2_data_management_both}{40} projects from the Data Management domain using both relational and non-relational DBMSs, whereas \nvarc{rq2_data_management_sql}{13} projects used only relational DBMSs and \nvarc{rq2_data_management_nosql}{6} used only non-relational DBMSs. The Data Management domain showed the highest adoption of using both models together. %As examples, the Flink project showed adoption of Cassandra, ClickHouse, DynamoDB, H2, HBase, MySQL, MarkLogic, MS SQL Server, Oracle, PostgreSQL, and Redis; the Flyway project showed adoption of Firebird, H2, Hbase, HyperSQL, IBM DB2, Ignite, Informix, MySQL, MS SQL Server, Oracle, PostgreSQL, SAP Adaptive Server, SapHana, Singlestore, SAP SQL Anywhere, Snowflake, SQLite, and Vertica; and the GeoTools project showed adoption of Google Cloud Datastore, H2, HyperSQL, IBM DB2, Informix, Ingres, MySQL, MongoDB, MS Access, MS SQL Server, Oracle, PostgreSQL, SAP Adaptive Server, SapHana, SAP SQL Anywhere, SQLite, and Teradata.

Combining DBMS models (see Section \ref{sec:dbcorpus}) surpass relational and non-relational models in the Data Management, Infrastructure Management, Security, Automation, Monitoring, Enterprise Resource Planning, Network, Collaboration, and File Management domains. Thus, out of the \nvarc{rq2_corpus_domains}{19} domains in our corpus, the combination of models in DBMSs is predominant in \nvarc{rq2_domain_predominant_total}{11}, corresponding to \nvarc{rq2_domain_predominant}{57.9\%} of the domains. This discovery reinforces the trend of using more than one distinct data model and that they can complement each other \citep{sahatqija2018comparison}. For instance, all projects from the Personal Management and Finances domains use only multi-model DBMSs. We also observed that no domain used only one DBMS model. Additionally, the \nvarc{rq2_hpc_ext}{three} projects in the High-Performance Computing domain showed no indication of database usage.

\vspace{+5mm}

{
\begin{footnotesize}
\begin{centering}
  \setlength{\fboxrule}{0.1em}
  \setlength{\fboxsep}{1em}
  \fbox{%
    \parbox{0.95\linewidth}{%
    \textbf{\\RQ2: How stable are the DBMSs during the projects' history?}\\
    
    \textbf{Answer: } MySQL, H2, and PostgreSQL are among the three most used relational DBMSs, while Redis and MongoDB are the most used non-relational DBMSs. Half of the projects adopted both relational and non-relational databases. This co-occurrence of models was especially prevalent in projects of the Data Management domain.\\
    %\textbf{Implications: }Certain DBMSs have consistently maintained a high level of adoption, making them strong candidates to consider when choosing a DBMS.
    }%
    }%
\end{centering}
\end{footnotesize}
}

\subsection{(RQ3) Which DBMSs are frequently replaced by others?}
\label{RQ3}

In this question, we look for migration patterns between DBMSs and quantify the frequency they occur throughout the projects' history. 

\begin{figure}[t]
\centering
\responsetoreviewer{rIIcXVinsertionsandremovalspI}{
\includegraphics[width=1.0\textwidth]{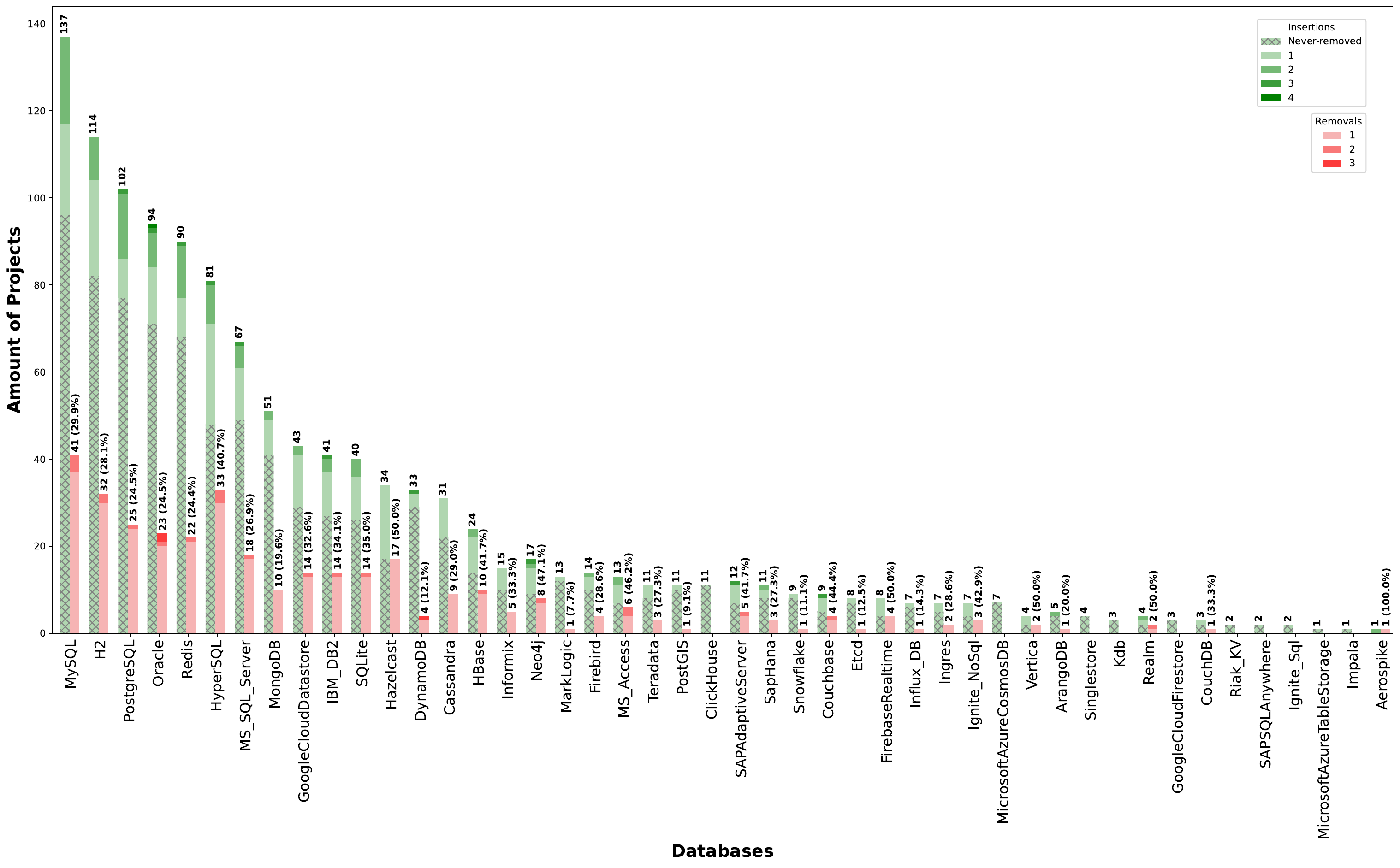}
}
\caption{\responsetoreviewer{rIIcXVinsertionsandremovalspII}{Distribution of DBMS insertions and removals in the projects. The removal percentages are calculated over the total number of insertions for each DBMS.}}
\label{fig:keptandremoved}
\end{figure}

Figure~\ref{fig:keptandremoved} presents the frequency of DBMSs being inserted or removed across the history of the projects. These numbers were calculated as follows. For each DBMS $A$, we iterate through each slice of each project to count how many times $A$ was inserted or removed. During the slice iteration, when a DBMS $A$ is found in one of the slices, we count it as the first insertion in the project. Then, we check the subsequent slices to see if $A$ is present there. If not, we consider that $A$ was removed from the project for the first time. We continue checking the subsequent slices, and if $A$ is found again, we update the insertion count to indicate that it was inserted twice. The process continues until we find the total number of insertions and removals per project. Then, we combine these numbers to analyze how many projects had 1, 2, 3, or 4 insertions of $A$ and how many projects had 1, 2, or 3 removals of $A$ throughout their histories. No projects in our corpus had more than 4 insertions or more than 3 removals. Finally, we also count how many projects inserted $A$ once and never removed it afterward.

To illustrate, remember that MySQL was found at some point in the history of \nvarc{rq2_mysql_total}{137} projects (see RQ2). This means that MySQL was inserted at least once in all of them. After the initial insertion, \nvarc{rq3_mysql_removal_total}{41} projects removed the DBMS (\nvarc{rq3_mysql_removal}{29.9\%} of the projects that used MySQL). After the removal, \nvarc{rq3_mysql_2nd_ins_total}{20} projects inserted MySQL for the second time,
and then, \nvarc{rq3_mysql_2nd_rem_total}{4} projects removed it again, leading to MySQL being present in the last slice of \nvarc{rq3_mysql_total}{112} projects. Note, however, that MySQL appears in the last commit of \nvarc{rq1_mysql_total}{113} projects (see RQ1). It occurs because MySQL was inserted into a project between the last slice and the last commit. Considering that MySQL was inserted twice for \nvarc{rq3_mysql_2nd_ins_total}{20} projects, \nvarc{rq3_mysql_once_total}{117} out of the \nvarc{rq2_mysql_total}{137} projects that adopted MySQL inserted it exactly once. Out of these \nvarc{rq3_mysql_once_total}{117} projects that inserted MySQL once, \nvarc{rq3_mysql_never_removed_total}{96} projects never removed it.

The removals happened for the majority of DBMSs we surveyed. Sometimes, more than once for the same project, with Oracle being removed 3 times and inserted 4 times in a single project, for instance. However, some DBMS were never removed once adopted by a project: \nvarc{rq3_no_removals}{ClickHouse (11 projects), MicrosoftAzureCosmosDB (7 projects), Singlestore (4 projects), Kdb (3 projects), GoogleCloudFilestore (3 projects), Riak\_KV (2 projects), SAPSQLAnywhere (2 projects), Ignite\_Sql (2 projects), MicrosoftAzureTableStorage (1 project), and Impala (1 project)}.

As one of our goals was to understand the frequency of DBMS replacements, the significant amount of removals we found had created the expectation that substitutions are frequent. To investigate this, we utilized sequential patterns mining \citep{fournier2017survey}. Each project in which we found a DBMS was transformed into a sequence, and each item in the sequence corresponds to a slice in the project's history. Thus, we have \nvarc{rq2_projects_with_dbms_history_total}{234} sequences, and the size of each sequence varies depending on the number of slices of each specific project. The sequences were coded according to the patterns defined in Section \ref{sec:replacement}. After generating the patterns, we filtered out those that met the established patterns to characterize the replacements, as discussed in Section~\ref{sec:replacement}.

In total, we found \nvarc{rq3_total_replacements}{296} replacement patterns across \nvarc{rq3_total_replacements_projects}{67} projects. These patterns breakdown as follows: \nvarc{rq3_relational_replacements_total}{89} replacement patterns among relational DBMSs (\nvarc{rq3_relational_replacements}{30.1\%} of the replacement patterns) across \nvarc{rq3_relational_replacements_projects_total}{44} projects (\nvarc{rq3_relational_replacements_projects}{65.7\%} of the projects with DBMS replacements), \nvarc{rq3_nonrelational_replacements_total}{66} replacement patterns among non-relational DBMSs (\nvarc{rq3_nonrelational_replacements}{22.3\%}) across \nvarc{rq3_nonrelational_replacements_projects_total}{21} projects (\nvarc{rq3_nonrelational_replacements_projects}{31.3\%}), and \nvarc{rq3_both_replacements_total}{141} replacement patterns involving both models (\nvarc{rq3_both_replacements}{47.6\%}) across \nvarc{rq3_both_replacements_projects_total}{43} projects (\nvarc{rq3_both_replacements_projects}{64.2\%}). 
Thus, in \nvarc{rq3_only_same_replacements_projects}{35.8\%} of the projects with replacements, the replacements only occur between DBMSs that follow the same data model. This may mean that data integrity is the relevant criterion when choosing a replacement. Replacements among non-relational DBMSs occur in only \nvarc{rq3_non_relational_only_replacements_projects}{12.5\%} of these cases. This indicates that the substitution between non-relational DBMSs is rarer, possibly because they provide different data types (graph, key-value, document, etc.), so replacing them is not an easy task. Finally, in \nvarc{rq3_both_replacements}{50.0\%} of the patterns, the replacements occur among distinct data models such as Cassandra replacing PostgreSQL in 3 projects. This may indicate that the project’s needs or data storage requirements have changed significantly in a given moment of the project's history, prompting a migration to a different data model. It may also suggest a growing trend toward adopting more diverse data models. This may occur due to some new desirable property that the DBMS in use could not provide, as reported by \citet{gessert2017nosql}.

\begin{table}[t]
\centering
\caption{DBMS replacement patterns.}
\begin{scriptsize}
\begin{tabular}{lr}
\hline
\textbf{Pattern} & \textbf{Support} \\
\hline
Cassandra $\rightarrow$ PostgreSQL$_{In}$ $\rightarrow$ Cassandra$_{Out}$    PostgreSQL & 3 \\
Cassandra $\rightarrow$ MS SQL Server$_{In}$ $\rightarrow$ Cassandra$_{Out}$    MS SQL Server & 3 \\
\hline
Couchbase $\rightarrow$ PostgreSQL$_{In}$ $\rightarrow$ Couchbase$_{Out}$    PostgreSQL & 3 \\
\hline
HBase $\rightarrow$ HyperSQL$_{In}$ $\rightarrow$ HBase$_{Out}$    HyperSQL & 4 \\
HBase $\rightarrow$ H2$_{In}$ $\rightarrow$ HBase$_{Out}$    H2 & 3 \\
HBase $\rightarrow$ MS SQL Server$_{In}$ $\rightarrow$ HBase$_{Out}$    MS SQL Server & 3 \\
HBase $\rightarrow$ Redis$_{In}$ $\rightarrow$ HBase$_{Out}$    Redis & 3 \\
\hline
Hazelcast $\rightarrow$ HyperSQL$_{In}$ $\rightarrow$ Hazelcast$_{Out}$    HyperSQL & 3 \\
\hline
HyperSQL $\rightarrow$ Redis$_{In}$ $\rightarrow$ HyperSQL$_{Out}$    Redis & 8 \\
HyperSQL $\rightarrow$ PostgreSQL$_{In}$ $\rightarrow$ HyperSQL$_{Out}$    PostgreSQL & 5 \\
HyperSQL $\rightarrow$ MySQL$_{In}$ $\rightarrow$ HyperSQL$_{Out}$    MySQL & 5 \\
HyperSQL $\rightarrow$ MongoDB$_{In}$ $\rightarrow$ HyperSQL$_{Out}$    MongoDB & 4 \\
HyperSQL $\rightarrow$ H2$_{In}$ $\rightarrow$ HyperSQL$_{Out}$    H2 & 3 \\
\hline
MySQL $\rightarrow$ Redis$_{In}$ $\rightarrow$ MySQL$_{Out}$    Redis & 3 \\
\hline
Oracle $\rightarrow$ MySQL$_{In}$ $\rightarrow$ Oracle$_{Out}$    MySQL & 3 \\
\hline
PostgreSQL $\rightarrow$ Oracle$_{In}$    PostgreSQL$_{Out}$ $\rightarrow$ Oracle & 3 \\
\hline
SQLite $\rightarrow$ H2$_{In}$ $\rightarrow$ SQLite$_{Out}$    H2 & 4 \\
SQLite $\rightarrow$ MS SQL Server$_{In}$ $\rightarrow$ SQLite$_{Out}$    MS SQL Server & 3 \\
\hline
\end{tabular}

\end{scriptsize}
\label{tab:patterns}
\end{table}

Table~\ref{tab:patterns} shows \nvarc{rq3_replacements}{18} sequential patterns that indicate DBMS replacements that occurred in three or more projects ($Support \geq 3$). We opted for this support threshold to remove incidental patterns that occurred only once or twice in our corpus. An example is the pattern \textbf{$PostgreSQL \rightarrow Oracle_{In}$ $PostgreSQL_{Out} \rightarrow$} \textbf{$Oracle$} with $Support = 3$. This indicates that three projects used PostgreSQL, then in a later slice started using Oracle and stopped using PostgreSQL, maintaining  Oracle in a later slice. The pattern \textbf{$HyperSQL \rightarrow Redis_{In} \rightarrow HyperSQL_{Out}$ $Redis$} with support = 8 indicates that eight projects replaced HyperSQL by Redis. These two examples present replacement patterns whose changes occur in different ways. In the first example, the usage of Oracle started at the same time that PostgreSQL left. In the second example, Redis enters at a given moment, and at a later time when HyperSQL leaves, Redis is maintained. Although the patterns shown in Table \ref{tab:patterns} confirmed the existence of replacements, they do not indicate that a specific substitution occurred frequently---since each pattern demonstrates that DBMSs were replaced in at least \nvarc{rq3_min_support}{3} and at most \nvarc{rq3_max_support}{8} projects (Support = \{\nvarc{rq3_support_values}{3, 4, 5, 8}\}).

\begin{table}[b]
\centering
\caption{Occurrences of DBMSs replacements and count of projects where other DBMSs replaced them.}
\begin{tabular}{lrrr}
\hline
\multicolumn{1}{l}{\textbf{Replaced DBMS}} & \multicolumn{1}{r}{\textbf{\# Replacer DBMSs}} & \multicolumn{1}{c}{\textbf{\# Projects}} & \textbf{\% Projects} \\
\hline
Couchbase & 13 & 3 & 33.3\% \\
SAP Adaptive Server & 10 & 4 & 33.3\% \\
Informix & 9 & 5 & 33.3\% \\
CouchDB & 5 & 1 & 33.3\% \\
Ingres & 8 & 2 & 28.6\% \\
Ignite-NoSql & 10 & 2 & 25.0\% \\
Realm & 4 & 1 & 25.0\% \\
HyperSQL & 18 & 19 & 23.5\% \\
Firebird & 4 & 3 & 21.4\% \\
HBase & 18 & 5 & 20.8\% \\
Hazelcast & 8 & 7 & 20.6\% \\
SQLite & 15 & 8 & 20.0\% \\
Teradata & 10 & 2 & 18.2\% \\
SapHana & 4 & 2 & 18.2\% \\
Neo4j & 4 & 3 & 17.6\% \\
IBM DB2 & 8 & 7 & 17.1\% \\
MS SQL Server & 13 & 11 & 16.4\% \\
Cassandra & 14 & 5 & 16.1\% \\
MS Access & 5 & 2 & 15.4\% \\
Influx DB & 1 & 1 & 14.3\% \\
Firebase Realtime & 2 & 1 & 12.5\% \\
Google Cloud Datastore & 12 & 5 & 11.6\% \\
Snowflake & 1 & 1 & 11.1\% \\
PostgreSQL & 14 & 11 & 10.8\% \\
H2 & 14 & 12 & 10.5\% \\
MySQL & 13 & 13 & 9.5\% \\
PostGIS & 1 & 1 & 9.1\% \\
Oracle & 11 & 8 & 8.5\% \\
MongoDB & 5 & 4 & 7.8\% \\
Redis & 6 & 7 & 7.8\% \\
MarkLogic & 3 & 1 & 7.7\% \\
DynamoDB & 5 & 1 & 3.0\% \\
\hline
\end{tabular}

\label{tab:substitutions}
\end{table}

As depicted in Table \ref{tab:patterns}, some DBMSs are replaced by several others, with frequent patterns indicating that a DBMS $A$ was replaced by a specific DBMS $B$ \new{in multiple projects. To further investigate the replacements, for each replaced DBMS $A$, we count the total number of replacer DBMSs $B$, and the total number of projects where $A$ was replaced. We display the results in Table~\ref{tab:substitutions}.} Our replacement counts consider the number of DBMSs that appeared at the same time or after a given DBMS was dropped on a given project. For example, HyperSQL was replaced \new{by \nvarc{rq3_repl_hypersql}{18} different DBMSs} in \nvarc{rq3_repl_hypersql_projects}{19} projects. In \nvarc{rq3_repl_hypersql_redis_ext}{eight} of those \new{projects, Redis figures among the replacers}. In the opposite direction, we did not find a pattern indicating HyperSQL replacing Redis. \new{Similar to HyperSQL, HBase was replaced by \nvarc{rq3_repl_hbase}{18} DBMSs. However, these replacements occurred in only \nvarc{rq3_repl_hbase_projects}{5} projects, and, in \nvarc{rq3_repl_hbase_hypersql}{four} of those projects, HyperSQL figures among HBase replacers.}

\new{In some situations, we did not find frequent patterns. For instance, MySQL was replaced by \nvarc{rq3_repl_mysql}{13} different DBMSs in \nvarc{rq3_repl_mysql_projects}{13} projects, but the replacements did no share a common replacer DBMS, overall.} The only MySQL replacement pattern that occurred frequently enough ($Support \ge 3$) to appear in Table \ref{tab:patterns} was its replacement by Redis, with no replacements in the opposite direction. 
Given these results and the expectations generated by the significant amount of removals (Figure~\ref{fig:keptandremoved}), we observe that substitutions between DBMSs are not frequent but do occur.

In summary, we have identified that HyperSQL is the DBMS mostly susceptible to replacements during the projects' history, since it is replaced in 19 different projects. Additionally, we observed that most replaced DBMSs have experienced more than one replacement in certain projects. This means that when a DBMS is removed, it may be replaced by more than one alternative DBMS, either in the same slice or in later slices of the project history. For example, in the Apache/Camel project, we discovered 14 distinct patterns of HBase substitutions for different DBMSs: \nvarc{rq3_apache_camel_hbase}{ArangoDB, Cassandra, Couchbase, Etcd, H2, HyperSQL, Ignite-NoSql, Influx DB, MS SQL Server, Microsoft Azure CosmosDB, MySQL, PostGIS, PostgreSQL, and Redis}.

\begin{table}[t]
\centering
\caption{Occurrences of DBMS substitutes and count of projects where they replaced other DBMSs.}
\begin{tabular}{lrrr}
\hline
\multicolumn{1}{l}{\textbf{Replacer DBMS}} & \multicolumn{1}{r}{\textbf{\# Replaced DBMSs}} & \multicolumn{1}{c}{\textbf{\# Projects}} & \textbf{\% Projects} \\
\hline
Microsoft Azure CosmosDB & 7 & 4 & 57.1\% \\
SAP SQL Anywhere & 2 & 1 & 50.0\% \\
Influx DB & 5 & 3 & 42.9\% \\
ArangoDB & 2 & 2 & 40.0\% \\
Etcd & 11 & 3 & 37.5\% \\
Cassandra & 18 & 11 & 35.5\% \\
Couchbase & 4 & 3 & 33.3\% \\
GoogleCloudFirestore & 2 & 1 & 33.3\% \\
Neo4j & 12 & 5 & 29.4\% \\
Ingres & 3 & 2 & 28.6\% \\
MS SQL Server & 18 & 19 & 28.4\% \\
MongoDB & 15 & 14 & 27.5\% \\
PostGIS & 5 & 3 & 27.3\% \\
HBase & 11 & 6 & 25.0\% \\
Ignite-NoSql & 2 & 2 & 25.0\% \\
Redis & 15 & 22 & 24.4\% \\
PostgreSQL & 17 & 24 & 23.5\% \\
MarkLogic & 5 & 3 & 23.1\% \\
Oracle & 17 & 19 & 20.2\% \\
DynamoDB & 10 & 6 & 18.2\% \\
ClickHouse & 5 & 2 & 18.2\% \\
Teradata & 2 & 2 & 18.2\% \\
H2 & 15 & 19 & 16.7\% \\
Hazelcast & 10 & 5 & 14.7\% \\
Google Cloud Datastore & 9 & 6 & 14.0\% \\
HyperSQL & 11 & 11 & 13.6\% \\
Informix & 4 & 2 & 13.3\% \\
MySQL & 13 & 18 & 13.1\% \\
Firebase Realtime & 2 & 1 & 12.5\% \\
IBM DB2 & 7 & 5 & 12.2\% \\
Snowflake & 3 & 1 & 11.1\% \\
SapHana & 1 & 1 & 9.1\% \\
SAP Adaptive Server & 2 & 1 & 8.3\% \\
SQLite & 3 & 3 & 7.5\% \\
\hline
\end{tabular}
\label{tab:substitutes}
\end{table}

Conversely, Table \ref{tab:substitutes} presents the DBMSs that replaced others. This analysis aimed to find out which DBMSs are mostly used as \new{replacements}, regardless of which DBMS they replaced. \new{Some DBMSs replaced many distinct DBMSs. For instance, MS SQL Server and Cassandra replaced, each, 18 distinct DBMSs in 19 and 11 projects, respectively. However, from the viewpoint of the number of projects where the replacement occurred,} the DBMSs most chosen to replace others with are Redis \new{(which replaced 15 DBMSs in 22 projects)} and PostgreSQL \new{(which replaced 17 DBMSs in 24 projects)}. 

\begin{figure*}
    \centering
    \includegraphics[width=\textwidth]{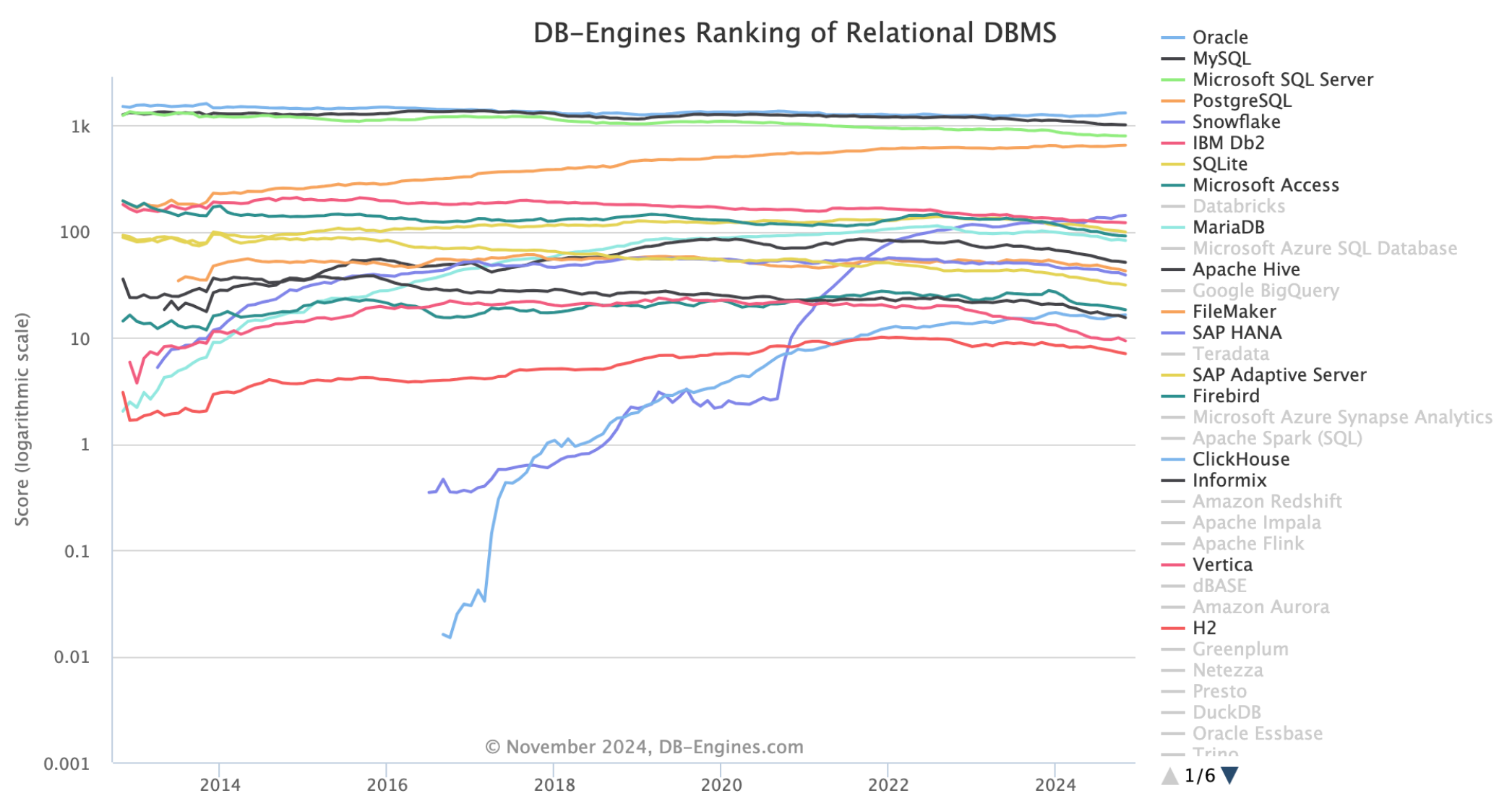}
    \caption{Relational DBMSs trend. Source: DB-Engines.}
    \label{fig:trend-relational}
\end{figure*}

\begin{figure*}
    \centering
    \includegraphics[width=\textwidth]{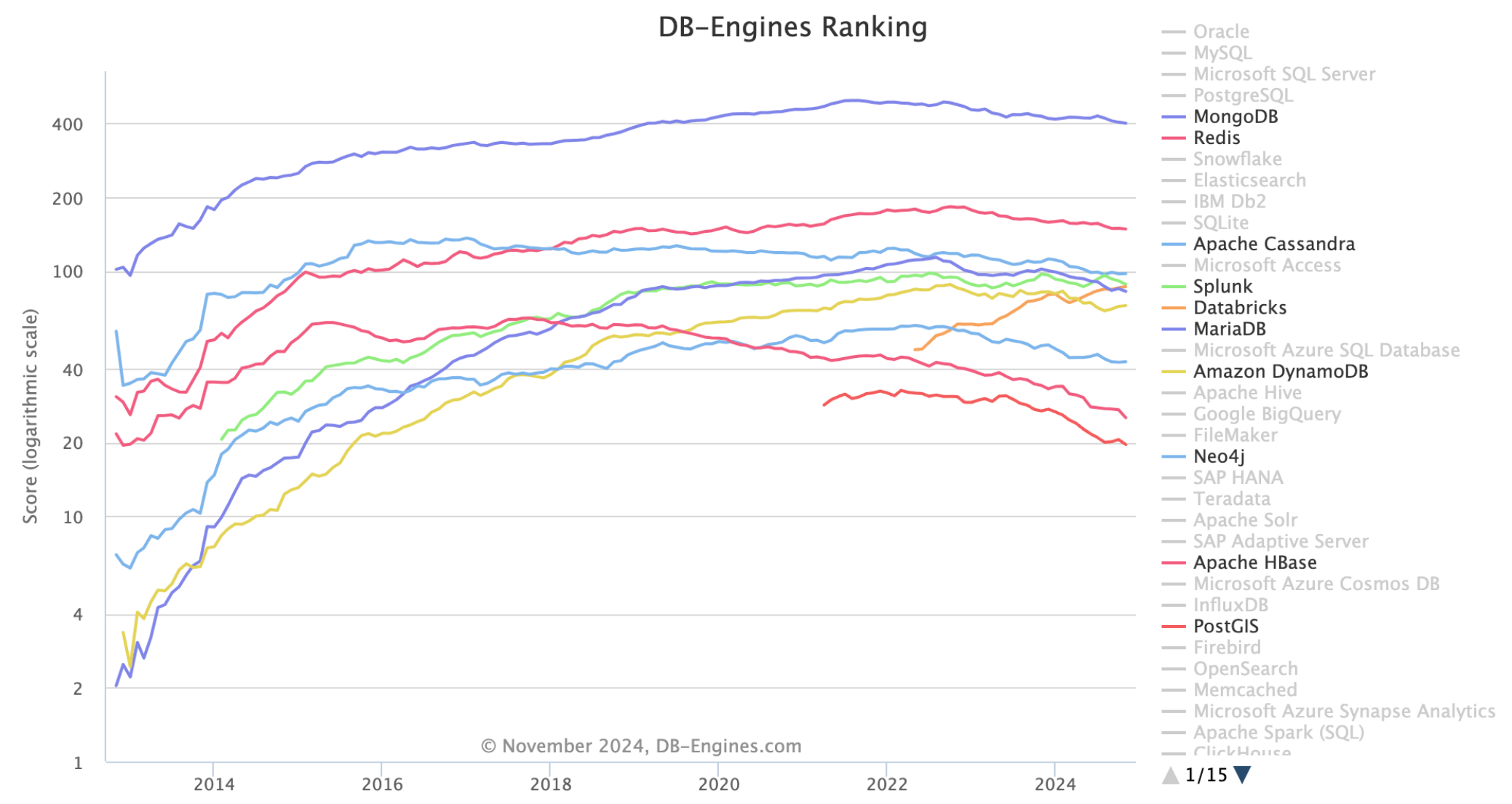}
    \caption{Non-relational DBMSs trend. Source: DB-Engines.}
    \label{fig:trend-non-relational}
\end{figure*}

The tendency for non-relational DBMSs to be used as replacements is also relevant. Figures~\ref{fig:trend-relational} and \ref{fig:trend-non-relational}, extracted from the DB-Engines website, present this trend by showing the evolution in the popularity of relational (Figure \ref{fig:trend-relational}) and non-relational (Figure \ref{fig:trend-non-relational}) DBMSs from 2014 to 2024 (note the logarithmic scale). Although a certain stability is observed in the popularity of relational DBMSs, the growth of non-relational DBMSs is notable. This trend reinforces some of Cattel's predictions \citep{cattell2011scalable} that NoSQL DBMSs would not be a ``passing fad" due to their simplicity, flexibility, and scalability. He predicted that developers would accept these advantages to the detriment of ACID transactions (ACID transactions are those that guarantee Atomicity, Consistency, Isolation, and Durability \citep{elmasri2010} -- relaxing consistency may lead to faster transactions, with the penalty that users may see old versions of the data at given points in time). Thus, this growing preference for non-relational solutions can be related to the business's need to deal with fast lookup or more complex querying capabilities that require a greater volume of data \citep{gessert2017nosql}, such as textual searches or choosing a standard for exchanging data.

\new{Comparing Tables~\ref{tab:substitutions} and \ref{tab:substitutes}, we observe that both HyperSQL and HBase were replaced by other \nvarc{rq3_repl_hypersql}{18} DBMSs and replaced other \nvarc{rq3_replby_hypersql}{11} DBMSs, indicating a decrease in their adoption.} 
On the other hand, Etcd (a distributed key-value DBMS) was not replaced but substituted \new{\nvarc{rq3_replby_etcd}{11} DBMSs}, suggesting a significant increase in its usage among the projects. This way, we can perceive which DBMSs tend to be discontinued and which tend to be most selected to replace others.

Nevertheless, despite discovering evidence of DBMS substitutions and some DBMSs being more prone to being replaced than others, we did not observe frequent patterns of a specific DBMS $A$ always being substituted by DBMS $B$. The growing co-occurrence among DBMSs, as we will show in Section~\ref{RQ4}, may influence the situation, making migrations from one DBMS to another unnecessary. This further supports the notion that DBMSs can complement each other, and the projects are utilizing multiple DBMSs to meet their specific needs.

\vspace{+5mm}

{
\begin{footnotesize}
\begin{centering}
  \setlength{\fboxrule}{0.1em}
  \setlength{\fboxsep}{1em}
  \fbox{%
    \parbox{0.95\linewidth}{%
    \textbf{\\RQ3: Which DBMSs are frequently replaced by others?}\\
    
    \textbf{Answer: }Our results reveal \nvarc{rq3_replacements}{18} patterns of DBMS replacements in projects. The most frequent involves HyperSQL. It was replaced in \nvarc{rq3_repl_hypersql_redis}{8} projects by Redis, \nvarc{rq3_repl_hypersql_postgresql}{7} projects by PostgreSQL, and \nvarc{rq3_repl_hypersql_mysql}{6} projects by MySQL. Overall, in \nvarc{rq3_same_replacements}{52.4\%} of the cases, the replacements occurred between DBMSs that have the same model. The remaining \nvarc{rq3_both_replacements}{47.6\%} of the replacements involve both data models. In a more comprehensive analysis, considering only the replaced DBMS, including the patterns that occurred in only 1 project, we found HyperSQL and HBase among the most susceptible to be replaced \new{-- each were replaced by \nvarc{rq3_repl_hypersql}{18} DBMSs in \nvarc{rq3_repl_hypersql_projects}{19} and 5 projects, respectively.}. Also, considering just the \new{number of projects}, we found Redis and PostgreSQL among the most used as a replacement.\\
    %\textbf{Implications: }Some DBMSs undergo frequent replacement by other DBMSs. Investigating the existence of replacement patterns allows for knowing the DBMS migration trends in the projects.
    }%
    }%
\end{centering}
\end{footnotesize}
}

\subsection{(RQ4) Which DBMSs are often used together?}
\label{RQ4}

\begin{figure}[t]%[width=0.5\textwidth]
    \centering
    \responsetoreviewer{rIIcXVIIfigIpI}{\includegraphics[width=1.0\textwidth]{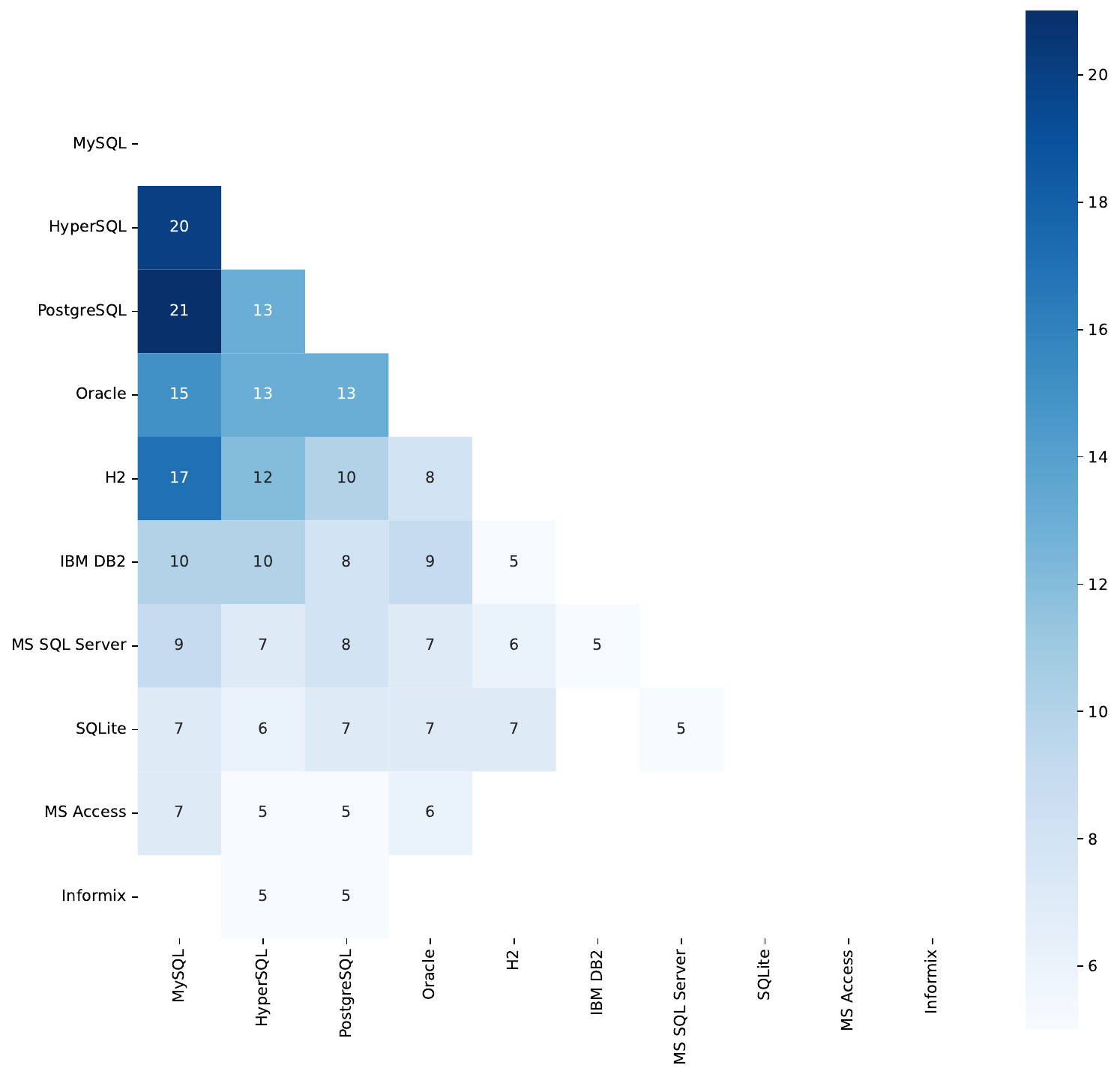}}
    \caption{\responsetoreviewer{rIIcXVIIfigIpII}{Correlation of the most frequent DBMSs at the beginning of the projects' history.}}
    \label{fig:heat_onlyrules_v1} 
\end{figure}

\begin{figure}[t]
    \centering 
    \responsetoreviewer{rIIcXVIIfigIIpI}{\includegraphics[width=1.0\textwidth]{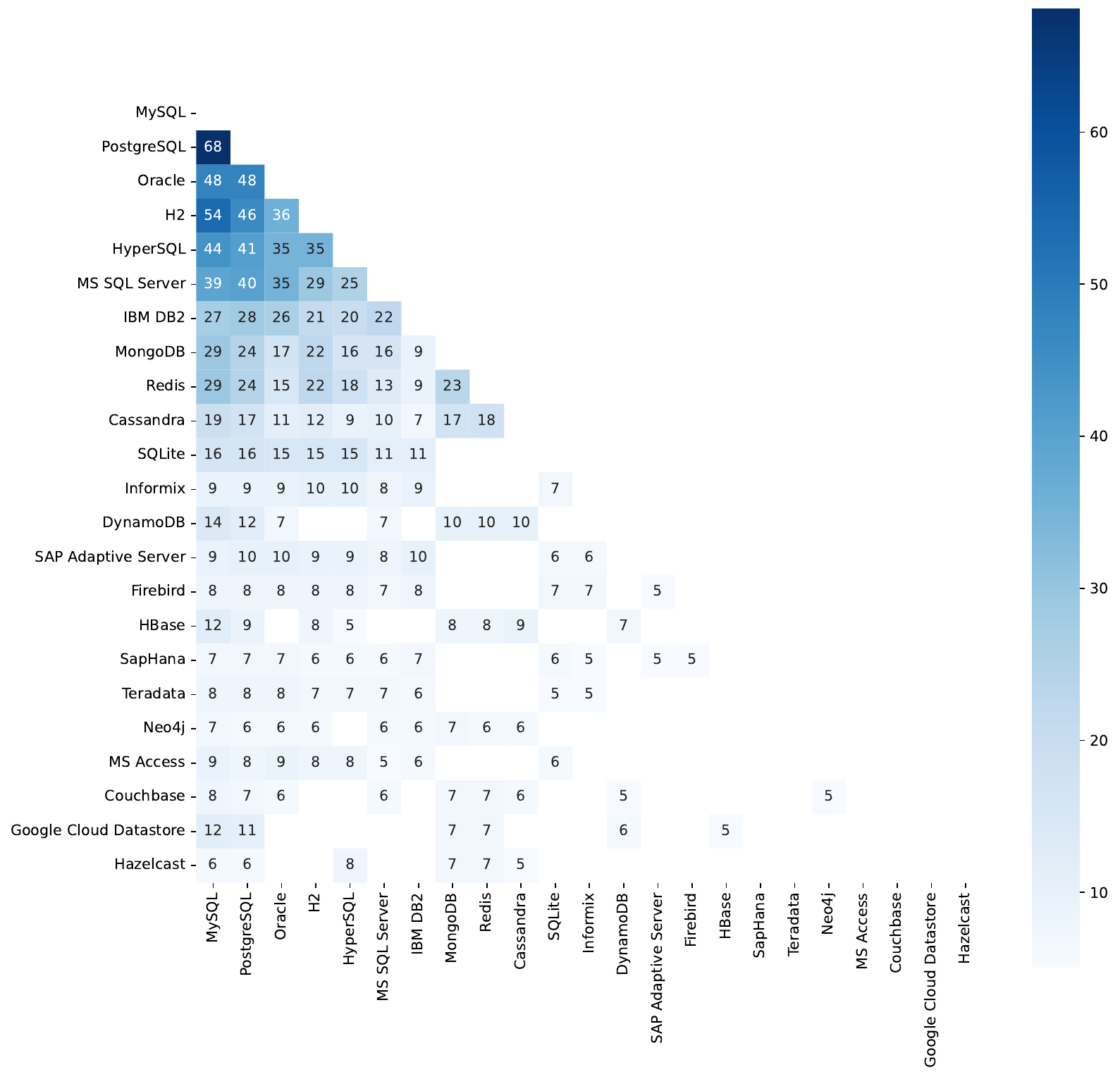}}
    \caption{\responsetoreviewer{rIIcXVIIfigIIpII}{Correlation of the most frequent DBMSs in the middle of the projects' history.}}
    \label{fig:heat_onlyrules_v5}   
\end{figure}

\begin{figure}[!htbp]
    \centering 
    \responsetoreviewer{rIIcXVIIfigIIIpI}{\includegraphics[width=1.0\textwidth]{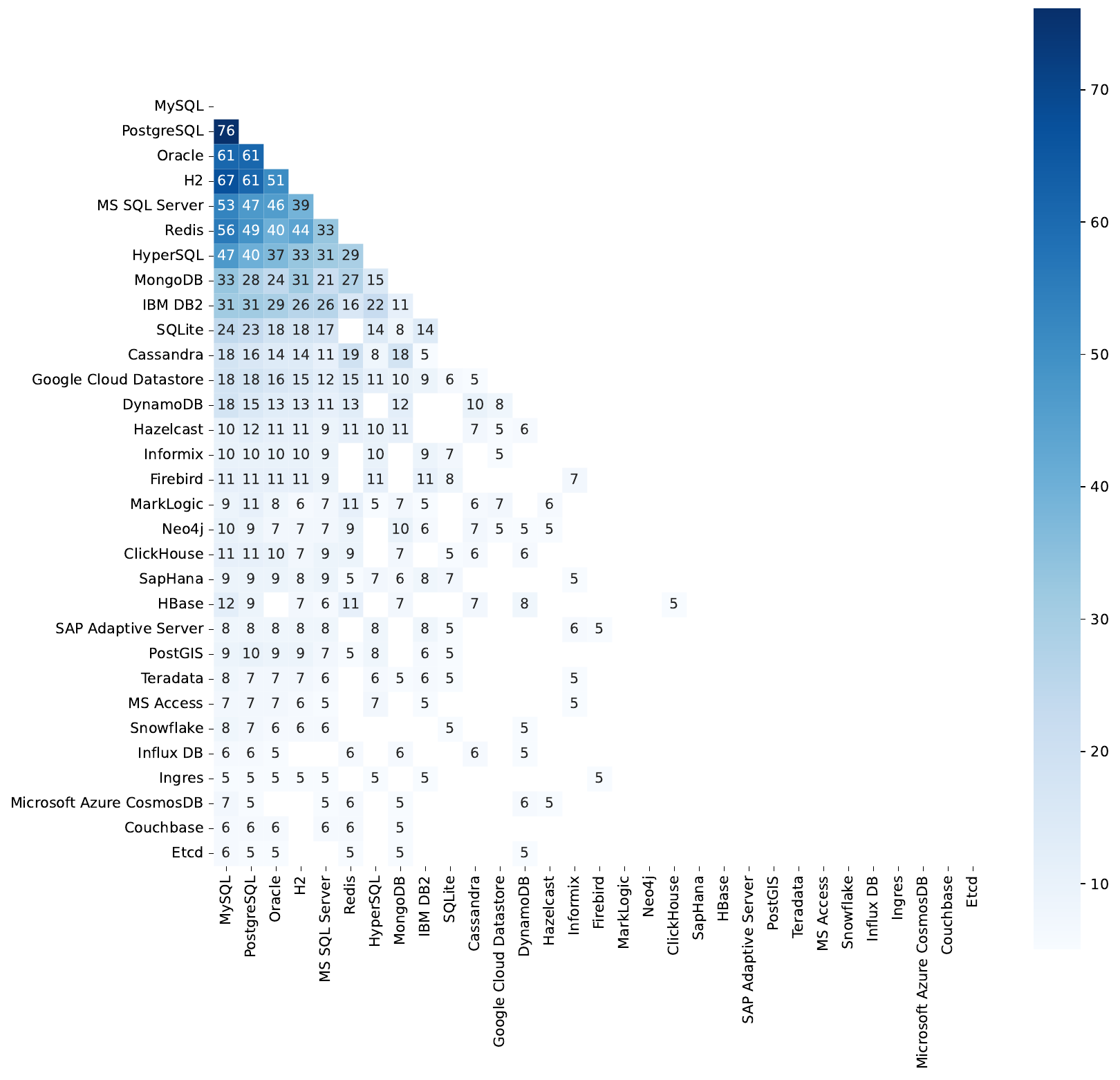}}
    \caption{\responsetoreviewer{rIIcXVIIfigIIIpII}{Correlation of the most frequent DBMSs at the end of the projects' history.}}
    \label{fig:heat_v10}     
\end{figure}

In this research question, we explored the synergy between the DBMSs. For this analysis, we adopted association rules to extract patterns that indicate the existence of concomitant use of DBMSs in our corpus. To perform our historical analysis, we took snapshots from the beginning, middle, and end of the projects' history and compared the results obtained in these three moments (slices). We mined the three slices by applying the Apriori algorithm and generated a heat map to represent the correlations between the DBMS in each slice, as shown in Figures~\ref{fig:heat_onlyrules_v1}, \ref{fig:heat_onlyrules_v5}, and \ref{fig:heat_v10}. We used a minimum frequency of five projects (minimum support of 5), which means we only considered the correlations that occurred in at least five projects to characterize a pattern. Note that in all of the heat maps, the upper diagonals of the heat maps were eliminated due to redundancy. In addition, the blank cells in the lower diagonal have a frequency below five projects.

Figure~\ref{fig:heat_onlyrules_v1} presents an overview of the synergy between 10 DBMSs early in the projects' history (corresponding to the slice that contains the first 100 commits of the projects, as explained in Section\ref{sec:research-method}). Among the most used DBMSs, we found MySQL and PostgreSQL in 21 projects, HyperSQL and MySQL in 20, and MySQL and H2 in 17. Notice that these are the same DBMSs most used individually, as we discussed in Section \ref{RQ2}. This demonstrates that certain DBMS combinations are preferred choices for developers and are widely adopted in various projects. Several factors can be attributed to the popularity of these combinations, such as complementary features provided by these DBMSs, performance characteristics, integration facility, and even developer familiarity. Thus, using these DBMSs together can provide more accurate solutions satisfying diverse requirements in different projects.

All of the 10 DBMSs for which we found synergies at the beginning of the projects' history are relational. No combinations involving non-relational DBMSs were observed in the projects, demonstrating that in the initial phase, the combinations between relational DBMSs are more commonly used than non-relational DBMSs. A possible explanation is the age of the projects and the fact that non-relational DBMSs have gained popularity recently.

Halfway through the history of the projects, the amount of combined use of DBMSs more than doubled when compared to the beginning. At this point, we found synergies among 23 DBMSs, as shown in Figure~\ref{fig:heat_onlyrules_v5}. This means that the projects employed more DBMSs together as they matured, pointing to a practice of polyglot persistence, where multiple databases are used to meet the distinct needs of a project \citep{Roy2022}. This increase may be related to the more complex database needs arising as projects undergo significant changes during their history. Despite the increase in the variety of DBMS combinations, the most frequent combinations remain similar to those observed at the beginning of the projects' history. MySQL and PostgreSQL, MySQL and H2, and MySQL and Oracle remain among the most used combinations, significantly increasing to 68, 54, and 48 projects, respectively. On the other hand, MySQL and HyperSQL, which was the second most popular combination at the beginning of the projects' history, are now not among the most popular combinations (although 44 projects use them together in the middle of the projects' history). We also highlight the increase in the frequency of combinations involving MS SQL Server and the emergence of combinations involving Cassandra and Redis on 18 projects. These may reflect the search for more customized solutions to the projects' database needs as they mature. 

Moreover, there are 9 combinations of non-relational and relational DBMS, indicating that non-relational  DBMSs are gaining force as complementary alternatives to relational DBMSs in various project scenarios. Examples include the combined use of Redis and PostgreSQL in 24 projects, MongoDB and H2 in 22 projects, and Cassandra and MySQL in 19 projects. However, despite the fact that we observed more combinations involving the two models and the emergence of combinations of non-relational DBMSs, the combinations involving only relational DBMSs remain predominant in this phase of the projects' history. 

According to Figure~\ref{fig:heat_v10}, the synergy between DBMSs increases at the end of the project history. However, the growth rate from the beginning to the middle (from 10 DBMSs to 23 DBMSs) is significantly higher than that from the middle to the end (from 23 DBMSs to 31 DBMSs). This demonstrates that until the middle, there is a wider exploration to meet the projects' demands. However, towards the end of the project history, the growth of the number of new DBMS combinations decreases, perhaps due to identifying more efficient and effective DBMS combinations leading to a more controlled growth.

MySQL, PostgreSQL, Oracle, and H2 remain among the most frequent combinations throughout the projects' history. Although HyperSQL lost its position to Redis, the combinations involving this DBMS remain popular in various projects. Interestingly, combinations involving  Redis became more frequent among projects (at this stage, we found combinations of Redis with 22 different DBMSs), indicating its increasing popularity and synergy at the end of the projects' history. According to \cite{cattell2011scalable}, Redis is a single-node key-value storage DBMS suitable for applications that search objects by a single attribute. Thus, this increase in popularity may mean it is being used for simple data manipulation. 

Combinations involving non-relational DBMSs---such as MongoDB, DynamoDB, and Google Cloud Datastore---become more frequent from the middle to the end of the analyzed project histories. The growth in the use of non-relational DBMSs is also reflected in combinations among the two models (for instance, MongoDB with IBM DB2 in 11 projects, DynamoDB with HBase in 8 projects, and Cassandra with HBase in 7 projects) and combinations containing only non-relational DBMSs, such as Redis with MongoDB in 27 projects, and Redis with Cassandra in 19 projects. These findings reaffirm an increasing trend of joint use of non-relational DBMSs in the projects' advanced stages.

Overall, these analyses indicate that the choice of DBMS combinations evolves over the projects' history, with certain DBMSs gaining popularity while others see fluctuations in their usage patterns. Non-relational DBMSs, in particular, became more prevalent in combinations as the projects progressed, possibly due to their advantages in handling certain data types and workloads.

It is worth noting two points here. First, the projects have different timelines, so the first slice of one project may comprise commits from 2010, while the first slice of a more recent project may comprise commits from 2024. Thus, the synergies we identified here are not related to specific versions of a DBMS since different projects may use the same combinations of DBMSs at different points in time (and thus, use different versions of those DBMSs). Second, since we use slice sizes of 100 commits, different projects have different numbers of slices. Thus, our analysis in Figure \ref{fig:heat_onlyrules_v5} may use slice 10 for a project that has 20 slices, and slice 40 for a project that has 81 slices. The same goes for Figure \ref{fig:heat_v10} -- it uses slice 20 for a project that has 20 slices, and slice 81 for a project that has 81 slices.

\begin{table}[t]
\centering
\caption{Top 10 rules with highest lifts in the first slice of the projects' history. MS SQL stands for MS SQL Server, DB2 stands for IBM DB2, Postg. stands for PostgreSQL, Access stands for MS Access, and HSQL stands for HyperSQL.}

\begin{tabular}{llrrrrrrr} 
\hline
         &  & \textbf{Sup} & \textbf{Sup} & \textbf{Sup} & \textbf{Conf} & \textbf{Conf} &  & \textbf{Lift}  \\ 
\textbf{A}           & \textbf{B} &  \textbf{(A)}& \textbf{(B)} & \textbf{(A$\rightarrow$B)} & \textbf{(A$\rightarrow$B)} & \textbf{(B$\rightarrow$A)} &  \textbf{Diff} & \textbf{(A$\rightarrow$B)}  \\ 
\hline
MS SQL&  DB2& 9& 10& 5& 0.55& 0.50& 0.05& 12.88\\
MS SQL&  SQLite& 9                & 12& 5& 0.55& 0.41& 0.14& 10.74\\
DB2& Oracle& 10& 20& 9& 0.90& 0.45& 0.45& 10.44\\
Informix& Postg. & 5& 23& 5& 1.0& 0.21& 0.79& 10.08\\
Access           & Oracle& 7& 20& 6& 0.85& 0.30& 0.55& 9.94\\
MS SQL& Oracle& 9& 20& 7& 0.77& 0.35& 0.42& 9.02\\
MS SQL& Postg. & 9& 23& 8& 0.88& 0.34& 0.54& 8.96\\
DB2& Postg. & 10& 23& 8& 0.80& 0.34& 0.46& 8.06\\
DB2& HSQL& 10& 32& 10& 1.0& 0.31& 0.69& 7.25\\
Informix& HSQL& 5& 32& 5& 1.0& 0.15& 0.85& 7.25\\
\hline
\end{tabular}
\label{tab:firstslice}
\end{table}

We further analyzed whether using a particular DBMS increases the chance of using another one. To do this, we filtered the rules with the highest lift values for the three moments in the project history, as shown in Tables~\ref{tab:firstslice},~\ref{tab:fifthslice}, and~\ref{tab:lastslice}.
As shown in Table~\ref{tab:firstslice}, the rule MS SQL Server $\rightarrow$ IBM DB2 has the highest lift we found early in the project history (12.88). This means that using MS SQL Server increases by 12.88 times the chance of using IBM DB2. The rule's confidence is 55\% (Conf(A$\rightarrow$B) = 0.55) when we have MS SQL Server as antecedent and IBM DB2 as consequent, and about 50\% (Conf(B$\rightarrow$A) = 0.50) the other way around. The slight difference of 5\% (Diff = 0.05) between the confidences shows that this rule has no majority direction. Thus, when MS SQL Server is used, there is a high chance of using IBM DB2 and vice versa. The next rule with a lift of 10.74 is MS SQL Server $\rightarrow$ SQLite with 55\% confidence (Conf(A$\rightarrow$B) = 0.55) in this direction and 41\% (Conf(B$\rightarrow$A) in the opposite direction. The 14\% difference (Diff = 0.14) between the confidence measures indicates is small. Thus, when MS SQL Server is employed, there is a significant likelihood of SQLite being used in conjunction, and vice versa. 

A few rules present significant differences, despite having low lifts: Informix $\rightarrow$ PostgreSQL with diff 0.79 and Informix $\rightarrow$ HyperSQL with diff 0.85. This demonstrates that these DBMSs have a strong dependency relationship, meaning that Informix has a higher chance of being adopted when other DBMSs are adopted. This indicates that Informix is not typically the initial choice for projects. Instead, it is combined with PostgreSQL and HyperSQL, suggesting Informix is a suitable add-on for these DBMSs, leading to their combined adoption.

\begin{table}[t]
\centering
\caption{Top 10 rules with highest lifts in the middle of the project history. Inf. stands for Informix, Couch. stands for CouchBase, SAP A. stands for SAP Adaptive Server, and Tera. stands for Teradata.}
\begin{tabular}{llrrrrrrr} 
\hline
         &  & \textbf{Sup} & \textbf{Sup} & \textbf{Sup} & \textbf{Conf} & \textbf{Conf} &  & \textbf{Lift}  \\ 
\textbf{A}           & \textbf{B} &  \textbf{(A)}& \textbf{(B)} & \textbf{(A$\rightarrow$B)} & \textbf{(A$\rightarrow$B)} & \textbf{(B$\rightarrow$A)} &  \textbf{Diff} & \textbf{(A$\rightarrow$B)}  \\ 
\hline
SapHana& Firebird& 7& 8& 5                & 0.71& 0.62& 0.09& 20.53\\
Firebird& Inf.& 8& 10& 7& 0.87& 0.70& 0.17& 20.12\\
Neo4j& Couch.& 8& 8& 5& 0.62& 0.62& 0.00& 17.97\\
SapHana& Inf.& 7& 10& 5& 0.71& 0.50& 0.21& 16.43\\
SapHana& SAP A.& 7& 10& 5& 0.71& 0.50& 0.21& 16.43\\
Firebird& SAP A.& 8& 10& 5& 0.62& 0.50& 0.12& 14.37\\
Tera.&  Inf.& 8& 10& 5& 0.62& 0.50& 0.12& 14.37\\
SAP A.&  Inf.& 10& 10& 6& 0.60& 0.60& 0.00& 13.80\\
Firebird& SQLite& 8& 22& 7& 0.87& 0.31& 0.56& 9.14\\
SapHana& SQLite& 7& 22& 6& 0.85& 0.27& 0.58& 8.96\\
\hline
\end{tabular}
\label{tab:fifthslice}
\end{table}

We noticed that during the middle of the projects' history, the lift values of the top 10 rules (Table~\ref{tab:fifthslice}) nearly doubled when compared to the lift values of the top 10 rules (Table~\ref{tab:firstslice}) identified at the beginning of the project's history. The increased lift suggests that specific DBMS correlations become stronger, indicating possible dependencies or synergies between them as the projects mature. For example, the rule SapHana $\rightarrow$ Firebird has a lift of 20.53 and 71\% confidence in this direction (Conf(A$\rightarrow$B) = 0.71), and 62\% confidence in the opposite direction (Conf(B$\rightarrow$A) = 0.62). Due to the small difference between the two confidences (Diff = 0.09), we cannot perceive a majority direction in this rule. This indicates a strong correlation between both DBMSs, meaning that when SapHana is chosen as a database, Firebird will likely be used in the same project, and the reciprocal is true.

Another interesting pattern we found is that adopting Neo4J increases the chance of adopting CouchBase by almost 18 times. A relevant aspect is the zero difference (Diff = 0.00) in confidence for both directions of this rule (Neo4j $\rightarrow$ CouchBase), which indicates they always appear together in the middle of the history of our projects.

\begin{table}[t]
\centering
\caption{Top 10 rules with highest lifts in the last slice of the projects' history. Access stands for MS Access, Cas. stands for Cassandra, Cosmos stands for Microsoft Azure CosmosDB, Hazel stands for Hazelcast, Influx stands for InfluxBD, Inf. stands for Informix, and SAP A. stands for SAP Adaptive Server.}
\begin{tabular}{llrrrrrrr} 
\hline
         &  & \textbf{Sup} & \textbf{Sup} & \textbf{Sup} & \textbf{Conf} & \textbf{Conf} &  & \textbf{Lift}  \\ 
\textbf{A}           & \textbf{B} &  \textbf{(A)}& \textbf{(B)} & \textbf{(A$\rightarrow$B)} & \textbf{(A$\rightarrow$B)} & \textbf{(B$\rightarrow$A)} &  \textbf{Diff} & \textbf{(A$\rightarrow$B)}  \\ 

\hline
Ingres&  Firebird& 5& 11& 5                & 1.00& 0.45& 0.55& 21.00\\
 SAP A.&  Inf.& 8& 10& 6& 0.75& 0.60& 0.15&17.32\\
Access           &  Inf.            & 7& 10& 5& 0.71& 0.50& 0.21& 16.50\\
Inf.& Firebird& 10& 11& 7& 0.70& 0.63& 0.07& 14.70\\
Teradata& Inf.& 8& 10& 5& 0.62& 0.50& 0.12& 14.43\\
 SAP A.& Firebird& 8& 11& 5& 0.62& 0.45& 0.17&13.12\\
SapHana& Inf.& 9& 10& 5& 0.55& 0.50& 0.05& 12.83\\
Influx& Cas.& 6& 22& 6& 1.00& 0.27& 0.73& 10.50\\
Teradata&  SAP A.& 8& 14& 5& 0.62& 0.35& 0.27& 10.31\\
Cosmos&  Hazel.& 7& 17& 5& 0.71& 0.29& 0.42& 9.70\\
\end{tabular}
\label{tab:lastslice}
\end{table}

Finally, in Table~\ref{tab:lastslice}, we present the top 10 rules with the highest lifts found at the end of the projects' history. We observe that lifts remain high towards the end of the projects' history, demonstrating strong co-occurrences in the adoption of different DBMSs. At this stage, we discovered two rules with 100\% confidence: Ingres $\rightarrow$ Firebird, and InfluxDB $\rightarrow$ Cassandra. Therefore, when the first of these DBMSs is chosen (antecedent), it highly influences the choice of the other (consequent). The rule with the highest lift is one of them (Ingres $\rightarrow$ Firebird). It has 100\% confidence and a lift of 21, meaning that the use of Ingres increases the likelihood of the use of Firebird by 21 times. In contrast, the confidences of the rule InfluxBD $\rightarrow$ Cassandra have a significant difference (Diff = 0.73), indicating a strong dependency of InfluxDB on Cassandra. This means that in 6 out of 22 of the cases where InfluxDB was adopted, Cassandra was also adopted, and in all of the cases (6 out of 6) where InfluxDB was adopted, Cassandra was adopted. The adoption of InfluxDB increases the likelihood of the adoption of Cassandra by about 10 times. Informix also appears to be used as a complementary DBMS, as it is always combined with other DBMSs, as already mentioned. According to Informix's documentation\footnote{https://www.ibm.com/products/informix}, it is a fast and scalable database server that manages traditional relational, object-relational, and dimensional databases. The IBM documentation\footnote{https://www.ibm.com/docs/en/informix-servers/12.10?topic=overview-getting-started} suggests that Informix is likely used as a complementary DBMS due to its strengths in handling unstructured and IoT data, efficient OLTP performance, and compatibility across various platforms. These characteristics make it well-suited for integration with other DBMSs in hybrid environments, where it can support specialized functions, such as real-time data processing or transactional workloads, alongside more conventional database systems.

We also noted a fluctuation in the lift values and confidence measures of some rules throughout the different stages of the project history. For example, the rule MS SQL Server $\rightarrow$ SQLite appears with a lift of 10.74 and confidences of 55\% (Conf(A$\rightarrow$B) = 0.55) and 50\% (Conf(B$\rightarrow$A) = 0.50) at the beginning of the project history. In the middle, the lift decreases to 2.80, and the $A \rightarrow B$ confidence decreases to 26\%. In the end, the lift decreases slightly to 2.46, and the $A \rightarrow B$'s confidence decreases at 12\%. As opposed to that, the rule SAP Adaptive Server $\rightarrow$ Informix, discovered in the middle of the project history, initially has a lift of 13.80 with 60\% (Conf(A$\rightarrow$B) and (Conf(B$\rightarrow$A) = 0.60) confidences, and the lift increases to 17.32, with confidences 75\% (Conf(A$\rightarrow$B) = 0.75) and 60\%(Conf(B$\rightarrow$A) = 0.60) at the end.
This fluctuation in lift values and confidence measures indicates that the associations and dependencies between certain DBMS combinations are not static and may vary during the project history. Various factors might have influenced these variations, such as changes in project requirements, technological advancements, or shifts in the development team's preferences. Despite the decrease in the initial lift values, the highlighted rules continue to represent significant associations between the aforementioned DBMS combinations. Thus, although the strength of the relationships may change, there is still some level of dependency or association between certain DBMSs that may become more or less prevalent as the projects mature.

We also observe that, in the early stage of the projects' history, the usage of a relational DBMS combined with another relational DBMS was more frequent. As observed in Table~\ref{tab:firstslice}, all of the DBMSs involved in the 10 most frequent rules are relational. However, this situation changes from the middle to the end of the project history, with non-relational DBMSs present in some of the top 10 rules highlighted in Tables~\ref{tab:fifthslice} and \ref{tab:lastslice}. However, it is important to consider that even among primarily relational DBMSs, there might have been cases where a secondary model of one of the DBMSs is non-relational. For instance, Sap Adaptive Server and IBM DB2 are primarily relational but support non-relational data models. Sap Adaptive Server also supports the Spatial Store data types, while IBM DB2 supports the Document Store, RDF Store, and Spatial Store types. This may suggest that the joint utilization of these DBMSs might be driven by the need for a specific data type not supported by the other DBMS, or to complement certain capabilities that one of the DBMSs lacks. %Additionally, the compatibility between the two DBMSs, both developed in C/C++, might have facilitated their combined usage.

%\begin{table}[t]
%\centering
%\caption{\revise{Similarities between DBMSs of the same vendors.}}
%\resizebox{6in}{!}{%
%
%\begin{tabular}{lllrrrrrrr} 
%\hline
%\textbf{Slice}          & \textbf{A} & \textbf{B} & \rotatebox{90}{\textbf{Sup(A)}} & \rotatebox{90}{\textbf{Sup(B)}} & \rotatebox{90}{\textbf{Sup(A$\rightarrow$B)}} & \rotatebox{90}{\textbf{Conf(A$\rightarrow$B)}} & \rotatebox{90}
%{\textbf{Conf(B$\rightarrow$A)}} &  \rotatebox{90}{\textbf{Diff}} & \rotatebox{90}{\textbf{Lift(A$\rightarrow$B)}}  \
%\hline
%\textbf{Last}  & PostGIS    & PostgreSQL & 10& 94& 10& 1.00                                  & 0.10                                  & 0.90                              & 2.40\
%\hline
%\end{tabular}
%
%\label{tab:similarities}
%\end{table}

%In our analysis, we observed the occurrence of rules where DBMSs from the same vendors are used together, as shown in Table \ref{tab:similarities}. For instance, SAP SQL Anywhere and SAP Adaptive Server are not used together in the projects' history. However, in the end, we noticed the co-occurrence of PostGIS and PostgreSQL.

Although not shown in Table \ref{tab:lastslice}, in the last slice of the projects' history we also found the expected co-occurrence between PostGIS and PostgreSQL. Since PostGIS is a spatial DBMS extension of PostgreSQL, it transforms PostgreSQL into a spatial DBMS by adding support for three features: spatial types, spatial indexes, and spatial functions\footnote{\url{https://postgis.net/}}. The rule with PostGIS $\rightarrow$ PostgreSQL, whose confidences are 100\% and 10\% demonstrates a strong dependency of PostGIS on PostgreSQL. In fact, in 100\% of the cases where PostGIS was used, PostgreSQL was also used, while in only 10\% of cases where PostgreSQL was used, PostGIS was also used. Furthermore, using PostGIS increases the chance of using PostgreSQL by 2.4 times. Considering that PostGIS depends on PostgreSQL to work, this 100\% confidence is expected. This information suggests that the projects that require spatial database functionality will probably adopt PostGIS and PostgreSQL together, reinforcing the idea that the co-occurrence between DBMSs is intended to meet the specific demands of the projects.

\vspace{+5mm}

{
\begin{footnotesize}
\begin{centering}
  \setlength{\fboxrule}{0.1em}
  \setlength{\fboxsep}{1em} 
  %\fcolorbox{silver}
  \fbox{%
    \parbox{0.95\linewidth}{%
    \textbf{\\RQ4: Which DBMSs are often used together?}\\
    
    \textbf{Answer: } We found co-occurrences involving 10 DBMSs at the beginning of the projects' history, prevailing the combinations among relational DBMSs: MySQL and PostgreSQL, HyperSQL and MySQL, and MySQL and H2. In the middle of the project history, the number of DBMSs that appear in co-occurrence increases to 23. Nine of them include non-relational DMBSs, such as Redis and MySQL, MongoDB and H2, and Cassandra and MySQL. The combined usage of PostgreSQL and MySQL, and MySQL and H2 get even more popular in the middle of the projects' history. At the end of the projects' history, the number of DBMSs used in combination with others increased to 31 DBMSs, with MySQL with PostgreSQL and MySQL with H2 among the most popular relational combinations. Our analysis also revealed that using some DBMSs increases the chance of adopting another in parallel. This is the case for MS SQL Server and IBM DB2 at the beginning of the project history, SapHana and Firebird in the middle, and Ingres and Firebird towards the end of the projects' history.\\
    %\textbf{Implications: } The results offer a valuable understanding of prevalent combinations and may help to perceive potential synergies or challenges in utilizing DBMSs together.
    }%
    }% 
\end{centering}
\end{footnotesize}
}

\subsection{(RQ5) How do applications interact with the DBMS?} 
\label{RQ5}

We split RQ5 findings into three aspects: ORM, Database-Related Files, and Queries. 

~

\begin{figure}
\centering
\responsetoreviewer{rIIcXVIrqIIdIpI}{
\includegraphics[width=0.6\linewidth]{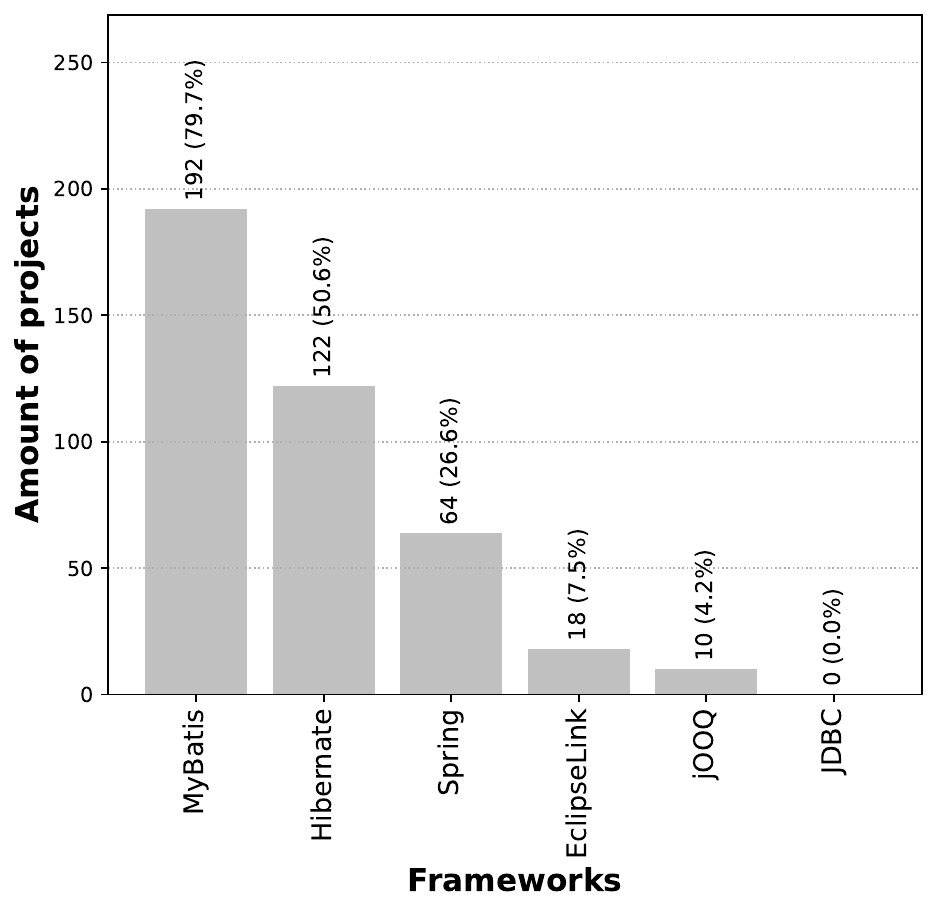}
}
\caption{\responsetoreviewer{rIIcXVIrqIIdIpII}{Distribution of ORM usage in our corpus.}}
\label{fig:fig_rq2.1}
\end{figure}

\noindent\textbf{ORM.} Figure \ref{fig:fig_rq2.1} shows our findings for the ORM aspect. Out of the \nvarc{rq2_projects}{362} projects that belong to our corpus, we found evidence of ORMs usage in \nvar{projects_with_orm}. Note that this number is higher than the number of projects for which we found evidence of DBMS usage (\nvarc{rq1_projects_with_dbms}{202} projects). This may indicate that there are projects that use some DBMSs that do not belong to the 50 popular DBMSs we searched for. 

MyBatis is the most popular ORM in our corpus. It is present in 192 (79.7\%) projects. Hibernate is the second in the ranking, corresponding to 50.6\% of projects. Next, we have Spring, which was found in 26.6\% of the projects, occupying the third place, and EclipseLink, which appears in 7.5\% of the projects, in fourth place. jOOQ appears in fifth place in the ranking, corresponding to 4.2\% of projects. Lastly, our heuristics did not find signs of usage of the JdbcMapper in any of the projects of our corpus.

We also analyzed the relative number of ORM files in each project by using the following formula for each project: \[ \dfrac{ORM_{Files}}{Project_{Files}}\] 

\begin{figure}
\centering
\includegraphics[width=\linewidth]{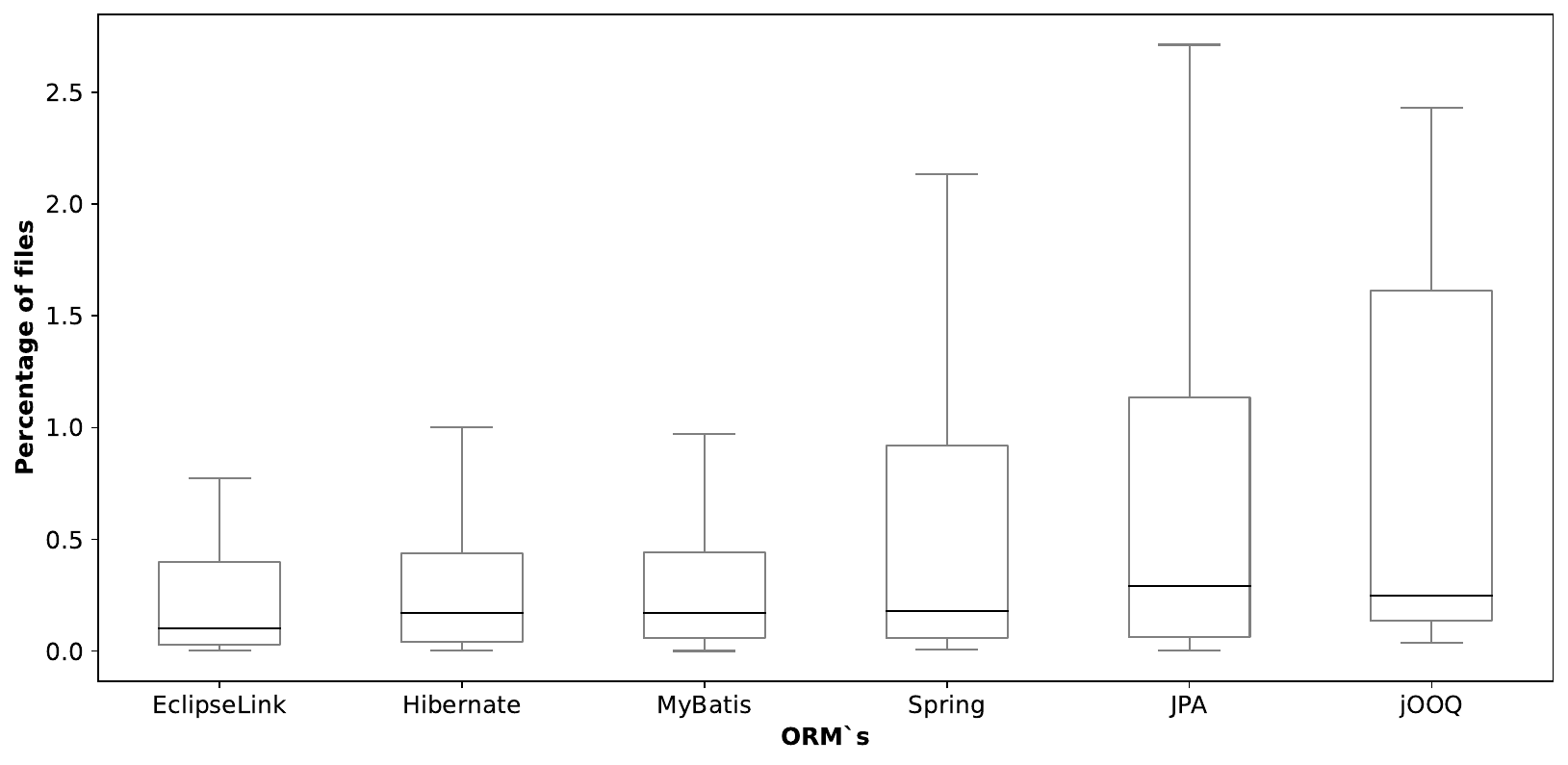}
\caption{Distribution of the percentage of ORM files by ORM (the outliers are not shown due to readability reasons).}
\label{fig:fig_rq2.1_2}
\end{figure}

In this formula, $Project_{Files}$ refers to the total number of files of the specific project, and $ORM_{Files}$ corresponds to the number of files related to a specific ORM found by our search in a specific project. This measures how pervasive the ORM is in the application source code. We then use a boxplot to analyze the distribution of the results (Figure \ref{fig:fig_rq2.1_2}). Note that JPA is present in Figure \ref{fig:fig_rq2.1_2}, but not in Figure \ref{fig:fig_rq2.1}. Although JPA is not a framework, we have included it in our research. JPA is a specification that other frameworks implement. Evaluating the results more deeply, we found that projects with evidence of using JPA sometimes have evidence of using more than one ORM. This makes it difficult to identify which ORM is implementing JPA. Thus, we chose not to merge the results and count JPA files separately.

The median of the percentage of files related to the ORMs EclipseLink, Hibernate, MyBatis, Spring, JOOQ, and JPA are close and vary from 0.10\% to 0.29\%. The high variation we encountered means that JPA and JOOQ are more verbose than the others. Note that the design of the projects that use these ORMs in our corpus may also have influenced these results. %The values found were 0.10, 0.17, 0.17, 0.18, 0.25, and 0.29, respectively. We observe that the medians found for the EclipseLink, Hibernate, MyBatis, and Spring frameworks are close. The proportion of files required to write software using these ORMs varies between 0.10\% and 0.29\%.

~

\noindent\textbf{Database-related Files.} For the Database-related Files aspect, we analyzed the relative number of files. To do this, we performed a calculation: \[ \dfrac{Type_{Files}}{Project_{Files}} \times 100 \] 
where $Project_{Files}$ refers to the total number of files of the specific project, and $Type_{Files}$ corresponds to the sum of files found by our searches in a specific project of a given $Type$ among Java DB-Code files, XML DB-Code files, or dependencies. 

\begin{figure}
\centering
\includegraphics[width=\linewidth]{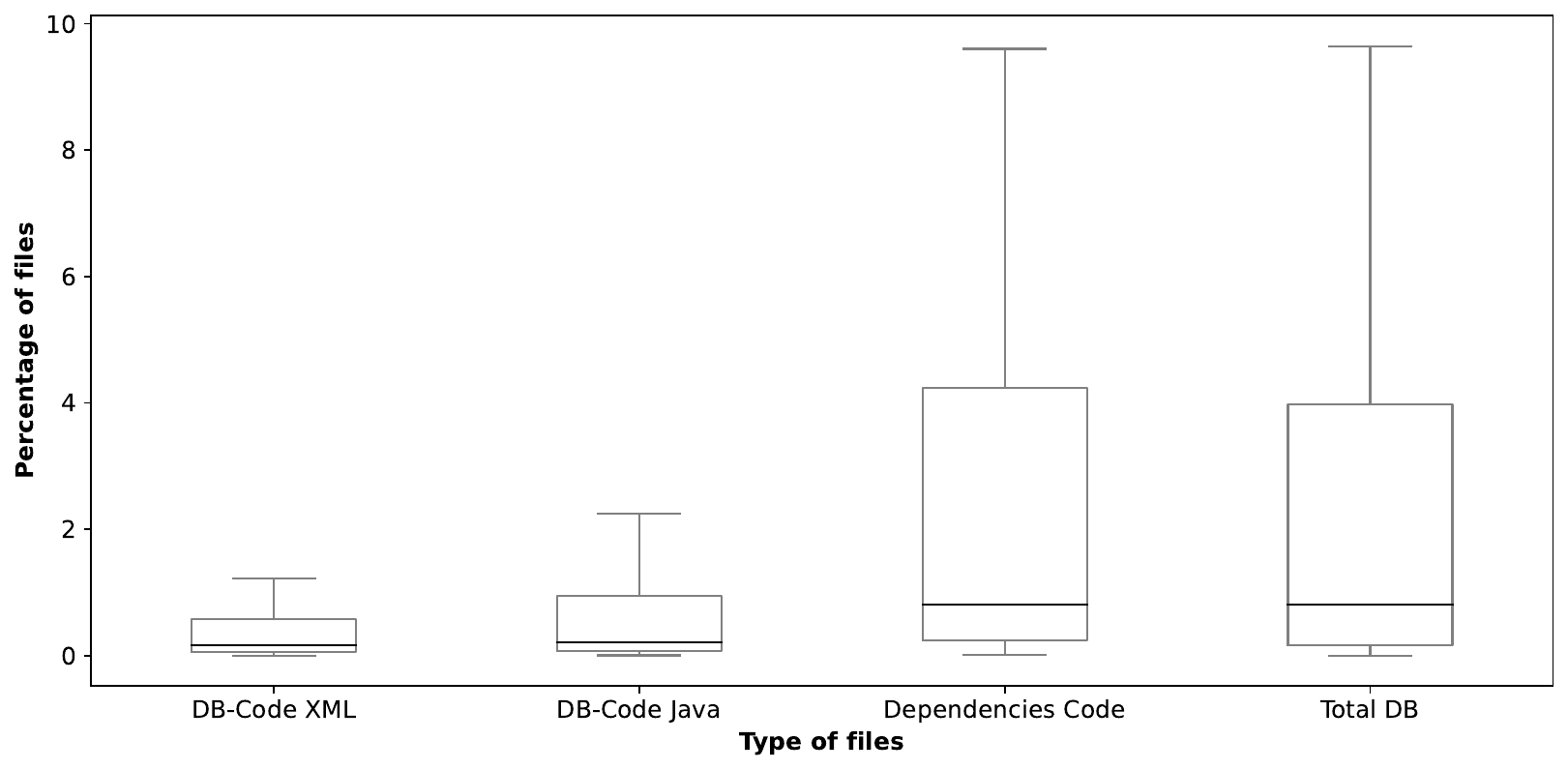}
\caption{Distribution of the percentage of files categorized by file type, namely: XML DB-Code, Java DB-code, Dependencies Code, and Total DB (the outliers are not shown due to readability reasons).}
\label{fig:fig_rq2.2}
\end{figure}

Figure \ref{fig:fig_rq2.2} shows the distribution of the percentage of files related to ORM categorized by file type found in our corpus. For the DB-Code files that are of the ``.xml'' type, the median is 0.17\%. When comparing the results of Java DB-Code files and XML DB-Code files, we obtained a lower result for XML DB-Code files, as not all ORM frameworks require implementation in XML files. %The two biggest outliers for XML DB-Code are 17.13\%, which belongs to the Spring-Integration project, and 11.62\%, which belongs to the Spring-Batch project. On GitHub, the Spring-Integration project is described as `Spring Integration provides an extension of the Spring programming model to support the well-known Enterprise Integration Patterns (EIP)''. For the Spring-Integration, we found evidence of the use of two ORMs: Spring and JPA. The project has 4,816 files. The use of XML in the Spring-Integration project stems from its historical role in configuring dependencies, messaging flows, and system integration. XML enables modularization, separation of concerns, and reusable configurations for managing complex applications. On Github, the Spring-Batch project is described as ``[...] a framework for writing batch applications using Java and Spring''. The Spring-Batch project has 2,468 files, and we found evidence of the use of three ORMs: Spring, JPA, and Hibernate. The fact that these projects use more than one ORM explains the fact that their number of DB-Code files is greater than the numbers we encountered for other projects.}

We also analyzed the distribution of the quantities found For DB-Code files that are of the ``.java'' type. For this file type, the median is 0.21\%, which is very low, meaning that, in general, a very small portion of the projects are related to database access functions. Of course, this also depends on the project's purpose. In fact, the biggest outlier we found (the BroadleafCommerce project)  has 14.20\% of DB-Code files (the outliers are not shown in Figure \ref{fig:fig_rq2.2} due to readability reasons). The project contains approximately 3880 files, and a significant portion of them are related to databases. This is expected since this is a tool focused on database usage. %On GitHub, the BroadleafCommerce project is described as ``an e-commerce framework written entirely in Java and leveraging the Spring framework. It is targeted at facilitating the development of enterprise-class, commerce-driven sites by providing a robust data model, services, and specialized tooling that take care of most of the heavy lifting work.'' Therefore, it is a tool focused on database usage.}

For the dependencies files (files that depend on or use Java DB-Code files), the median is equal to 0.81\%, which is four times higher than the one we found for the DB-Code files. For this group of files, we found much more variation than we did for the DB-Code files and XML-Code files. This is mainly influenced by the application design. %For this group, we found thirteen outliers ranging from 10.2\% to 32.26\%. The project with the largest outlier (32.26\%) is Onedev. On GitHub, the Onedev project is described as ``Git Server with CI/CD, Kanban, and Packages. Seamless integration.'' The project has a large number of files, around 5,000, and it has a high number of files that use other files that contain some mention of the database (approximately 2,000).}

Finally, when we measure everything together, the large variation we found for the dependency files is reflected in our findings for the Total-DB files. The median in this case is 0.80\%. %Data dispersion is found between the values 0.0006\% and 70.04\%. The project with the largest outlier (70.04\%) here is the one that has a smaller number of files, causing its proportion to be high (Ebean).}

%From these results, we can infer that within our corpus, projects have a proportion of \revise{1.09\%} DB-related files, which includes Java DB-Code files, XML DB-Code files, and Dependencies files.

\begin{figure}
\centering
\includegraphics[width=0.4\linewidth]{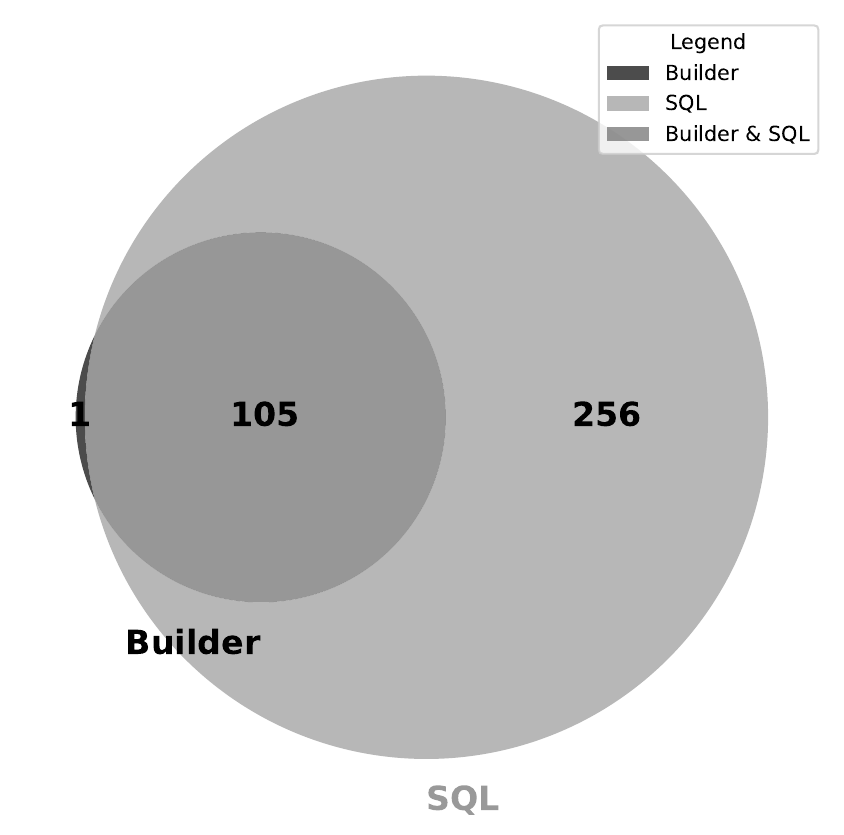}
\caption{Use of Builders and SQL as query alternatives in our corpus.}
\label{fig:fig_rq2.3}
\end{figure}
%https://github.com/gems-uff/db-mining/blob/master/src/results_database_characterization.ipynb
~

\noindent\textbf{Queries} We analyzed the source code of each project of our corpus to extract information about how queries are performed. In other words, we investigated whether the projects write SQL statements directly in the code or use Builders to generate SQL statements. Figure \ref{fig:fig_rq2.3} shows that we found 256 projects that use only SQL, while Builders are used in 105 projects. Note that almost all projects that use Builder also use SQL. Only one project, Classgraph, uses only Builders. This result is interesting when we contrast it to the results we obtained for RQ1. There, we found evidence of DBMS usage in \nvarc{rq1_projects_with_dbms}{202} projects, while here, we found 256 projects that use SQL. This means that 54 projects use relational DBMSs that are not in the list of 25 relational DBMSs we searched for.

When we analyzed the number of files that contained SQL queries per project, we found projects with values that vary between 0 and 8,269, with an average of 123.2 files per project and a standard deviation of 653.9. For the Builder files, we find projects with values between 0 and 389 files, with an average of 51.5 files.

\vspace{+5mm}

\begin{footnotesize}
\begin{centering}
  \setlength{\fboxrule}{0.1em}
  \setlength{\fboxsep}{1em}
  \fbox{%
    \parbox{0.95\linewidth}{%
    \textbf{\\RQ5: How do applications interact with the DBMS?}\\
    
    \textbf{Answer: } MyBatis is the most popular ORM in our corpus, followed by Hibernate and Spring. We can observe that projects that use EclipseLink, Hibernate, and MyBatis need fewer database-related files in practice than projects that use other ORMs. The average of files related to ORM found per project was 4.95\%. SQL is more used than Query Builders to query the DBMS.  %\\
    %\textbf{Implications: } The rates we found for the ORMs vary little from each other, which may mean that no ORMs are more verbose than others. In our corpus, projects have a proportion of 1.09\% DB-related files, which means that in case of an ORM change, the developer will have to change around 1.09\% of the project files.}}%
    }%
}
\end{centering}
\end{footnotesize}

\subsection{Results Summary}
\label{sec:summary}

\responsetoreviewer{rIIcIVpI}{
Table \ref{tab:rq_findings} presents the main results and findings for each research question, structured according to relevant aspects. The complete list is available at our companion website\footnote{\url{https://tinyurl.com/2nyt757h}}.}

\begin{table}[t]
    \centering
    \caption{\responsetoreviewer{rIIcIVtIpI}{Main results and findings for each RQ.}}
    \label{tab:rq_findings}
    \responsetoreviewer{rIIcIVtIpII}{ 
    \begin{scriptsize}
    \begin{tabular}{p{11,4cm}}
        \hline
        \textbf{RQ1: Which DBMS are the most popular across software projects?} \\
        \hline
        \begin{itemize}
        \item Relational DBMSs were used in 82.2\% of the projects, non-relational DBMSs in 68.3\%, and multi-model DBMSs in 4.0\%.
        \item We found that 103 projects (51.0\%) adopt both relational and non-relational models.
        \item MySQL is used in 55.9\% of the projects that use DBMS, followed by PostgreSQL (46.0\%) and H2 (44.6\%).
        \item MySQL is the most used DBMS in the Software Development, Data Management, and Infrastructure Management domains.
         \item MySQL and Redis are frequently employed within the same domains.
        \end{itemize} \\
        \hline
        \textbf{RQ2: How stable are the DBMSs during the projects’ history?} \\
        \hline
        \begin{itemize}
        \item 32 projects in our corpus stopped using a DBMS or switched to one not included in our list of 50 DBMSs.
        \item Removals happened for the majority of the DBMSs we surveyed.
        \item Some DBMSs were never removed once adopted, such as ClickHouse, Microsoft Azure Cosmos DB, and SingleStore.
        \item Of the projects using MySQL, 41 (29.9\%) removed it initially, 20 reintroduced it, and 4 removed it again, leaving MySQL in 112 projects in the final analysis.
        \end{itemize}\\
        \hline
        \textbf{RQ3: Which DBMSs are frequently replaced by others?} \\
        \hline
        \begin{itemize}
        \item  We found 296 replacement patterns in total across 67 projects.
        \item  In 5.8\% of the projects with replacements, the replacements only occur between DBMSs that follow the same data model. 
        \item  In 50.0\% of the patterns, the replacements occur among distinct data models. 
        \item  HyperSQL is the DBMS mostly susceptible to replacements during the projects’ history since it has been replaced in 19 different projects.
        \end{itemize}\\
        \hline
        \textbf{RQ4: Which DBMSs are often used together?} \\
        \hline
        \begin{itemize}
        \item  At the start of the projects, we observed co-occurrences involving 10 DBMSs, with the most common combinations being MySQL and PostgreSQL, HyperSQL and MySQL, and MySQL and H2. 
        \item Halfway through the history of the projects, the amount of combined use of DBMSs more than doubled when compared to the beginning. 
        \item MySQL and PostgreSQL, MySQL and H2, and MySQL and Oracle remain among the most used combinations in the middle of the project’s history.
        \item MySQL, PostgreSQL, Oracle, and H2 remain among the most frequent DBMS combinations until the end of the projects' history. 
        \end{itemize}\\
        \hline
        \textbf{RQ5: How do the applications interact with these DBMS?} \\
        \hline
        \begin{itemize}
        \item MyBatis is the most popular ORM, appearing in 79.6\% of projects. Hibernate ranks second, and is present in 50.6\% of projects, followed by Spring, found in 26.5\% of the projects. 
        \item The median of the percentage of files related to the ORMs EclipseLink, Hibernate, MyBatis, Spring, JOOQ, and JPA are close and vary from 0.10\% to 0.29\%.
        \item JPA and JOOQ are more verbose than the others. 
        \item The median of database-related files is 0.80\%.
        \item Almost all projects that use Builder also use SQL.
        \end{itemize}\\
        \hline
    \end{tabular}
    \end{scriptsize}}
\end{table}

 \subsection{Qualitative Analysis}
\label{sec:qualitative}

Identifying the reasons for the inclusion, removal, and substitution of technologies is a highly complex task. Some projects are over 20 years old, and software architectures have evolved significantly over the past two decades to address scalability, flexibility, and deployment needs. Early architectures were typically monolithic, bundling all components into a single deployable unit. Additionally, many projects have migrated repositories, and commits often lack clear objectives for changes, complicating the understanding of these modifications.

To complement our investigation, we conducted a qualitative analysis using ten randomly selected projects from our corpus: Activiti, Bisq, CAS, CommaFeed, JMeter, ModelDB, Nifi, Ngrinder, SpotBugs, and Zeppelin. The goal was to validate our findings and understand the reasons behind the addition, removal, or migration of specific DBMS in each project. For the validation, we analyzed the GitHub repositories of each project to investigate the motivations for these changes.

\responsetoreviewer{RIICII}{We conducted a keyword search for each project from the sample using GitHub's Code Search feature.  We search for the specific names of DBMSs we identified in RQ1 and RQ2 for that specific project, as well as the terms \emph{Database} and \emph{DB}. For example, in the project ModelDB, we identified the usage of H2, SQLite, MongoDB, PostgreSQL, MySQL, and MS SQL Server. As mentioned in Section \ref{sec:databaseHeuristics}, our heuristics do not differentiate the usage of MySQL and MariaDB, and PostgreSQL and CockroachDB. Anytime one of these DBMS is detected, we also search for its sibling. Thus, for the ModelDB project, we searched for the keywords \emph{H2, MariaDB, Maria, MySQL, PostgreSQL, Postgre, CockroachDB, Cockroach, MongoDB, Mongo, MS SQL Server, MSSQLServer, MSSQL, SQLServer, Microsoft Azure SQL Database, Microsoft Azure, Azure SQL Database, Database,} and \emph{DB}. We analyzed the obtained results in the following categories: code, issues, pull requests, discussions, and commits. Two authors carried out the search, and the results were discussed and reviewed by both authors. A third author then validated the findings to ensure there was no bias in the search process.}

\responsetoreviewer{rIIcIII}{Table \ref{tab:qualitative} shows a snippet of the DBMS usage patterns we found for each project. The complete list is available at our companion website\footnote{\url{https://tinyurl.com/5dbvdkju}}. In our website, the patterns marked in green correspond to those for which supporting evidence was identified through our qualitative analysis, such as commits, issues, and pull requests, which demonstrate the existence of the pattern or the occurrence of related events (e.g., addition and removal) in sequence. The patterns displayed on our website also include those for which no evidence was found in the project, as determined by our qualitative analysis. The patterns may not have been found due to several factors. In our qualitative analysis, we specifically searched for the exact names of DBMSs. Therefore, commits, issues, and pull requests that do not align with these patterns would not be returned. Furthermore, some projects may have migrated to a different version control system over time, as they are older projects, which could also impact the availability and retrieval of historical data. Our analysis revealed several key influential factors for DBMS migrations, including maturity, stability, licensing, maintainability, performance, and modularity. We discuss the results in the following.}

\begin{table}
 \centering
    \caption{\responsetoreviewer{rIIcIIItIpI}{DBMS usage patterns for the projects in our sample. MySQL stands for MySQL or MariaDB, PostgreSQL stands for PostgreSQL or CockroachDB, SQL Server stands for MS SQL Server or Microsoft Azure SQL Database, Filestore stands for Google Cloud Filestore, and Datastore stands for Google Cloud Datastore.}}
    \label{tab:qualitative}
    \responsetoreviewer{rIIcIIItIpII}{
    \begin{scriptsize}
    \begin{tabular}{p{1.5cm}|p{9.3cm}}
    \hline
    \textbf{Project} & \textbf{DBMS Replacement Patterns} \\
    \hline
    \multirow{3}{*}{Activiti} 
    & {Hazelcast $\rightarrow$ HyperSQL$_{In}$ $\rightarrow$ Hazelcast$_{Out}$ HyperSQL} \\
    & {HyperSQL $\rightarrow$ PostgreSQL$_{In}$ $\rightarrow$ HyperSQL$_{Out}$  PostgreSQL}\\
    & ... \\
    \hline 
    \multirow{6}{*}{CAS} 
    & {Couchbase $\rightarrow$ H2$_{In}$ $\rightarrow$ Couchbase$_{Out}$ H2} \\
    & {Couchbase $\rightarrow$ Cassandra$_{In}$ $\rightarrow$ Couchbase$_{Out}$ Cassandra}\\
    & {Couchbase $\rightarrow$ InfluxDB$_{In}$ $\rightarrow$ Couchbase$_{Out}$ InfluxDB} \\
    & {Couchbase $\rightarrow$ SQL Server$_{In}$ $\rightarrow$ Couchbase$_{Out}$ SQL Server} \\
    & {Couchbase $\rightarrow$ Filestore$_{In}$ Couchbase$_{Out}$ $\rightarrow$ Filestore} \\
    & {PostgreSQL $\rightarrow$ Hazelcast$_{In}$ $\rightarrow$ PostgreSQL$_{Out}$ Hazelcast}\\
    & ... \\
    \hline
    \multirow{1}{*}{CommaFeed} 
    & {Redis $\rightarrow$ H2$_{In}$ $\rightarrow$ Redis$_{Out}$ H2} \\
    & ... \\
    \hline
    \multirow{1}{*}{JMeter} 
    & {MySQL $\rightarrow$ PostgreSQL$_{In}$ MySQL$_{Out}$ $\rightarrow$ PostgreSQL}\\
    \hline
    \multirow{4}{1,5cm}{ModelDB} 
    & {PostgreSQL $\rightarrow$ H2$_{In} \rightarrow$ PostgreSQL$_{Out}$ H2}\\
    & {SQLite $\rightarrow$ MongoDB$_{In}$ $\rightarrow$ SQLite$_{Out}$ MongoDB}\\
    & {SQLite $\rightarrow$ MySQL$_{In}$ SQLite$_{Out}$ $\rightarrow$  MySQL} \\
    & {SQLite $\rightarrow$ SQL Server$_{In}$ SQLite$_{Out}$ $\rightarrow$  SQL Server} \\
    & ... \\
    \hline
    \multirow{1}{*}{Ngrinder} & {SQLite $\rightarrow$ H2$_{In}$ $\rightarrow$ SQLite$_{Out}$ H2}\\
    \hline
    \multirow{8}{*}{Nifi} 
    & {Cassandra $\rightarrow$ PostgreSQL$_{In}$ $\rightarrow$ Cassandra$_{Out}$ PostgreSQL}\\
    & {Cassandra $\rightarrow$ Hazelcast$_{In}$ $\rightarrow$ Cassandra$_{Out}$ Hazelcast}\\
    & {Couchbase $\rightarrow$ Hazelcast$_{In}$ $\rightarrow$ Couchbase$_{Out}$ Hazelcast}\\
    & {Couchbase $\rightarrow$ PostgreSQL$_{In}$ $\rightarrow$ Couchbase$_{Out}$ PostgreSQL}\\
    & {Ignite NoSql $\rightarrow$ PostgreSQL$_{In}$ $\rightarrow$ Ignite NoSql$_{Out}$ PostgreSQL}\\
    & {Ignite NoSql $\rightarrow$ Hazelcast$_{In}$ $\rightarrow$ Ignite NoSql$_{Out}$ Hazelcast}\\
    & {PostgreSQL $\rightarrow$ Cassandra$_{In}$ $\rightarrow$ PostgreSQL$_{Out}$ Cassandra}\\
    & {PostgreSQL $\rightarrow$ Couchbase$_{In}$ $\rightarrow$ PostgreSQL$_{Out}$ Couchbase}\\
    & ... \\
    \hline
    \multirow{1}{*}{SpotBugs} 
    & {Datastore $\rightarrow$ Oracle$_{In}$ $\rightarrow$ Oracle$_{Out}$ Datastore}\\
    \hline
    \multirow{3}{*}{Zeppelin} 
    & {Ignite NoSql $\rightarrow$ Neo4j$_{In}$ $\rightarrow$ Ignite NoSql$_{Out}$ Neo4j}\\
    & {SQL Server $\rightarrow$ Redis$_{In}$ SQL Server$_{Out}$ $\rightarrow$ Redis} \\
    & {MySQL $\rightarrow$ Redis$_{In}$ MySQL$_{Out}$ $\rightarrow$ Redis} \\
    & ... \\
    \hline   
    \end{tabular}
\end{scriptsize}}
\end{table}

\responsetoreviewer{rIIcXIX}{In the Activiti project, Hazelcast was initially used as a distributed cache, which was later replaced by HyperSQL (pull request \href{https://github.com/Activiti/Activiti/pull/330}{\#330}), confirming a replacement rule we found in our analysis. Activiti also adopted HyperSQL as its embedded and testing database, favoring it over H2 due to HyperSQL's reputation for stability and maturity within Java environments (pull request \href{https://github.com/Activiti/Activiti/pull/478}{\#478}). Eventually, HyperSQL was replaced by PostgreSQL as a test DBMS, as documented in issue \href{https://github.com/Activiti/Activiti/issues/1368}{\#1368} of the project, though no specific reasons for this switch from HyperSQL to PostgreSQL were identified in the project documentation.

}

The CAS project has integrated various databases and modules to meet specific functional needs. The project's README page specifically mentions Cassandra, Memcached, Apache Ignite, MongoDB, Redis, and DynamoDb. We also found evidence of the use of H2 (commit \href{https://github.com/apereo/cas/commit/403eb56}{403eb56}), Cassandra (pull request \href{https://github.com/apereo/cas/pull/2650}{\#2650}), InfluxDB (pull request \href{https://github.com/apereo/cas/pull/4092}{\#4092}), Hazelcast (pull request \href{https://github.com/apereo/cas/pull/1054}{\#1054}), MS SQL Server (commit \href{https://github.com/apereo/cas/commit/b15a7dc}{b15a7dc}), and Filestore (commit \href{https://github.com/apereo/cas/commit/ffe979c}{ffe979c}). However, no explicit reason for the use of these DBMSs was found. Since CAS is an authentication service, the support of many different DBMSs is expected. Couchbase's removal was likely driven by reliability issues related to errors like \emph{Could not store} and \emph{OperationTimeOut}, especially with concurrent access (pull request \href{https://github.com/apereo/cas/pull/371}{\#371}).

The CommaFeed project adopted Redis for caching. It was later removed since the need for caching ceased to exist in the project (commit \href{https://github.com/Athou/commafeed/commit/0446944}{0446944}). \new{Our heuristics found evidence of the use of PostgreSQL and MySQL, but they did not appear in our replacement rules because they were present since the }beginning of the project and were never replaced. The README of the project explicitly mentions support for four DBMSs: H2, PostgreSQL, MySQL, and MariaDB. We also found evidence of the use of MS SQL Server (issue \href{https://github.com/Athou/commafeed/pull/396}{\#396}). In this case, the replacement rule we found for H2 replacing Redis seems to be a coincidence, since H2 was not introduced with the specific aim of replacing Redis. \responsetoreviewer{rIIcXXI}{While Redis was, in fact, removed, H2, in this case, was introduced as a default DBMS to be used in small instances. Although H2 is also suitable for larger instances, the project supports other databases like PostgreSQL, MySQL, and MariaDB for more robust use cases.

}

For the JMeter project, we identified that MySQL (in this specific case, MariaDB) was replaced by PostgreSQL. This change occurred in Slice 3, in August 2002. At that time, the project did not use GitHub as its version control system -- instead, it used SVN. We went to the original SVN repository, but could not find that specific version there. The developers kept there only a later version from 2011. We searched for keywords in both repositories. While we did find evidence of usage of both MariaBD (issue \href{https://github.com/apache/jmeter/issues/5104}{\#5104}) and PostgreSQL (commit \href{https://github.com/apache/jmeter/commit/4fe05e3}{4fe05e3}), we found no documented reasons for the switch to PostgreSQL.

The VertaAI ModelDB project has implemented several changes regarding database support. It removed PostgreSQL (pull request \href{https://github.com/VertaAI/modeldb/pull/3485}{\#3485}), migrated from SQLite to MongoDB due to issues related to fast recovery from fails (issue \href{https://github.com/VertaAI/modeldb/issues/221}{\#221}), and worked on making the system compatible with MySQL (issue \href{https://github.com/VertaAI/modeldb/issues/274}{\#274}), MariaDB (pull request \href{https://github.com/VertaAI/modeldb/pull/2729}{\#2729}), and MS SQL Server (pull request \href{https://github.com/VertaAI/modeldb/pull/2265}{\#2265}). These adjustments highlight the evolution of ModelDB to better align with production environments, including improved compatibility with a range of database systems.

In the NGrinder project, Hazelcast (commit \href{https://github.com/naver/ngrinder/commit/c3aef8c}{c3aef8c}) was introduced as a solution for sharing data among clusters, reducing the load on the central database. Previously, Ehcache was used for distributed caching and data sharing between clustered controllers. The primary goal of this change was to improve data management efficiency in distributed environments by offloading repetitive read-and-write operations from the central database and distributing them through Hazelcast (pull request \href{https://github.com/naver/ngrinder/pull/365}{\#365}). This shift focuses on scalability and performance in environments with multiple controllers. However, we did not detect this migration since Ehcache is not on the list of DBMSs we searched for. In addition to the introduction of Hazelcast, we identified the removal of SQLite (commit \href{https://github.com/naver/ngrinder/commit/e950267}{e950267}), which occurred on Oct 20, 2012, due to its lack of support for altering columns.

In Apache NiFi, we found several removals. Support to Cassandra was removed in anticipation of a new implementation with updated drivers (commit \href{https://github.com/apache/nifi/commit/d9bcc8b}{d9bcc8b}). Couchbase (issue \href{https://issues.apache.org/jira/browse/NIFI-13150}{\#NIFI-13150}) was removed due to being outdated and unmaintained. Apache Ignite (issue \href{https://issues.apache.org/jira/browse/NIFI-11582}{\#NIFI-11582}) support was phased out due to outdated dependencies and configuration challenges. For PostgreSQL (issue \href{https://issues.apache.org/jira/browse/NIFI-9845}{\#NIFI-9845}), older versions were dropped, while PostgreSQL 14 is now supported. Hazelcast (issue \href{https://issues.apache.org/jira/browse/NIFI-7549}{\#NIFI-7549}) was added to enhance distributed caching. We notice that in this project, most of the replacement patterns we found are, in fact, the removal of old versions and additions of new versions of a DBMS. We found 37 replacement patterns, but lots of them fit the pattern remove DBMS $A$, add DBMS $B$, and then another rule adds DBMS $A$ back and removes DBMS $C$. We believe that in such cases, this simply indicates the code was updated to support a new version of DBMS $A$. The addition of DBMS $B$ and the removal of DBMS $C$ (which we also find in other rules) is probably the real replacement in this case. 

In SpotBugs, we found evidence of Oracle being used on several files, as well as evidence of usage of GoogleCloudDatastore (commit \href{https://github.com/spotbugs/spotbugs/commit/ff9a199}{ff9a199}), \new{but we found no documentation of the reason for the migration to Oracle.}

Apache Zeppelin is a web-based notebook for data analytics. Support for Redis (pull request \href{https://github.com/apache/zeppelin/pull/1497}{\#1497}) was introduced in Apache Zeppelin to enhance functionality for notebook users working with Redis-stored data. The Zeppelin project has undergone notable changes in its database support over time. Key developments include the removal of MySQL and MS SQL Server (pull request \href{https://github.com/apache/zeppelin/pull/211}{\#211}) drivers due to licensing conflicts, and the upgrade of the H2 database to improve security (pull request \href{https://github.com/apache/zeppelin/pull/4284}{\#4284}). Furthermore, the project removed Ignite (commit \href{https://github.com/apache/zeppelin/commit/6cf0252}{6cf0252}) and added support for Neo4j (pull request \href{https://github.com/apache/zeppelin/pull/1582}{\#1582}), emphasizing a shift towards more adaptable and secure database solutions. These changes reflect ongoing adjustments to improve performance, compatibility, and compliance with licensing standards.

One of the most significant drivers behind the switch from MySQL to MariaDB relates to licensing issues. While MySQL operates under a GPL license with an open-source exception, concerns over the potential risks of audits and compliance led to the adoption of MariaDB, which is distributed under the LGPL license. The LGPL license is perceived as more permissive and reduces legal and compliance-related risks in production environments. We identified that the switch was mentioned due to the license in the Zeppelin project. The switch also occurred in the ModelBD project. However, since our heuristics do not differentiate MySQL from MariaDB, we do not detect these migrations. 

For the Bisq project, our heuristics did not find evidence of the use of a DBMS. Our search in the repository code and issues also did not uncover any evidence of DBMS usage, so this is a true negative. 

In summary, our qualitative analysis of several cases shows that the decision to remove or replace a DBMS was directly tied to the specific needs of each project. Our findings indicate that the choice of DBMS in software projects is influenced by a combination of factors, including licensing, performance, scalability, and maintainability. %These results emphasize the need for careful consideration of technical, legal, and performance-related factors in the selection and management of DBMS in Java open-source software projects.

\section{Discussion \label{sec:discussion}} 

This section discusses the practical implications of our work for different stakeholders.

\paragraph{Software Engineers.} One key takeaway for software engineer practitioners regards the understanding of DBMS usage and stability. Popular DBMSs (RQ1) tend to have more learning resources and a larger community of users asking and answering questions, which makes the process of adopting the DBMS easier. DBMSs that have been in use for a long time (RQ2) or are more stable---replaced less frequently---(RQ3) should also be considered. For example, in Figure~\ref{fig:rq1}, we see that HyperSQL is a popular choice. However, frequent migration patterns, as described in RQ3 (Section~\ref{RQ3}), show that HyperSQL is often replaced by other DBMSs (Table \ref{tab:substitutions}), which means it may not be a good choice after all. By selecting a DBMS with a strong history of stability, engineers can avoid costly migrations in the later stages of their projects.

%Knowing about the high adoption of certain DBMSs can greatly aid in choosing which one to use. Popular DBMSs (RQ1) tend to have more learning resources and a larger community of users asking and answering questions, which makes the process of adopting the DBMS easier. Software Engineers can also use our results to factor the domains of their applications (RQ1), the stability of the DBMSs (RQ3, RQ5), and the synergies among different DBMSs (RQ4). 

%From another perspective, considering the DBMS that have been in use for a longer time (\textbf{RQ3}) or have been replaced the least (\textbf{RQ5}) can also be beneficial, as it indicates which DBMSs are more stable. Using a stable DBMS reduces potential barriers throughout the project and avoids costly DBMS migrations.

%
Additionally, knowing which DBMSs are commonly used together, as outlined in RQ4 (Section~\ref{RQ4}), can assist engineers in streamlining database integration processes, particularly in complex systems that may require multiple data models. For instance, our study reveals that combinations of DBMSs, such as Redis and MySQL, are often used together at different points in the project history, informing engineers on potential pairing options.
Finally, understanding which DBMSs are more commonly used in specific domains allows for adopting those that are likely more suited to the characteristics of the data.
We also found that the use of ORM is concentrated in a small percentage of files (RQ5). Hence, software engineers can use this information to estimate the effort and choose simpler ORM solutions. 
%
%Finally, knowing which DBMSs are often used together allows to identify which ones work well together and streamline integration efforts.

\paragraph{Educators.} 
Educators can leverage our findings to adjust their curricula and align with industry trends. For example, Figure \ref{fig:rq1} lists the most frequently used DBMSs, such as MySQL, PostgreSQL, H2, Oracle, and Redis, which would be important for students to learn to ensure their skills are applicable in the industry (Section \ref{RQ1} (RQ1)). On top of that, the same considerations we made for Software Engineers are valid. The more widely used a DBMS is, the more learning resources and applications are usually available. 

%The same considerations that Software Engineers have for choosing a DBMS are valid for educators: the availability of learning resources, domain of applications, and frequency of usage (RQ1) can aid the decisions on which DBMSs to teach and how to teach them.  Educators may use these characteristics both to choose the most frequently used DBMS or the least frequently used ones. %In the first case, the high availability of learning material reduces the of effort of preparing new material. In the second case, the effort will be higher, but it increases the potential of creating a unique material.

In a more general sense, the rise of non-relational DBMSs (especially in domains like \emph{software development} and \emph{infrastructure management}) suggests that educators should put a stronger emphasis on NoSQL in their curricula. Traditional database courses often focus heavily on relational models, but given the findings in RQ1, RQ2, and RQ4, courses should expand to include a balanced introduction to the different types of DBMSs, particularly NoSQL systems like Redis, Cassandra, and MongoDB.

In a general sense, database courses often teach about the usage of DBMSs in isolation, indicating which type of DMBS suits better specific situations. However, we found that different DBMSs are often used together and may be able to tackle a broader set of situations. With this information, educators can adapt their courses to present the possibilities and show how to integrate multi-model databases to cover a broader spectrum of solutions.

Finally, \textbf{RQ3} highlights the frequent occurrence of DBMS migrations, emphasizing the need for educators to prepare students for managing transitions and replacements. Educators could include hands-on projects that simulate database migrations and highlight practical scenarios from the study, such as the replacement of SQLite with H2.

%Finally, educators can also adapt their courses to prepare students for potential DBMS migrations in their future careers. We found the these migrations occur, and often multiple times per project (RQ5).

\paragraph{Researchers.} 
Researchers can benefit from the insights provided in this paper. Understanding the distribution of DBMSs (RQ1 and RQ2) allows researchers to focus their efforts on studying those that will have the most significant impact. 
Moreover, the methodology defined in this study can be reused in other contexts, offering a valuable framework for further research.

This work also raises a set of questions that researchers can investigate in future work. Since the use of ORMs is concentrated in specific files (RQ5), there is an opportunity to investigate how ORM files are organized within projects, such as whether a consistent architecture like MVC is used or if other design patterns emerge. There is also room to research the reasons why projects switch DBMSs (RQ3) -- whether due to technical challenges, policy changes, or social factors. Researchers can explore why certain DBMSs dominate specific domains (RQ1), the reasons behind particular combinations of DBMSs (RQ4), and the impact of switching DBMSs on project costs and outcomes (RQ3). More specifically, it would be interesting to understand how DBMS adoption impacts project success and sustainability metrics, such as code complexity, maintenance overhead, number of new contributors, or long-term stability.

%Moreover, they can propose and evaluate methods for facilitating these migrations. 
\responsetoreviewer{rIIcXXIII}{While the present study focuses on open-source projects, researchers could explore how DBMS trends are reflected in widely adopted technologies like AI/ML, blockchain, and IoT.} These domains may have unique database needs that differ from traditional software domains, and the adoption trends in these spaces could provide further insights into the future of DBMS technology.

\paragraph{DBMS Vendors.} This paper provides valuable insights into market dynamics, which can benefit DBMS vendors. Identifying that their DBMSs are being replaced by others over time (RQ3) can prompt them to investigate the underlying issues and take corrective actions to reverse this trend and increase customer retention. 

Additionally, vendors can benefit by developing migration tools that facilitate transitions from the more widely used DBMSs to their own offerings. Still, the common co-use of multiple DBMSs (RQ4) may provide opportunities for vendors to improve interoperability with other popular systems. For instance, ensuring seamless integration with widely used DBMSs such as Redis and MySQL could offer a competitive advantage by simplifying implementation for engineers working with diverse technology stacks.

%Finally, vendors can observe the trends on DBMS adoption, identify how the applications interact with the other DBMS (RQ2) and change their products to support different use cases and domains (RQ1).

\section{Threats to Validity}
\label{sec:validity}

As with any empirical work, our study has limitations due to design decisions and the nature of this research. In this section, we discuss the threats to the validity of our study. 

\paragraph{External.}
% External
Our corpus may not be representative of all Java open-source projects that are popular, mature, and active. We only considered open-source projects hosted on GitHub. Moreover, we avoided toy projects and inactive projects by filtering out projects with less than 1,000 stars, no pushes in the last 3 months (relative to September 2024), more than 10 contributors, and more than 1,000 commits in the main line of development. Despite these efforts, our filtering process might not cover all relevant projects, potentially limiting the generalizability of the results. 

%We collected metadata in March 2021 and cloned the repositories in September 2024. The state of the projects during this period may have changed, as projects could have become inactive or received new commits. Thus, the generalizability of the results over time is limited, and our analysis may not reflect the current state of DBMS usage.

% External
We used the DB-Engines ranking and the JRebel survey to select DBMSs and ORMs, respectively. Popular Java-specific DBMSs and ORMs might have been omitted from our analysis since we did not use Java popularity as a criterion. This threatens the generalizability of our findings to other tools or contexts. 

\paragraph{Internal.}
% Internal
Although we inspected the list of projects in our corpus to exclude irrelevant ones (e.g., toy projects, ORMs, or databases), there is a risk of misclassified projects in our corpus. Even though the analysis was conducted by two authors and revised by two others, this could affect the accuracy of our conclusions, as systematic errors may have influenced the corpus selection.

% Internal
In the analysis of DBMS co-occurrence, the presence of multiple DBMSs in a project’s repository may not necessarily indicate that they are used together. This could mean that the project was designed to support multiple DBMSs, where users can choose one at installation time. A deeper source code analysis would be needed to confirm actual co-usage, which could lead to incorrect conclusions about DBMS synergy across projects.

\paragraph{Construct.}

% Construct
\responsetoreviewer{rIIcIpI}{The domain of each project was determined with the aid of ChatGPT. Although we achieved a high Cohen's Kappa, there may be misclassified projects in our dataset. To diminish this threat, a large portion of our corpus was also manually classified. Since we conducted some analysis based on this classification, the results might change if the projects are misclassified.}

% Construct
\responsetoreviewer{rIIcIpII}{The heuristics we designed to detect DBMS connections and ORM usage may not capture all possible patterns, leading to false positives or negatives. Although we validated the heuristics on a subset of projects, the accuracy of our results might still be affected by undetected errors, threatening the validity of our conclusions. The effect of using incorrect heuristics would be the presence of false positives (when we detect something erroneously) and false negatives (when we don't detect a DBMS or ORM that is, in fact, used in the project) in our results. }

% Construct
\responsetoreviewer{rIIcIpIII}{We divided each project's history into slices of 100 commits. Although this decision accommodates different project sizes, it might not accurately reflect the complete timeline of DBMS usage and migrations. We acknowledge that from one snapshot to the other, DBMSs may have been added, replaced, or removed more than once. As a side effect, we may not have a complete record of DBMS usage and migrations in our analysis (false negatives). However, using a fixed number of slices instead of a fixed slice size may lead to projects with very large slices, which would increase the chances of false negatives in our analysis.
In summary, using a different slice size may impact the results we obtained. The smaller the slice size, the less chances of false negatives. Ideally, one should analyze every commit in the projects' histories, but this would be unfeasible since it would require lots of processing time.  }

% Construct
Our heuristics do not differentiate MySQL from MariaDB, PostgreSQL from CockroachDB, MS SQL Server from Microsoft Azure SQL Database, and Sybase Adaptive Server Enterprise from SAP Adaptive Server. Due to that fact, we do not detect possible migrations involving these pairs of DBMS. This is evidenced in our qualitative analysis (Section \ref{sec:qualitative}), where we detected migrations from MySQL to MariaDB that were not captured by our heuristics.

% Construct
The 50 popular DBMS of DB-Engines and the JRebel-based ORM selections might not perfectly represent the DBMSs and ORMs most relevant to Java projects, which could affect how well our operational definitions reflect the actual constructs of ``Java DBMS usage" and ``ORM adoption" in the field. The lack of Java popularity as a criterion introduces a potential mismatch between the theoretical construct and its operational definition in our study. However, using a pre-compiled list of Java DBMSs (or Java ORM) would possibly hide new insights, such as finding usage of a DBMS that is possibly out of the list (if such a list existed). We believe the results of our analysis can be used to build such a list. 

% Construct
The segmentation of each project’s history into slices of 100 commits does not represent equivalent time periods across projects of different ages, which may introduce bias in interpreting temporal trends. This threatens the construct validity of our study since trends over time may be misrepresented. Future analysis might delve deeper into the temporal dimensions of the evolution of the DBMSs to more accurately capture the progression and nuances of DBMS usage over time.

\paragraph{Conclusion.}
% Conclusion
To ensure the effectiveness of our heuristics, we performed a validation step with five projects by manually inspecting their websites, looking for false positives and negatives. However, the limited sample size for validation may reduce the confidence in our statistical conclusions. \new{To mitigate this threat, we made several adjustments in how we built our heuristics based on those results and reflected them in the remaining heuristics to guarantee a high level of precision and recall for the analyzed DBMSs and ORMs.}

\section{Related Work}
\label{relatedWork}

Undoubtedly, the work of \cite{Lyu_2017} is the most similar to ours, since they search for database usage in a corpus of projects. In their work, they studied Android apps to find out how mobile apps use local DBMSs. Local DBMSs are those that are installed directly on mobile devices. To conduct this research, they used a corpus composed of 1,000 apps on the Google Play app store, sampled from 34 categories. They found that Android apps use eleven different DBMSs: SQLite, Oracle, Realm, Couchbase, MongoDB, SnappyDB, LevelDB, Waze, InterBase, UltraLite, and UnQLite. To discover which database each project used, they used Soot to analyze the bytecode and searched for database API package names (e.g., \emph{android.database.sqlite}). They found that a large portion of the apps (40.8\%) do not use any DBMS. For those that use local DBMSs, SQLite is the most common (93.1\%). Interestingly, Oracle appears next with 4.6\%. While they focus on Android mobile apps, our focus is on a broader corpus of open-source Java applications. Besides, their focus is on a single snapshot of the project, while in our work, we also analyze the history of DBMS usage in our corpus: how the usage trends evolve and which DBMS is replaced during the projects' history. 

Few studies focus on inspecting applications to gain knowledge related to how DBMSs are used in practice -- which ORMs are used, and how queries are performed \citep{Goeminne_2015,Linares_2015, Lyu_2017,yan2017,yang2018}. %We address these works in detail in the next paragraphs.
After finding out that the most popular DBMS in mobile Android apps is SQLite, \cite{Lyu_2017} focused on apps that use SQLite to investigate issues such as how queries are performed, misuse of the database access APIs that could lead to SQL injection attacks, and performance issues related to using database transactions in loops. 
\cite{Linares_2015} studied 3,113 Java projects from GitHub to understand how they document database usage. The query they performed to select the projects included only projects that used JDBC to access the database. They also surveyed 147 developers of such projects. They concluded that most database access methods (77\%) are not documented.

The works of \cite{Goeminne_2015,yan2017,yang2018} specifically focus on ORM and how the DBMS is accessed by the applications. 
\cite{Goeminne_2015} studied 3,707 Java projects from GitHub to find out how they use database frameworks to access relational databases. To determine the usage of a framework, they looked at the imports and configuration files. Their analysis focuses on five database frameworks (JDBC, Spring, JPA, Vaadin-GWT, and Hibernate). For those, they studied the frameworks' co-occurrence and survival rates. Although JDBC does not provide abstractions of the database schema since it is not an ORM framework, it is the most used framework (it appears in 2,271 projects of their corpus). \responsetoreviewer{rIIcXXV}{While they focus on whether certain database frameworks co-occur frequently, and whether some database frameworks get replaced over time by others, they do not perform that same kind of analysis for the DBMSs. Thus, their work is complementary to ours.

} 

Since ORM frameworks such as Hibernate, SQLAlchemy, and others hide query generation and execution, they may become the reason for database performance degradation in applications that deal with large amounts of data and require quick response times. In that context, \cite{yang2018} extend their previous work \citep{yan2017} to find performance anti-patterns in real-world web applications that use ORM. They profile the latest version of twelve open-source web applications and study their bug-tracking systems. Then, they classified the causes of inefficiency into three categories: ORM API misuses, database design, and application design. About half of these performance problems were due to ORM API misuses. In total, they found nine anti-patterns that affect the performance of web applications. To show their importance, they applied corrections to the performance issues and obtained median speed-ups of 2$\times$ (with a maximum of 39$\times$). Although they study ORM, the focus of their study is completely different from ours. 

We also analyzed the literature that studies coevolution of DBMS, some looking at schema evolution, some looking at code and schema coevolution \citep{qiu2013empirical,Goeminne_2014,scherzinger2020empirical,dimolikas2020study,vassiliadis2021profiles}. Each of these works brings a different contribution regarding both the methods they applied and the findings about the influence of source code changes in DBMS changes and vice versa. However, none of them study which DBMSs are used throughout the projects' history.

\cite{qiu2013empirical} conducted an empirical study on ten applications using popular DBMSs to understand the coevolution between database schemas and the source code of these applications. To this end, they analyze the entire history of these changes over time, concluding that schemes often evolve and involve many changes. These changes lead to the need to co-evolve the source code. Furthermore, they observed that each atomic schema change generates modifications in 10 to 100 source lines. Regarding a valid database revision, the number of changed source lines grows exponentially from about 100 to 1,000, reaffirming the influence of evolving database schemas on the source code of applications.

\cite{Goeminne_2014} conducted a study to analyze coevolution between source code and related activities in a large, data-intensive open-source system, the OSCAR application repository\footnote{\label{Oscar-Aplicattion}https://github.com/scoophealth/oscar}. They concluded that there is a strong coevolution between source code file changes and database-related file changes. However, the changes in the database technology used over the project's lifetime do not significantly impact the source code's evolution. They also observed that all contributors changed the source code and the database-related files, indicating that responsibilities were not distinctly separated between the different contributors.

\cite{scherzinger2020empirical} analyzed the coevolution of NoSQL DBMS schemas by investigating entity class declarations in the commit history of ten projects. These projects were selected from a pool of 1,200 open-source Java projects hosted on GitHub, specifically focusing on those with the largest DBMS schemas. By tracking the growth of the schemas and the nature of their changes over these projects' history, they found that denormalization, a technique used to improve query performance in NoSQL databases, was common in these schemas. They also found evidence of evolutionary changes in all the analyzed projects. Moreover, they noted that the turnover rate in these schemas is higher than in studies dealing with the evolution of relational schemas.

\cite{dimolikas2020study} conducted a study on the evolution of DMBS schema tables, with a focus on the foreign key structures of the related tables. They performed historical analysis on six relational schemas to extract information about table births, table deaths, intra-table updates, and their foreign key relationships. Then, they introduced a concise taxonomy of topological graph patterns through their analysis to characterize table relationships. They discovered that the topological complexity hierarchy significantly influences the tables' behavior during evolution. Therefore, evolutionary behavior depends on this hierarchy.

\cite{vassiliadis2021profiles} conducted a historical analysis on relational DBMS sche\-mas across 195 open-source projects to understand how these schemas evolve. They analyzed the frequency of the changes in the project's commit history and identified schema families with similar evolution characteristics. They used this to define schema evolution patterns as well as evolution measures. Given this, they discovered that schema evolution is absent in most projects, except those with active schema maintenance profiles, refuting the belief that schema evolution is extensive.

\begin{table}[t]
\centering
\begin{footnotesize}
\caption{Comparative analysis between related work and this study. We abbreviate the Related Work names due to space constraints: (Q)iu et al. (13), (G)oeminne et al.  (14), (G)Goeminne et al. (15), (L)inares-Vásquez et al. (15), (L)yu et al. (17), (Y)ang et al. (18), (S)cherzinger and Sidortschuck (20), (D)imolikas et al. (20), (V)assiliadis (21) and (M)aximino et al. (2024), this work.}
\label{tab:relatedwork}
\begin{tabular}{lcccccccccc} 
%\hline
%\multicolumn{1}{c}{\textbf{Analysis}}         & \textbf{Qiu et al.  (2013)}                  & \textbf{Goeminne et al. (2014)}             & \textbf{Scherzinger and Sidortschuck (2020)} & \textbf{Dimolikas et al. (2020)}            & \textbf{Vassiliadis (2021)}                 & \textbf{Maximino et al. (2023) This study}   \\ 
\multicolumn{1}{c}{\textbf{Analysis}}         & \rotatebox{90}{\textbf{Q13}}& \rotatebox{90}{\textbf{G14}}&  \rotatebox{90}{\textbf{G15}} & \rotatebox{90}{\textbf{L15}}&\rotatebox{90}{\textbf{L17}}& \rotatebox{90}{\textbf{Y18}}&\rotatebox{90}{\textbf{S20}}& \rotatebox{90}{\textbf{D20}}& \rotatebox{90}{\textbf{V21}}& \rotatebox{90}{\textbf{M24}}\\ 
\hline
\textbf{DBMS adoption}                        & \textcolor{red}{X}                          & \textcolor{red}{X}                          &   \textcolor{red}{X}& \textcolor{red}{X}&\textcolor{blue}\checkmark& \textcolor{red}{X}&\textcolor{red}{X}                           & \textcolor{red}{X}                          & \textcolor{red}{X}                          & \textcolor{blue}\checkmark \\
\textbf{Relational DBMS}                      & \textcolor{blue}\checkmark & \textcolor{blue}\checkmark &   \textcolor{red}{X}& \textcolor{red}{X}&\textcolor{blue}\checkmark& \textcolor{red}{X}&\textcolor{red}{X}                           & \textcolor{blue}\checkmark & \textcolor{blue}\checkmark & \textcolor{blue}\checkmark  \\
\textbf{Non-relational DBMS}                  & \textcolor{red}{X}                          & \textcolor{red}{X}                          &   \textcolor{red}{X}& \textcolor{red}{X}&\textcolor{blue}\checkmark& \textcolor{red}{X}&\textcolor{blue}\checkmark  & \textcolor{red}{X}                          & \textcolor{red}{X}                          & \textcolor{blue}\checkmark  \\
\textbf{DBMS interaction (ORM)}& \textcolor{red}{X}& \textcolor{red}{X}& \textcolor{blue}\checkmark&  \textcolor{blue}\checkmark&\textcolor{red}{X}& \textcolor{blue}\checkmark& \textcolor{red}{X}& \textcolor{red}{X}& \textcolor{red}{X}&\textcolor{blue}\checkmark\\
\textbf{Historical analysis}                  & \textcolor{blue}\checkmark & \textcolor{blue}\checkmark &   \textcolor{blue}\checkmark& \textcolor{red}{X}&\textcolor{red}{X}& \textcolor{red}{X} &\textcolor{blue}\checkmark  & \textcolor{blue}\checkmark & \textcolor{blue}\checkmark & \textcolor{blue}\checkmark  \\
\textbf{Coevolution}                     & \textcolor{blue}\checkmark & \textcolor{blue}\checkmark &   \textcolor{red}{X}& \textcolor{red}{X}&\textcolor{red}{X}& \textcolor{red}{X}&\textcolor{blue}\checkmark  & \textcolor{blue}\checkmark & \textcolor{blue}\checkmark & \textcolor{blue}\checkmark  \\
\textbf{DBMS schemas}  & \textcolor{blue}\checkmark & \textcolor{blue}\checkmark &   \textcolor{red}{X}& \textcolor{red}{X}&\textcolor{blue}\checkmark& \textcolor{red}{X}&\textcolor{blue}\checkmark  & \textcolor{blue}\checkmark & \textcolor{blue}\checkmark &  \textcolor{red}{X}\\
\textbf{Size of corpus (\textgreater{}=100)} & \textcolor{red}{X}                          & \textcolor{red}{X}                          &   \textcolor{blue}\checkmark& \textcolor{blue}\checkmark&\textcolor{blue}\checkmark& \textcolor{red}{X}&\textcolor{red}{X}                           & \textcolor{red}{X}                          & \textcolor{blue}\checkmark & \textcolor{blue}\checkmark  \\
\hline
\end{tabular}
\end{footnotesize}

\end{table}

Table \ref{tab:relatedwork} shows a comparison of the related work. We include our work (M24) in the table as a comparison reference. The most similar work to ours is that of \cite{Lyu_2017}. It is the only one that investigates which DBMSs are \textbf{adopted }in a set of projects, and the only one (besides ours) that considers both data models (\textbf{relational} and \textbf{non-relational}) in their work. Their target, however, is much more specific than ours -- they study the adoption of DBMSs in mobile Android apps while we investigate a broader set of open-source Java projects. Also, they focus on a set of 11 DBMSs while we search for usage clues of 50 DBMSs in the source code of our corpus. While in their context, the most popular DBMS is SQLite (which is natural, since SQLite is light and adequate for mobile apps), in our case, the most popular DBMS is MySQL.

Besides our work, three other study DBMS usage through \textbf{ORM} \citep{Goeminne_2015, Linares_2015,yang2018}. Two of them focus on a single ORM framework: JDBC \citep{Linares_2015} and Rails \citep{yan2017,yang2018}. \cite{Goeminne_2015} examines a larger set of frameworks (19), but found evidence of wide usage of only 5, including JDBC. In our work, we consider JDBC to be a database access library instead of an ORM framework, and this is why JDBC is not included in our ORM analysis. Our results differ from their findings. While the most popular ORM framework (not including JDBC) in their corpus was Spring, in our case, it was MyBatis. It is worth noting that their corpus is larger than ours (13,307 Java projects) and was obtained using a start set proposed by \cite{mil2013} and excluding the projects that were not available in March 2015. Their corpus selection process did not consider any constraint (e.g., start, number of commits, contributors, etc.) except for the projects not being forks. Thus, their corpus probably contains many small, non-popular, and personal projects, which may explain the different results we obtained in our analysis. Besides, their work has been conducted almost a decade ago. 

Most of the related work performs a \textbf{historical} investigation to study the \textbf{coevolution} of source code changes with DBMS changes \citep{qiu2013empirical,Goeminne_2014,scherzinger2020empirical,dimolikas2020study,vassiliadis2021profiles}, and in that sense, they are complementary to our study. Differently from them, we look at the coevolution by detecting DBMS replacement patterns. The remaining study that performs \textbf{historical} analysis \citep{Goeminne_2015} focuses on ORM. The authors use a different technique: statistical survival analysis. It is not clear, however, how the death events (finding out when a framework stopped being used) were detected. 

Differently from our work, a set of works analyzes changes on the \textbf{database schema} \citep{qiu2013empirical,Goeminne_2014,Lyu_2017,scherzinger2020empirical,dimolikas2020study,vassiliadis2021profiles}. In that sense, they are complementary to our study.

Finally, only four works, besides ours, analyze a \textbf{corpus size} of more than 100 projects. \cite{Goeminne_2015} investigates 13,307 Java projects, but the selection criteria of the corpus was very open. This means it may contain toys and personal projects. \cite{Linares_2015} use a corpus of 3,113 projects, but their focus is completely different from ours: they investigate database documentation on the source code. \cite{Lyu_2017} investigates 1,000 mobile apps, which have a different nature than the more general projects we consider in our corpus. Finally, \cite{vassiliadis2021profiles} study database schema evolution on a corpus of 195 projects. All other studies use a corpus of twelve or fewer projects (12, 10, 10, 6, and 1, exactly). When we consider that we look at the history of the projects in several snapshots (slices), we go from a corpus of \nvarc{rq2_projects}{362} projects to a corpus of \nvar{total_slices} snapshots (or project versions), which, when compared to what is used in most the literature we discuss here, we consider to be significant. Note that from the works that conduct historical analysis only \cite{Goeminne_2015,vassiliadis2021profiles} and this work use a corpus larger than 100 projects. The corpus used by \cite{vassiliadis2021profiles} is smaller than ours, and  \cite{Goeminne_2015} does not clarify if they analyze every version of the projects on their corpus.

\section{Conclusion}
\label{final_considerations}

This paper presented a comprehensive investigation of the adoption of DBMSs over a corpus of \nvarc{rq2_projects}{362} Open-Source Java projects. First, we investigate DBMS adoption and usage using the latest snapshot of each project in our corpus. Then, we conducted a historical analysis, considering the history of the projects. To do so, we sliced the project history into slices of 100 commits and applied heuristics to indicate the presence of each DBMS in each slice. This information was processed to analyze the stability, migration patterns, and synergy between DBMSs. 

We could observe that MySQL, H2, and PostgreSQL are among the three most used relational DBMSs, while Redis and MongoDB are the most used non-relational DBMSs. Half of the projects adopted both relational and non-relational databases. This co-occurrence of models was especially prevalent in projects of the Data Management domain.  %When comparing the results we obtained when analyzing only the last version of the project (RQ1) with the results we obtained analyzing the project history (RQ2), we notice that 32 projects that use a DBMS in their history do not show signs of DBMS usage in its last version. This may mean that they stopped using DBMS completely. 

The concomitant use of DBMSs grows as the projects mature. In the first slice of the project, we observed pair combinations of 10 DBMSs, prevailing combinations among relational DBMSs (e.g., MySQL and PostgreSQL in 21 projects, HyperSQL and MySQL in 20, and MySQL and H2 in 17). In the middle of the project history, we observed combinations of 23 DBMSs. At this point, we start to see combinations of relational and non-relational DBMSs (e.g., Redis and PostgreSQL appear in 24 projects, MongoDB and H2 in 22 projects, and Cassandra and MySQL in 19 projects). In the last slice of the project history, the number of DBMS pair combinations increased to 31. At that point, combinations with MySQL and PostgreSQL were still popular, but combinations involving Redis became more frequent. 

Additionally, we discovered 296 patterns that indicate that DBMSs were replaced in projects, but only 18 of these patterns occur in more than three projects. The most frequent replacement occurred in only eight projects, with HyperSQL being replaced by Redis. \new{HyperSQL and HBase were also the DBMSs that underwent the most replacements: each was replaced by 18 DBMSs in 19 and 5 projects, respectively. Redis was the DBMS that replaced HyperSQL the most. HyperSQL was the DBMS that replaced HBase the most.} Furthermore, 52.4\% of the replacements occur between DBMSs of the same model, of which only 12.5\% is non-relational. We observed that, although DBMS replacements are not so frequent, they occur. We suppose the low frequency in the DBMS replacements may be directly related to the current tendency to use more than one DBMS together. According to \cite{sahatqija2018comparison}, one possible explanation is that non-relational DBMSs were not created to substitute the relational ones but to complement them. This diminishes the need to migrate from one DBMS to another since two DBMSs of distinct models can be used concurrently. 

Our findings offer strategic insights for organizations considering migration between DBMSs---understanding the prevalent frequencies and patterns of DBMS adoption can significantly inform such decisions. For professionals aspiring to specialize in the area, understanding how DBMSs are adopted can shape their educational and career paths. Educational institutions and training programs can utilize our results to refine their curriculum, emphasizing the most extensively adopted DBMSs and prevalent migration scenarios. Finally, our findings can guide projects seeking to enhance their integration and compatibility options, as well as DBMS tool developers, directing their efforts to align with what happens in practice.

We envision some interesting future work. One of them is confirming whether two DBMSs are, in fact, used together by investigating which entities are stored in each DBMS. Using more than one DBMS in a given project may mean the project can work with several DBMSs, but the user chooses a single one at installation time. To make sure more than one DBMS is used, we would need a deeper analysis of the source code. Another interesting study would be to find out why certain DBMSs are used together through qualitative analysis, for instance, by conducting interviews with the developers of the projects. 

We also plan to extend our analysis to explore the different characteristics of the DBMSs. For example, we can investigate the correlation between the scale of projects and the type of database management systems (DBMS) they utilize \new{(e.g., proprietary versus open source)} or analyze the popularity trends of lightweight versus heavy DBMSs across different project categories and their respective use cases.
Finally, we plan to extend this analysis to other programming languages and specific domains to determine how they influence the DBMS choice.

\paragraph{Acknowledgements.}
The authors would like to thank the National Science Foundation (NSF) grants 2247929, 2303042, 2303612, and 2303612; CNPq grants 305020/ 2019-6, 311955/2020-7, and 309410/2023-1; CNPq/MCTI/FNDCT grant 408812/2021-4;  MCTIC/CGI/FAPESP grant 2021/06662-1; and FAPERJ grants E26/201.038/2021 and E-26/210.478/2024, Fundação Araucária grants PRD2023361000043 for the financial support.

\paragraph{Statements and Declarations}
This paper has immensely benefited from the comments and suggestions of the three anonymous reviewers, to whom we are deeply thankful. We also acknowledge the use of Grammarly and ChatGPT 4.o for the improvement of spelling, grammar, vocabulary, and style of the text. We also utilized ChatGPT 4.o to speed up the writing of Python code and to classify projects, as mentioned earlier. All suggestions were carefully examined, tested, and often corrected by us, whereby we take full responsibility for the form and content of the paper.

\paragraph{Competing Interests} The authors declare that there are no financial or non-financial interests that are directly or indirectly related to this work.

\paragraph{Data Availability} All data and code used in our analysis are publicly available in our GitHub repository \url{https://github.com/gems-uff/db-mining}.

~

\bibliographystyle{spbasic}
\bibliography{bibliography}

%\iffalse
\iftrue

\newenvironment{myquote}%
  {\list{}{
    \leftmargin=0.5in%
    \rightmargin=0in%
  }\item[]}%
  {\endlist}

\newcommand{\reviewer}[4][R]{\textbf{#1#2.#3:} \itshape #4 \upshape}
\newcommand{\answer}[4][R]{\textbf{Answer to #1#2.#3:}\xspace#4\vspace{4mm}}
\newcommand{\answerquote}[1]{\begin{myquote}\setlength{\parskip}{1.5mm}{\fontsize{9pt}{9pt}\selectfont\color{blue}{#1}}\end{myquote}}
\newcommand{\revref}[3][R]{\textbf{#1#2.#3}}
\newcommand{\revsection}[2][\normalsize]{\vspace{3mm}\hrulefill\hspace{2mm}{#1\uppercase{#2}}\hspace{2mm}\hrulefill\vspace{3mm}}

\renewcommand{\responsetoreviewer}[3][black]{#3}

\clearpage
\pagestyle{empty}
\setlength{\parindent}{0cm}
\setlength{\parskip}{1.5mm}
\newgeometry{margin=1in}
\normalsize

\begin{center}\LARGE\bfseries
Analyzing the Adoption of Database Management Systems Throughout the History of Open Source Projects

Response to Reviewers\end{center}
\raggedbottom

Dear Editor(s),\\

We would like to thank the reviewers for their valuable work. We have made the changes requested in this Minor Review. Below we detail how we addressed each of the comments. 

Kind regards, \\

Camila A. Paiva, Raquel Maximino, Frederico Paiva, Rafael Accetta Vieira, Nicole Espanha, João Felipe Pimentel, Igor Wiese, Marco Aurélio Gerosa, Igor Steinmacher, Leonardo Murta, Vanessa Braganholo.

\revsection[\large \bfseries]{Answers to Reviewer R2}

\reviewer{2}{1}{I am somewhat surprised by the new findings obtained for some RQs, which differ from the findings of the paper's 1st version. The most significant finding changes are for RQ3 (Which DBMSs are frequently replaced by others?), RQ4 (Which DBMSs are often used together?), and RQ5 (How do applications interact with the DBMS?). I understand these new findings are explained by the new slicing approach, the new set of considered DBs, the new categorization of projects, etc. The paper discusses the threats to validity associated with the research methodology, but I am wondering if the authors can comment on the impact of such threats on the results. I mean, we know that some methodology decisions, if changed, can lead to different results, but the question is \textbf{how much they can change the results}. I suggest trying to discuss the impact on the results of each validity threat. This is very important for the reader because, as the new discussion section mentions, the results could
benefit multiple people: if the results are misleading, it could have a huge impact on the reader that guides their actions based on the paper results (e.g., teaching future software engineers about most potentially misleading common databases, co-occurring DBs, DB replacement patterns, etc.)

}

\answer{2}{1}{There are indeed several aspects that could influence the results. For some types of threats, such as the ones related to the corpus selection and classification, as well as the choice of using DB-Engines, there was already a discussion in the threats to validity section. For the remaining ones, we added a discussion as requested. 

\answerquote{\rIIcIpI}

\answerquote{\rIIcIpII}

\answerquote{\rIIcIpIII}

}

\reviewer{2}{2}{The qualitative analysis is very informative, but I would suggest providing more details about it. How thorough was the analysis? how did the methodology mitigate interpretation bias by the researcher(s) who did the analysis?}

\answer{2}{2}{Two authors used GitHub's Code Search feature to search for the DBMS names we identified in RQ1. Once they found mentions of the addition, removal, or migration of a DBMS, they discussed with each other and reviewed each other's results. Finally, another author validated the findings by inspecting the mentions and the results to ensure there was no bias in the process. We changed the paragraph to better describe the process:
\answerquote{\RIICII}
}

\reviewer{2}{3}{Also, the qualitative analysis found interesting reasons for replacements, introductions, and removals of DBMs... how many patterns are consistent with these findings and how many are not? I see some explanations about this for some projects, but it would be great to provide details about pattern consistency for each project, when possible.}

\answer{2}{3}{%ToDo
We have addressed this request by including the identified patterns and those that were not found in Section \ref{sec:qualitative}, which we reproduce below. This section provides a detailed analysis of pattern consistency for each project, highlighting the alignment between our qualitative findings and the observed patterns.
\answerquote{\rIIcIII}
\answerquote{
    Table~\ref{tab:qualitative}. \rIIcIIItIpI
    
    \rIIcIIItIpII}
}

\reviewer{2}{4}{Since there are a lot of results and findings for each RQ, and the boxes that summarize the answers for RQ may not include all the relevant results/findings, I would suggest adding a table showing the most important and specific findings for each RQ. My suggestion is grounded by the potential need of the reader to quickly know the most common databases, co-occurring DBs, DBs that are replaced, etc. In this way, the table would act as a reference guide that, for example, educators and other roles can quickly check. This can also be added to the github repository.}

\answer{2}{4}{
We have addressed this request by incorporating a table that summarizes the most important and specific findings for each research question. These tables highlight key aspects such as the most common databases, co-occurring DBs, and database replacements, serving as a reference guide for educators and other stakeholders. Additionally, we have included the complete list in our GitHub repository for easy access.
\answerquote{\rIIcIVpI}

\answerquote{
    Table~\ref{tab:rq_findings}. \rIIcIVtIpI
    
    \rIIcIVtIpII

%    Table~\ref{tab:rq2_findings}. \rIIcIVtIIpI
    
%    \rIIcIVtIIpII

%    Table~\ref{tab:rq3_findings}. \rIIcIVtIIIpI
    
%    \rIIcIVtIIIpII

%    Table~\ref{tab:rq4_findingsI}. \rIIcIVtIVpI
    
%    \rIIcIVtIVpII

%    Table~\ref{tab:rq4_findingsII}. \rIIcIVtVpI
    
%    \rIIcIVtVpII

%    Table~\ref{tab:rq5_findings}. \rIIcIVtVIpI
    
%    \rIIcIVtVIpII
    
}
}

\reviewer{2}{5}{``proprietary versus free'' --> ``proprietary versus open source''}

\answer{2}{5}{We have made the change on the conclusion as requested. 
%\answerquote{\RIICV}
}

\reviewer{2}{6}{``This not only simplifies the development process but also ensures that changes in the database layer are seamlessly propagated to the application layer, reducing the likelihood of errors and improving maintainability'' --> propagation of changes can go both ways}

\answer{2}{6}{We changed the text to indicate that the propagation of changes can go both ways.

\answerquote{\RIICVI}}

\reviewer{2}{7}{``Using sequential pattern mining, we uncovered recurring sequences in which certain DBMSs are replaced by others, either concomitantly or within a transition interval.'' --> ``either concomitantly or within a transition interval'' is not clear}

\answer{2}{7}{We rephrased the sentence to make it clear.
\answerquote{\RIICVII}
}

\reviewer{2}{8}{``revealing polyglot persistence'' --> polyglot might not be the right term to use here}

\answer{2}{8}{According to~\cite{fowler2011polyglot}, polyglot persistence occurs when a variety of different storage technologies is used for different kinds of data within the same application or enterprise. We added a citation to the term. %ToDo
\answerquote{\RIICVIII}
}

\reviewer{2}{9}{``Finally, after gaining a deeper understanding of the adoption of DBMSs in the projects, we investigated in detail how applications interact with these DBMSs (RQ5)'' --> ``…interact with these DBMSs via Object-Relational Mapping (RQ5)''}

\answer{2}{9}{We also investigated how applications interact with DBMSs through direct SQL queries -- in addition to ORM. We updated the text accordingly.

\answerquote{\rIIcIX}
}

\reviewer{2}{10}{``However, towards the end of the data collection, MySQL, PostgreSQL, H2, and Oracle were more frequently found together, …'' --> ``However, towards the end of the project lifespan as analyzed by our methodology, …''}

\answer{2}{10}{As requested, we changed it to ``However, towards the end of \new{the project \textbf{history} as analyzed by our methodology, …}''. 
}

\reviewer{2}{11}{``with EclipseLink requiring the fewest files to adopt in a project.'' --> ``… to adopt it…''}

\answer{2}{11}{We adjusted the text to correct the grammar.
}

\reviewer{2}{12}{Clarify what stability means when explaining RQ2 in section ``3.1 Research Questions''}

\answer{2}{12}{We changed the description of RQ2 to clarify the meaning of stability in this context.

\answerquote{\rIIcXII}
}

\reviewer{2}{13}{``We consider a given project to adopt a certain DBMS when at least one of the regular expressions of its respective heuristics is found on that project source code.'' --> ``… least one of the regular expressions of its respective heuristics produces one or more hits on the project's source code''}

\answer{2}{13}{We updated the text according to your suggestion.}

\reviewer{2}{14}{Figure 4 (and possibly other figures): it is good that it shows the \% (y-axis) but also show counts for each project}

\answer{2}{14}{We added both the count and the \% to several figures.

\answerquote{
    \resizebox{0.92\textwidth}{!}{\rIIcXIVrqIpI}

    Fig.~\ref{fig:rq1} \rIIcXIVrqIpII

    \resizebox{0.92\textwidth}{!}{\rIIcXIVprojectsperDBpI}

    Fig.~\ref{fig:projectsperDB} \rIIcXIVprojectsperDBpII
}
}

\reviewer{2}{15}{Some figures (Figure 8 and others) are hard to read: I need to zoom in to understand their text. Increase the font of the figures and if the text (e.g., axis tick labels) get too big, consider using abbreviations or better placement}

\answer{2}{15}{We increased the font of Figures~\ref{fig:rq1}, \ref{fig:projectsperDB}, (see \revref{2}{14}), \ref{fig:domain_dbmodels} and \ref{fig:keptandremoved}.

\answerquote{

    \resizebox{0.92\textwidth}{!}{\rIIcXVdomaindbmodelspI}

    Fig.~\ref{fig:domain_dbmodels} \rIIcXVdomaindbmodelspII

    \resizebox{0.92\textwidth}{!}{\rIIcXVinsertionsandremovalspI}

    Fig.~\ref{fig:keptandremoved} \rIIcXVinsertionsandremovalspII
}
}

\reviewer{2}{16}{Figure 15 (and possibly other figures): use the counts on top of each bar (it was like this before the revision)}

\answer{2}{16}{We added both the count and the \% to Figure~\ref{fig:fig_rq2.1}.

\answerquote{

    \resizebox{0.6\textwidth}{!}{\rIIcXVIrqIIdIpI}

    Fig.~\ref{fig:fig_rq2.1} \rIIcXVIrqIIdIpII
}
}

\reviewer{2}{17}{Some figures are black and while now, but before they were colorful. I like them colorful more.}

\answer{2}{17}{We changed the palette of some figures to make them colorful again.

\answerquote{
    \resizebox{0.6\textwidth}{!}{\rIIcXVIIfigIpI}

    Fig.~\ref{fig:heat_onlyrules_v1} \rIIcXVIIfigIpII

    \resizebox{0.6\textwidth}{!}{\rIIcXVIIfigIIpI}

    Fig.~\ref{fig:heat_onlyrules_v5} \rIIcXVIIfigIIpII

     \resizebox{0.6\textwidth}{!}{\rIIcXVIIfigIIIpI}

    Fig.~\ref{fig:heat_v10} \rIIcXVIIfigIIIpII
}
}

\reviewer{2}{18}{``we conducted a keyword search'' --> how? Via GitHub's search engine?}

\answer{2}{18}{As stated in \revref{2}{2}, we used GitHub's Code Search. We updated the text to better describe the methodology.
}

\reviewer{2}{19}{In the qualitative analysis section, there are references to commits, issues, pull requests, and more. I strongly suggest adding the web link to each of these referenced documents.}

\answer{2}{19}{We added the links as suggested. Each commit number, pull request, and issue is now clickable in the PDF and redirects to the corresponding web address. We provide an example of the change below. The change was consistently made throughout the Qualitative Analysis Section.  

\answerquote{\rIIcXIX}

}

\reviewer{2}{20}{``... but they did not appear in our replacement rules since they were present since the ... '' --> ``... replacement rules because they were present since ...''}

\answer{2}{20}{We changed the text as suggested. 
}

\reviewer{2}{21}{``We also found evidence of the use of MS SQL Server (issue \#396).
In this case, the replacement rule we found for H2 replacing Redis seems to be a coincidence, since H2 was not introduced with the specific aim of replacing Redis. H2, in this case, was introduced as a default DBMS to be used in small instances.'' --> was Redis removed? was it used for larger instances?
}

\answer{2}{21}{Redis was, in fact, removed, and the project has other DBMS options for larger instances. We changed the text to make this clearer. 

\answerquote{\rIIcXXI}

}

\reviewer{2}{22}{``…but no documentation of the reason for the migration to Oracle.'' --> ``… but we found no documentation of the reason for the migration to Oracle.''
}

\answer{2}{22}{We made the change as requested. 
}

\reviewer{2}{23}{``how DBMS trends manifest in emerging technologies like AI/ML, blockchain, and IoT.'' --> I don't think these technologies are emerging anymore. 
}

\answer{2}{23}{You are completely right. We changed the sentence to reflect that. 

\answerquote{\rIIcXXIII}
}

\reviewer{2}{24}{``To reduce this threat, we made several adjustments in how we built our heuristics based'' --> ``To mitigate this threat, we made …''
}

\answer{2}{24}{We altered the text as suggested. 
}

\reviewer{2}{25}{``While they focus on whether certain database frameworks co-occur frequently, and whether some database frameworks get replaced over time by others, they do not perform the same kind of analysis for the DBMSs themselves. Thus, their work is complementary to ours.'' --> I don't understand this: ``they do not perform the same kind of analysis for the DBMSs themselves.'' Clarify.
}

\answer{2}{25}{We rewrote the sentence to make it clearer. 

\answerquote{\rIIcXXV}
}

\fi

\end{document}